\documentclass[aps,prb,twocolumn,citeautoscript, superscriptaddress]{revtex4-1}
\usepackage{graphicx}
\usepackage{mathrsfs}
\usepackage{amsmath}
\usepackage{amssymb}
\usepackage{amsbsy}
\usepackage{color}
\usepackage[nice]{nicefrac}
\usepackage{float}
\usepackage{lipsum}
\usepackage[a4paper,lmargin=2cm,rmargin=1.2cm,tmargin=2.2cm,bmargin=2.2cm]{geometry}
\graphicspath{{./figures/}}
\usepackage{subcaption}
\usepackage{mwe}
\usepackage{tikz}
\usepackage{braket}

\usepackage{color}
\definecolor{darkred}{rgb}{0.6,0.,0.}
\definecolor{darkgreen}{rgb}{0.,0.5,0.}
\definecolor{darkblue}{rgb}{0.,0.,0.6}

\usepackage[
breaklinks=true,
colorlinks=true,
linkcolor=darkred,
citecolor=darkgreen,
urlcolor=darkblue]{hyperref}

\usepackage[inactive,tightpage]{preview}
\PreviewMacro[*[[!]{\inputgnuplottex}
\PreviewMacro[*[[!]{\includegraphics}

\def\be{\begin{eqnarray}}
\def\ee{\end{eqnarray}}

\def\ba{\begin{align}}
\def\ea{\end{align}}

\def\comment#1{(see comment in source)}

\renewcommand{\-}{\,-\,}

\newcommand{\br}{\mathbf{r}}

\newcommand{\normord}[1]{:\mathrel{#1}:}

\let\oldmarginpar\marginpar
\renewcommand\marginpar[1]{\-\oldmarginpar[\raggedleft\tiny #1]%
{\raggedright\tiny #1}}

\begin{document}

\title{Stability of fractional Chern insulators in the effective continuum limit of Harper-Hofstadter bands with Chern number $|C|>1$}

\author{Bartholomew Andrews}
\affiliation{TCM Group, Cavendish Laboratory, University of Cambridge, Cambridge CB3 0HE, United Kingdom}
\author{Gunnar M\"oller}
\affiliation{Functional Materials Group, School of Physical Sciences, University of Kent, Canterbury CT2 7NZ, United Kingdom}

\date{\today}

\begin{abstract}

%{\bf Better title?}

We study the stability of composite fermion fractional quantum Hall states in Harper-Hofstadter bands with Chern number $|C|>1$. From composite fermion theory, states are predicted to be found at filling factors $\nu = r/(kr|C| +1)$, $r\in\mathbb{Z}$, with $k=1$ for bosons and $k=2$ for fermions. Here, we closely analyze these series in both cases, with contact interactions for bosons and nearest-neighbor interactions for (spinless) fermions. In particular, we analyze how the many-body gap scales as the bands are tuned to the effective continuum limit of Chern number $|C|$ bands, realized near flux density $n_\phi=1/|C|$. Near these points, the Hofstadter model requires large magnetic unit cells that yield bands with perfectly flat dispersion and Berry curvature. We exploit the known scaling of energies in the effective continuum limit in order to maintain a fixed square aspect ratio in finite-size calculations. Based on exact diagonalization calculations of the band-projected Hamiltonian for these lattice geometries, we show that for both bosons and fermions, the vast majority of finite-size spectra yield the ground-state degeneracy predicted by composite fermion theory. For the chosen interactions, we confirm that states with filling factor $\nu=1/(k|C|+1)$ are the most robust and yield a clear gap in the thermodynamic limit. For bosons with contact interactions in $|C|=2$ and $|C|=3$ bands, our data for the composite fermion states are compatible with a finite gap in the thermodynamic limit. We also report new evidence for gapped incompressible states stabilized for fermions with nearest-neighbor interactions in $|C|>1$ bands. For cases with a clear gap, we confirm that the thermodynamic limit commutes with the effective continuum limit within finite-size error bounds. We analyze the nature of the correlation functions for the Abelian composite fermion states and find that the correlation functions for $|C|>1$ states are smooth functions for positions separated by $|C|$ sites along both axes, giving rise to $|C|^{2}$ sheets; some of which can be related by inversion symmetry. We also comment on two cases which are associated with a bosonic integer quantum Hall effect (BIQHE): For $\nu=2$ in $|C|=1$ bands, we find a strong competing state with a higher ground-state degeneracy, so no clear BIQHE is found in the band-projected Hofstadter model; for $\nu=1$ in $|C|=2$ bands, we present additional data confirming the existence of a BIQHE state.
\end{abstract}


\maketitle

%{\color{red}
%Paragraph 1:
%General background of the field, motivate why it is interesting
%some general intro from my last paper is found, below. Additional points to mention: defects in higher Chern bands can carry non-Abelian statistics.
%}

New realizations of artificial gauge fields can be achieved by light-matter coupling in cold atoms~\cite{Lin:2009us, Aidelsburger:2013ew, 2013PhRvL.111r5302M, Jotzu:2014kz, Tai:2017ji, Dalibard:2011gg, Goldman:2014bv, Goldman:2016faa}, by more general Floquet systems with periodically modulated Hamiltonians~\cite{Oka:2009kc, Goldman:2014bva, GoldmanCooperArtificialFields}, or possibly by exploiting spin-orbit coupling in two-dimensional materials~\cite{Kane:2005hl, Tang:2011by}. In conjunction with repulsive interactions, they provide exciting opportunities to observe interesting flavors of fractional quantum Hall physics~\cite{Kol:1993wv, Palmer:2006km, 2009PhRvL.103j5303M, Sterdyniak:2013du, Moller:2015kg, Spanton:2017vf}. 
The recurring motif in these systems, called ``fractional Chern insulators'' (FCIs)~\cite{Regnault:2011bu}, is the existence of topological flat bands with nonzero Chern numbers that mimic the topological properties of the lowest Landau level (LLL) of particles in a magnetic field~\cite{Haldane:1988gh, Cooper:2008hx, Tang:2011by,Neupert:2011db,Sun:2011dk, Sheng:2011ku, Regnault:2011bu, Bergholtz:2013ey, Parameswaran:2013br}. 
%The study of partially-filled Chern insulators, which also exhibit a fractional quantum Hall effect, has now been of sustained interest to theorists for several years. Much of this motivation is due to the wide range of new possibilities offered by such ``Fractional Chern Insulators" (FCIs), and most notably, due to the fact that FCIs can extend the study of fractional quantum Hall states, previously restricted to continuum Landau levels, to configurable lattice-based systems. 
Although, for unit Chern number, the physics of FCIs is continuously connected to Landau level physics~\cite{Scaffidi:2012dx,Wu:2012ky,Wu:2013ii}, this connection is no longer possible for $|C|>1$, resulting in a series of \emph{lattice-specific} fractional quantum Hall states~\cite{Palmer:2006km, 2009PhRvL.103j5303M, Sterdyniak:2013du, Moller:2015kg}. 
Furthermore, FCIs in higher Chern number bands have the potential for exotic physical phenomena, such as hosting lattice defects carrying non-Abelian statistics~\cite{Barkeshli:2012kw, Barkeshli:2013da, Liu:2017eb}.

The Harper-Hofstadter model~\cite{Harper:1955bj, Azbel:1964tk, Hofstadter:1976wt} has played a special role in the study of quantum Hall effects. It was the first model in which the Chern number was identified as the topological invariant determining the quantization of the Hall conductance in integer quantum Hall states~\cite{Thouless:1982kq}. 
The first theory of FCIs, or fractional quantum Hall states on lattices, was formulated by Kol and Read in the context of the Hofstadter mode~\cite{Kol:1993wv}, generalizing early notions~\cite{Kalmeyer:1987uh, Fradkin:1989gg, Fradkin:1990hj} and using the framework of composite fermion theory~\cite{Jain:1989tq}. Furthermore, the Hofstadter model has provided the basis for the first proposals for FCIs in optical lattice realizations of cold atomic gases~\cite{Sorensen:2005bt,Palmer:2006km,Hafezi:2008un, 2009PhRvL.103j5303M}. 
More recently, the Hofstadter model represents one of the first examples for experimental realizations of artificial gauge fields in cold atomic gases~\cite{Aidelsburger:2013ew, 2013PhRvL.111r5302M, Tai:2017ji, Aidelsburger:2013ew, Aidelsburger:2015hm}, although access to highly entangled low-temperature phases will require further advances in cooling or adiabatic state preparation~\cite{Barkeshli:2015iv, He:2017gc, Motruk:2017eo}.

The Harper-Hofstadter model provides bands of any Chern number $C\in\mathbb{Z}$, with varying magnitudes of the single-particle gap.
% of different Chern bands and has been extensively studied, both analytically in the development of quantum Hall theory\cite{Thouless:1982kq, Kol:1993wv}, and recently in experiments\cite{Dean:2013bv, 2013PhRvL.111r5302M, Tai:2017ji, Aidelsburger:2013ew, Aidelsburger:2015hm}. 
In this model, it is well understood how to construct isolated Chern bands of any Chern number that can support fractional quantum Hall liquids~\cite{Moller:2015kg}. However, numerical studies have been challenging, since finite-size systems have to simultaneously satisfy several integer relations between the number of particles, the number of flux quanta and the number of sites---which are incommensurable in general. Hence, having chosen a specific flux density and filling factor, one is led to study a series of systems with varying aspect ratios. Here, we would instead like to take a proper two-dimensional thermodynamic limit for the system while keeping the aspect ratio fixed and square, since it is expected that square Hofstadter lattices are especially stable~\cite{Scaffidi:2014gf}. It is possible to identify finite-size geometries which are exactly or almost square, by considering large magnetic unit cells (MUCs)~\cite{Moller:2015kg, Bauer:2016ju}. Moreover, the limit of large MUCs is appealing, as it provides Chern bands with a flat dispersion and additionally a perfectly flat band geometry~\cite{Moller:2015kg, Bauer:2016ju}. Hence, the Hofstadter model allows one to optimize the criteria of band flatness and flat geometry, shown to be correlated with the stability of fractional Hall liquids for the $|C|=1$ cases~\cite{Parameswaran:2012cu, Goerbig:2012cz, 2014PhRvB..90p5139R, Jackson:2015fv, Claassen:2015ea, Bauer:2016ju, Lee:2017gq}. We refer to the limit of $n_\phi\to 1/|C|$ as the effective continuum limit, to distinguish it from the continuum limit $n_\phi\equiv p/q\to 0$. The continuum limit is expected to exist, as $n_\phi \to 0$ implies that the magnetic length $\ell_0 \gg a$, where $a$ is the lattice constant, so the discreteness of the lattice should be irrelevant and the continuum physics is recovered. For our numerical analyses, we further define the thermodynamic (effective) continuum limit as the (effective) continuum limit subsequently taken to large particle number ($N,q\to\infty$).\footnote{In this paper, the (effective) continuum limit at fixed aspect ratio is denoted as $\lim_{N,q\to\infty}(q\Delta)=\lim_{N\to\infty}(\lim_{q\to\infty}(q\Delta))$, and is distinguished from the limit at fixed flux density: $\lim_{q,N\to\infty}(q\Delta)=\lim_{q\to\infty}(\lim_{N\to\infty}(q\Delta))$}

%The Hofstadter model provides a plethora of different Chern bands, and it is well understood how to construct isolated Chern bands of any Chern number that can support fractional quantum Hall liquids.\cite{Moller:2015kg} However, numerical studies have been challenging, as finite-size systems have to satisfy simultaneously several integer relations, which are incommensurable in general: Therefore, having chosen a specific flux density and filling factor, one is led to study series of systems with varying aspect ratio. Instead, we would like to take a proper two-dimensional thermodynamic limit for the system while keeping the aspect ratio constant. Additionally, it is expected that Hofstadter lattices that are exactly square are especially stable.\cite{Scaffidi:2014gf} It is possible to identify finite-size geometries which are exactly or almost square, by considering large magnetic unit cells.\cite{Jackson:2015fv,Moller:2015kg} The limit of large magnetic unit cells is appealing, as it provides Chern bands with flat dispersion, and that additionally have a perfectly flat band geometry,\cite{Jackson:2015fv,Moller:2015kg} which was shown to be correlated with the stability of fractional Hall liquids for the $|C|=1$ cases.\cite{2014PhRvB..90p5139R,Jackson:2015fv,Jackson:2015fv}

%{\color{red}
%Paragraph 3:
%How we address it / what do we do?
%}

In this paper, we study the stability of quantum Hall states of the Abelian composite fermion series $\nu=r/(|kC|r+1)$~\cite{Moller:2015kg}, with $k=1(2)$ for bosons (fermions), in Chern bands with $|C|=1,2,3$ in the Hofstadter model, focusing on finite-size square systems. To find such configurations, we vary the flux density while moving within a series of single-particle bands with fixed Chern number, allowing us to find finite-size configurations with an aspect ratio of (approximately) one, as well as matching a target filling factor. The results from such different realizations of Chern bands can be combined into a single measure for the stability of the phase, owing to the known scaling of the many-body gap with the number of sublattices~\cite{Bauer:2016ju}. 

The ground-state degeneracy of these states agrees with the predictions of composite fermion theory. As expected, we find that states with filling factor $\nu=1/(k|C|+1)$ are the most robust, with the effective continuum limit remaining approximately independent of $N$ and inversely proportional to $|C|$. Our results show considerable finite-size effects for most other filling fractions, leaving the behavior in the thermodynamic limit indeterminate. While taking the thermodynamic effective continuum limit does not generally alleviate these finite-size effects, we find some system sizes where competing states are eliminated when square geometries are considered. % (see Appendix~\ref{sec:fixed_n_phi}, Fig.~\ref{fig:fixed_comparison}). 

To further characterize the target composite fermion states or their competing phases, we analyze their two-particle correlation functions, and for select examples also their particle entanglement spectra (PES). In our microscopic model, we find that correlation functions are modulated with a period of $|C|$ sites along both the $x$ and $y$ axes of the square Hofstadter model, yielding the visual appearance of $|C|^{2}$ smooth correlation functions.

This paper is organized as follows: In Sec.~\ref{sec:model}, we introduce the Harper-Hofstadter Hamiltonian and explain how to obtain finite-size geometries with approximately square aspect ratio for the desired filling factors. In Sec.~\ref{sec:results}, we present our numerical evidence for FCI phases of bosons in $|C|=1,2,3$ Hofstadter bands and fermions in $|C|=1,2$ Hofstadter bands, including many-body spectra, ground-state correlation functions, and particle entanglement spectra. In Sec.~\ref{sec:limit}, we comment on the overall trends regarding the thermodynamic effective continuum limits and analyze the role that the Chern number plays in the scaling. Finally, in Sec.~\ref{sec:conc}, we provide conclusions on the stability of FCI phases in the effective continuum limit of $|C|>1$ Harper-Hofstadter bands and suggest avenues for future research.

\section{Model}
\label{sec:model}
\subsection{Single-Particle Harper-Hofstadter Hamiltonian}

The single-particle Hamiltonian for the Harper-Hofstadter model~\cite{Harper:1955bj} was obtained as the tight-binding representation of a single-orbital lattice model subject to Peierls' substitution for a homogeneous magnetic field $\mathbf{B} = B \mathbf{e}_z$ (with $\mathbf{B} = \nabla \times\mathbf{A}$), giving
\be
\label{eq:Hamiltonian}
H_0 = - \sum_{i,j} t_{ij} e^{\phi_{ij}} c_j^\dagger c_i + \text{H.c.},
\ee
with complex hoppings of phase $\phi_{ij}$ relating to the vector potential $\mathbf{A}$ such that
\be
\label{eq:TightBindingPhases}
\phi_{ij} = \frac{e}{\hbar} \int_{\br_i}^{\br_j} \mathbf{A}\cdot d\mathbf{l} + \delta\phi_{ij}.
\ee
In the Landau gauge $\mathbf{A} = B x \mathbf{e}_y$, and for rational flux density $n_\phi= B a^2= p/q$ (with $p$ and $q$ coprime), the phases $\phi_{ij}$ naturally repeat under translations $T_{qa\mathbf{e}_x}$ by $q a\mathbf{e}_x$, and also under the translation $T_{a\mathbf{e}_y}$. This corresponds to a MUC of $q\times 1$ sites. (For simplicity, we set $a=1$, below.) However, other choices for MUC geometries $l_x \times l_y= q$ with the same area can be made, and thus enclosing the same number of magnetic flux quanta. These choices correspond to a gauge freedom in the problem, encoded in terms of additional phase factors $\delta\phi_{ij}$ occurring in the tight-binding model. These phase factors can be thought of as an additional phase generated by magnetic translations for hopping terms crossing the MUC boundary, or alternatively the tight-binding parameters can be expressed in terms of a vector potential in a periodic gauge, with $\mathbf{A}(\mathbf{r}+l_\mu \mathbf{e}_\mu)=\mathbf{A}(\mathbf{r})$ \cite{2013PhRvB..88l5426H}. 
For an explicit construction of the periodic gauge, see Appendix \ref{sec:RectangularLandau}. We further note that the choice of the MUC affects the definition of the momenta for single-particle eigenstates, and in Appendix \ref{sec:PeriodicLandauTransform}, we provide an additional note expanding on how the states are remapped throughout the Brillouin zone under such gauge transformations.

Within the Hofstadter spectrum, band gaps of any cumulative Chern number can be found. 
In order to facilitate our numerical work, we closely examine specific flux densities at which the lowest band has Chern number $|C|$ and remains well separated from higher excited bands of the Harper-Hofstadter model. Following M\"oller and Cooper~\cite{Moller:2015kg}, such cases are realized when the density of states $n_s(n_\phi=p/q)=1/q$, corresponding to flux densities
\be
\label{eq:SingleBandCases}
n_\phi = \frac{p}{|C|p - \text{sgn}(C)}\equiv\frac{p}{q},\quad p \in \mathbb{N}.
\ee
In this paper, we will focus on the cases (\ref{eq:SingleBandCases}) in the limit of large $q$ and consider flux densities in the close vicinity of points $n_\phi=1/|C|$.  At other nearby flux densities, we would find a low-energy manifold made up of several bands with the same cumulative Chern number $C$. However, we do not explore such cases here in order to maximize the number of $k$ points in the Brillouin zone in our numerics.

\subsection{Hofstadter models and Chern insulators}

Given some hesitations in the literature, let us discuss whether (partially) filled bands of the Harper-Hofstadter Hamiltonian (\ref{eq:Hamiltonian}) should be considered (fractional) Chern insulators. One of the superficial reasons why Chern insulators might be dissociated from the Harper-Hofstadter model is that the latter represents a homogeneous magnetic field, or constant flux per plaquette, while the former is simply defined as having bands with finite Chern numbers arising from complex hopping phases. It could further be argued that the Hofstadter model is characterized by a finite flux per MUC, while the flux averages to zero in Chern insulators like the Haldane model \cite{Haldane:1988gh}. 
However, in existing cold-atom realizations of the Hofstadter model, there is no physical magnetic field present---rather, these experiments realize the model directly as a tight-binding lattice with complex hopping terms induced by laser-assisted hoppings or more general time-modulated Floquet-Hamiltonians~\cite{Goldman:2014bv, Eckardt:2017hc} in order to mimic the Aharonov-Bohm effect of a magnetic field.

The overall flux threading the lattice is also not a good distinction of Hofstadter models from generic cases, given that flux is defined only modulo the flux quantum $\Phi_0$ in a lattice geometry and so any integer number of flux quanta can be inserted within a given plaquette of the lattice without altering the physics. 
A finite-size realization of the Hofstadter model requires an integer number of flux quanta per MUC; thus it can always be interpreted as a model without net flux: Any excess flux can be neutralized by adding an opposite flux through one of the plaquettes in a unit cell~\cite{2012PhRvL.108y6809H}. 
Indeed, one could argue that any finite-size implementation of the complex hopping phases (\ref{eq:TightBindingPhases}) that is compatible with periodic boundary conditions effectively corresponds to such an insertion of neutralizing flux, which gives rise to the $\delta\phi_{ij}$ term in Eq.~(\ref{eq:TightBindingPhases}).

Comparing the translational symmetries of Hofstadter models with other general tight-binding models with Chern bands~\cite{Tang:2011by, Neupert:2011db, Sun:2011dk, Regnault:2011bu}, we can finally find one formal distinction between these cases. The translational symmetry group of the Harper-Hofstadter Hamiltonian is smaller than the translation symmetry of the underlying lattice potential due to the commensurability of the two length scales in the problem. By contrast, generic models typically have a full translational symmetry group identical to that of the lattice potential. Inversely, we could say that the translational symmetry group of the Hofstadter lattice can be enhanced if we allow simultaneous translations and gauge transformations (effectively translating the origin of the MUC), while no additional symmetries can be found in generic models.

As the terminology of (fractional) Chern insulators focuses on the topological properties, it seems natural to include those states realized in the Chern bands of the Hofstadter model. Conversely, since the Hofstadter model is closely related to physical magnetic fields, the terminology of lattice fractional quantum Hall states is also appropriate for these models.

\subsection{Many-Body Hamiltonian}

We study the many-body physics of interacting particles in the Harper-Hofstadter model, described by the Hamiltonian
\be
\label{eq:FullHamiltonian}
H = H_0 + P_{\text{LB}}\left[\sum_{i<j} V(\br_i - \br_j) \normord{\rho_i  \rho_j}\right]P_{\text{LB}},
\ee
where $P_{\text{LB}}$ denotes the lowest band projection operator and $\normord{\rho_i  \rho_j}$ indicates the normal ordering of the density operators, with site labels $i$, $j$. 

In this paper, we extend the work of M\"{o}ller and Cooper~\cite{Moller:2015kg} on bosonic contact interactions ($V_{ij}=U\delta_{ij}$) as well as considering the case of fermions with nearest-neighbor (NN) interactions ($V_{ij}=V\delta_{\langle i,j\rangle}$). In both cases, we target a number of known candidate phases for incompressible quantum Hall states. 

We explore the spectrum of the many-body Hamiltonian (\ref{eq:FullHamiltonian}) using exact diagonalization calculations and identify incompressible states by means of their ground-state degeneracy, many-body gap $\Delta$, as well as the correlations and entanglement properties of the corresponding ground-state wave functions. The incompressible phases which we find show a clear quasidegenerate ground state, such that the gap $\Delta$ to the excited states is much larger than the splitting between states in the ground-state manifold or the typical level spacing among higher lying excitations. We quote the gap in units of the interaction strength $U$ ($V$), implicitly setting $U=V=1$, below.

\subsection{Target FCI Phases in General Chern Bands}

%{\color{red}
%Review some of the theory on the predicted phases.
%
%Chern-Simons theory \cite{Kol:1993wv}, or equivalently the lattice composite fermion picture \cite{2009PhRvL.103j5303M}
%}

Several families of incompressible quantum Hall states have been proposed to occur in Chern bands with higher Chern numbers $|C|>1$, most importantly including generalizations of the Jain states~\cite{Wang:2012kv, Liu:2012ek, Sterdyniak:2013du}, 
an Abelian series of states arising from the composite fermion construction~\cite{Kol:1993wv, 2009PhRvL.103j5303M} and generalizations of the non-Abelian Read-Rezayi states~\cite{Read:1999wx} to higher Chern bands~\cite{Sterdyniak:2013du}. There were also reports of states that simultaneously break translation symmetries while displaying a quantized Hall response~\cite{Spanton:2017vf}.

The incompressible character of these phases is expressed by a preferred density of particles, measured in terms of the number density per unit area of accessible single-particle states. While this manifold of single-particle states is trivially given by the continuum Landau levels in continuum fractional quantum Hall states, the relevant low-energy subspace for Chern insulators is set by a single-particle gap of the single-particle dispersion in a tight-binding model that is large compared to the dispersion of the low-lying band(s). 

%{\color{red}
%(describe this in some more detail)
%}

From composite fermion theory, one predicts a series of Abelian quantum liquids~\cite{Kol:1993wv, 2009PhRvL.103j5303M} at filling factors
\be
\label{eq:FillingFactorJainEquiv}
\nu(k, r, C) = \left|\frac{n}{n_s}\right| = \frac{r}{ |kC|r + 1},
\ee
where $k$ is the number of flux quanta attached to the particles, $|r|$ is the number of bands filled in the composite fermion spectrum, and its sign indicates the relative sign of the Chern number $C^*$ for the composite fermion band relative to the Chern number $C$ of the low-energy manifold~\cite{Moller:2015kg}. The states (\ref{eq:FillingFactorJainEquiv}) carry a ground-state degeneracy of $d= |kCr + 1|$~\cite{Kol:1993wv, Moller:2015kg}.

Where required, we consider a number of other competing phases. Prominently, this includes the states of the bosonic Read-Rezayi series, found at filling factors $\nu=\kappa/2$ in $C=1$ bands~\cite{Read:1999wx} (with $\kappa \in \mathbb{Z}^+$) which carry a ground-state degeneracy $d_\text{RR}=\kappa+1$. The generalizations of the Read-Rezayi states to higher Chern bands~\cite{Sterdyniak:2013du} do not generically occur at the same filling factors as the composite fermion states in Eq.~(\ref{eq:FillingFactorJainEquiv}), so we do not encounter them explicitly. At sufficiently weak interactions, generically one may find competition with condensed phases~\cite{2010PhRvA..82f3625M, Natu:2016fp, Hugel:2017kt, Kozarski:2017te}. These may survive up to large values of the interaction at time-reversal symmetric points of the Hofstadter spectrum, but are likely less competitive elsewhere. Further instabilities include density-wave or crystalline orders. These were found to be stabilized in related time-reversal symmetric flat band models~\cite{2012PhRvL.108d5306M}, and are generically expected to be among the competing phases in Chern insulator models.

%It may be interesting to also look into the limit of filling many CF Landau-levels, $r\to \infty$, which represents the equivalent of the half-filled Landau-level, and converges to
%\be
%\label{eq:LimitingFilling}
%\lim_{r\to\infty} \nu^{C^*=rC} = \frac{1}{kC}.
%\ee
%At these points, the composite fermion spectrum resembles a Fermi sea, as the band gaps between the composite fermion levels decrease as $1/r$ and evolve into a quasi-continuum. In analogy to the half-filled continuum Landau levels, one may ask whether this filling can be susceptible to the equivalent of a CF pairing instability, or possibly more exotic states. In the $C=1$ case, the possibility of a Moore-Read state at $\nu=1$ is well known \cite{Sterdyniak:2012jo}.
%For the $C=2$ band near $n_\phi=1/2$, such a phase has  been described in a related continuum model \cite{2012PhRvL.108y6809H, Moller:2014fa}, though the model does not provide a quantitative description of $C=2$ bands \cite{Harper:2014fq}. 

\subsection{Scaling to the Continuum Limit at Fixed Aspect Ratio}
\label{subsec:lattice_geometries}

Finite-size geometries of the square Harper-Hofstadter model are determined by the number of sites in the $x$ and $y$ directions $N_x$, $N_y$, and the total number of flux quanta $N_\phi$ piercing the system. Each geometry may allow an additional gauge choice for the shape of the MUC given by $l_x$ and $l_y$ sites, and the number of repetitions $L_x$ and $L_y$ of the MUC within the simulation cell along these axes. In this paper, we shall be looking at square systems in terms of the total number of sites, hence systems with a unit aspect ratio:
\begin{equation}
R=\frac{N_x}{N_y}=\frac{L_x l_x}{L_y l_y}=1.
\end{equation}
Note that the spectra are gauge invariant and depend only on the total system size, but not on the shape of the MUC. However, the definition of momentum depends on the choice of MUC, as further discussed in Appendix \ref{sec:PeriodicLandauTransform}.

We examine the filling factors (\ref{eq:FillingFactorJainEquiv}) for Chern bands $|C|=1,2,3$ with $|r|=1,2,3$ and a finite particle number $N$ set by the available Hilbert space sizes, typically ranging up to about 10--20 particles. For each case, we generate a sample of finite-size systems including all possible square geometries in a specified range of effective flux densities $n_\phi=p/q$ subject to Eq.~(\ref{eq:SingleBandCases}). A restriction is placed on the numerator such that $2\lesssim p\lesssim 1000$, where low $p$ value configurations are excluded because they may correspond to band gaps with a lower Chern number, and the upper bound is determined by the computational cost and numerical accuracy of our calculations of the single-particle spectrum.

For certain cases, the restriction on the numerator of the effective flux density does not yield any square configurations. In these situations, we look for approximately square configurations, with a fixed maximum error $\epsilon\approx1\%$ such that
\begin{equation}
\delta R = \left|\frac{N_x}{N_y}-1\right|\leq\epsilon,
\end{equation}
taking $N_x > N_y$ by convention.
In practice, the allowable deviation of the aspect ratio from one is adjusted slightly so that we obtain a comparable sample size, or number of geometries, for each filling factor $\nu$.

In Ref.~\onlinecite{Bauer:2016ju}, Bauer \emph{et al.} undertook a similar study for models with short-range repulsive interactions in the $C=1$ bands of the Hofstadter model. Using geometric considerations for the Laughlin and Moore-Read states for $N=8$ particles, they found that the many-body gap scales as $\Delta\sim1/q$ for bosons and $\Delta\sim1/q^2$ for fermions. They also found approximate continuum limits for $n_\phi\to 0$ for the three phases they considered.

As shown below, we find that the (effective) continuum limit is helpful for examining the stability of candidate incompressible Hall states, as it improves the effectiveness of finite-size scaling analyses. For comparison, we also undertake the conventional finite-size scaling at fixed flux density $n_\phi$, as previously considered by M{\"o}ller and Cooper~\cite{Moller:2015kg}. An extract of our additional data for this thermodynamic limit is shown in Appendix~\ref{sec:fixed_n_phi}.

\section{Results}
\label{sec:results}

We present results on fractional quantum Hall states in Harper-Hofstadter bands with higher Chern number. However, to establish our methodology, we first review the case of $|C|=1$ bands. Our results reproduce the known continuum limit, i.e., the well-known quantum Hall physics of the LLL of a homogeneous magnetic field. Our results go beyond previous studies on the Hofstadter lattice in that we study fermionic quantum Hall states in addition to bosonic ones, and we consider the thermodynamic limit in addition to the continuum limit.

\subsection{FCIs in $|C|=1$ Harper-Hofstadter Bands}
\label{subsec:FCI_C1}

%{\color{red}
%Main aims: 
%\begin{enumerate}
%\item establish the methodology, explore limits to scaling in the size of the MUC due to the numerical accuracy of the single-particle eigenstates
%\item demonstrate the extent to which the continuum limit + finite-size scaling of the system on the lattice reproduces the continuum physics.
%\item comparison to the continuum problem on taking the limit $q\to\infty$.
%\end{enumerate}
%}

\subsubsection{Bosonic States}
\label{subsubsec:FCI_C1_bosons}

We first review the case of bosons in Chern number $|C|=1$ bands, which are well known to support fractional quantum Hall states in the continuum LLL~\cite{Cooper:2001gy, Regnault:2003ws, Cooper:2008hx}. We consider states of the Jain series (\ref{eq:FillingFactorJainEquiv}) with $|C|=1$, $|r|=1,2,3$ and we restrict ourselves to Hilbert space dimensions $\dim\{\mathcal{H}\} < 10^7$, which typically allows us to consider particle numbers $N\lesssim 12$.\footnote{Here we neglect trivially small particle numbers and cases where finite-size effects are clearly dominant.} Furthermore, there are no states corresponding to $r=-1$, as Eq.~(\ref{eq:FillingFactorJainEquiv}) is undefined for this value. 
Overall, we have considered 24 different combinations of particle size and filling factor, with an average of $\sim 38$ different geometries for each, giving a total of 921 different exact diagonalization calculations.
\begin{figure}
\begin{minipage}[b]{\linewidth}
\begin{minipage}[b]{.5\linewidth}
\begin{tikzpicture}
\node at (0,0) {\centering\includegraphics{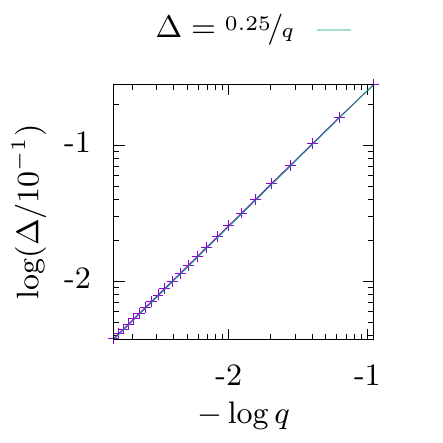}};
\node[overlay] at (-2,2) {(a)};
\phantomsubcaption
\label{fig:boson_plots_C1_1}
\end{tikzpicture}
\end{minipage}%
\begin{minipage}[b]{.5\linewidth}
\begin{tikzpicture}
\node at (0,0) {\centering\includegraphics{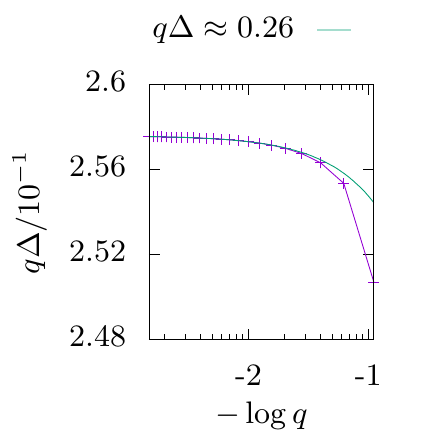}};
\node[overlay] at (-2,2) {(b)};
\phantomsubcaption
\label{fig:boson_plots_C1_2}
\end{tikzpicture}
\end{minipage}%
\end{minipage}
\caption{Magnitude of the gap for the bosonic 12-particle $\nu=3/4$ state in the $|C|=1$ band, as a function of MUC size, $q$. (a) Log-log plot of $\Delta$ vs $q^{-1}$. (b) Scaling of $q\Delta$ to a constant value in the continuum limit $n_\phi\to 0$.}
\label{fig:boson_plots_C1}
\end{figure}
Apart from exceptional cases that we discuss in detail below, we found that all candidate states show a degenerate ground-state manifold composed of $|krC+1|$ states, in line with the predictions of composite fermion theory.
For each system that we examine, we have plotted the energy gap above the ground-state manifold, $\Delta$, against the MUC size, $q$, to test whether the expected \emph{reciprocal} scaling~\cite{Bauer:2016ju} is realized. An example scaling for an $r=3$ state with $\nu=3/4$ is shown in Fig.~\ref{fig:boson_plots_C1_1}. 

Next, we test the scaling hypothesis and extract the coefficient for $\Delta\propto q^{-1}$ at large MUC size, shown in Fig.~\ref{fig:boson_plots_C1_2}. For $|C|=1$, this is the continuum limit. Theoretically, the large-$q$ limit of $q \Delta$ should be independent of $q$, while finite-size effects imply variations for small $q$. When establishing the continuum limit, we thus neglect small-$q$ outliers to take account of this fact. Notice that, as a result, the line of best fit in Fig.~\ref{fig:boson_plots_C1_1} does not exactly correspond to the $q \Delta$ limit in Fig.~\ref{fig:boson_plots_C1_2}. Using this procedure, we find agreement with the value of the many-body gap for the bosonic Laughlin state calculated by Bauer \emph{et al.}~\cite{Bauer:2016ju} and we now go beyond their work by considering the thermodynamic limit, as well as the continuum limit, for a large sample of systems.
\begin{figure}
\begin{center}
\includegraphics{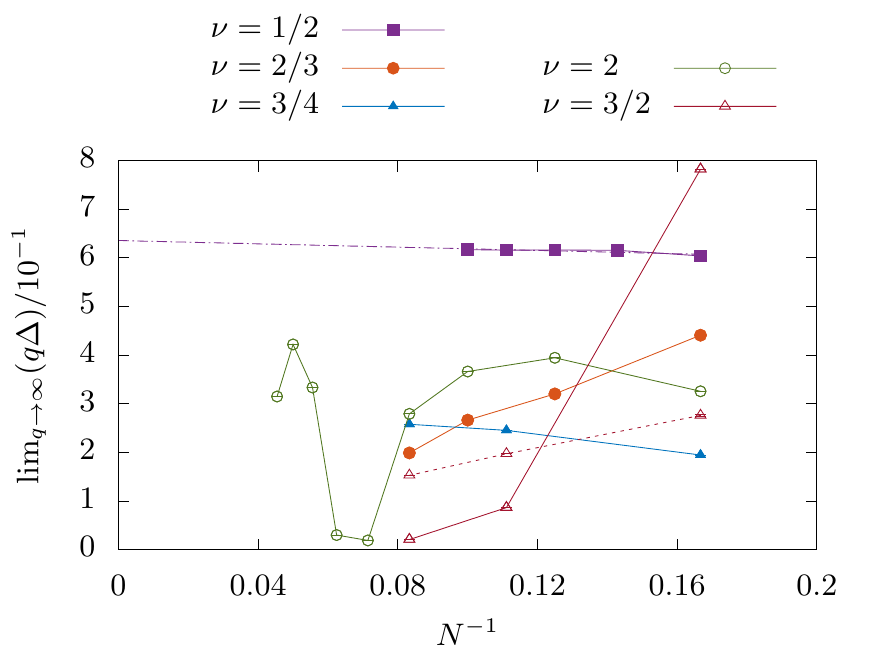}
\caption{Finite-size scaling of the gap to the thermodynamic continuum limit at fixed aspect ratio, for bosonic states in the $|C|=1$ band. The extrapolation to the $y$ axis is shown for the robust $\nu=1/2$ states. The dashed line for the $\nu=3/2$ series corresponds to the scaling behavior given a ground-state degeneracy of 4, as predicted by Read-Rezayi theory. Squares, circles, and triangles denote states with $|r|=1,2,3$, respectively, where the filled (hollow) symbols correspond to positive (negative) $r$. All of the error bars are smaller than the data points on the scale of the plot.}
\label{fig:bosons_final_plot_C1}
\end{center}
\end{figure}
\begin{figure}
\centering
\begin{minipage}[b]{.49\linewidth}
\begin{tikzpicture}
\node at (0,0) {\includegraphics[width=\linewidth]{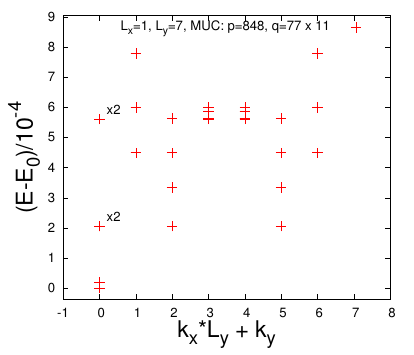}};
\node[overlay] at (-2,1.6) {(a)};
\phantomsubcaption
\label{fig:boson_spectra_C1_1}
\end{tikzpicture}
\end{minipage}
\begin{minipage}[b]{.49\linewidth}
\begin{tikzpicture}
\node at (0,0) {\includegraphics[width=\linewidth]{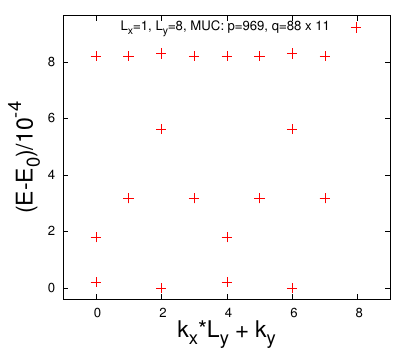}};
\node[overlay] at (-2,1.6) {(b)};
\phantomsubcaption
\label{fig:boson_spectra_C1_2}
\end{tikzpicture}
\end{minipage}
\caption{Energy spectra for bosonic states in the $|C|=1$ band. (a) The 14-particle $\nu=2$ state with $p=848$, resolved to $n=6$ states per sector. (b) The 12-particle $\nu=3/2$ state with $p=969$, resolved to $n=3$ points per sector.}
\label{fig:boson_spectra_C1}
\end{figure}

We collect the continuum limit of $q \Delta$ for the various filling factors under consideration at all available particle numbers ($6\leq N\lesssim 12$). In Fig.~\ref{fig:bosons_final_plot_C1}, we plot how the limiting value $\lim_{q\to\infty} q\Delta(N,q)$ varies with particle number for each filling factor. We attempt to proceed with a scaling extrapolation to the thermodynamic continuum limit on the basis of an inverse regression for $\lim_{q\to\infty} q\Delta$ against $N^{-1}$. Using this approach, we examine, for each filling factor, the $\lim_{q\to\infty} q\Delta$ limit as $N^{-1}\to 0$.

Figure~\ref{fig:bosons_final_plot_C1} shows the continuum limits for the five filling factors under consideration. Note that the error bars due to the extrapolation in $q\to \infty$ are negligible for these points on the scale of the plot. We have also verified that these data agree with many-body gaps of the corresponding states in the continuum LLL on the torus. We include data on a comparison of the correlation functions, below.
% as shown in Appendix~\ref{sec:fixed_n_phi}. %\ref{sec:CorrelationsAccuracy}. (??)

The results are insightful in that they illustrate the caveats of interpreting data on finite-size geometries. Composite fermion theory suggests that the stability of states in the Jain series decreases with $|r|$. However, this is only partially borne out by the data.

First, we find that the Laughlin state corresponding to $r=1$ has the largest gap and has negligible finite-size corrections for the gap in the continuum limit. In this case, we also see close agreement of the limiting value obtained from a finite-size scaling at fixed flux density. The corresponding data are shown in Appendix \ref{sec:fixed_n_phi}, Fig.~\ref{fig:bosons_C1r1_individual_plot}. We extrapolate a thermodynamic continuum limit of $\lim_{N,q\to\infty}(q\Delta)=0.64\pm0.01$ in this case, where the error given is the asymptotic standard error in the linear regression\cite{Note1}.

For the next states in the series, we find that the $r=2$, $\nu=2/3$ state has a gap that reduces approximately linearly with inverse system size, while the $r=3$, $\nu=3/4$ state appears more stable. By contrast, analyses of continuum quantum Hall states in the LLL (on the sphere) show that both of these states are stable and the latter has the smaller gap~\cite{Regnault:2003ws}. Owing to the higher symmetry, continuum calculations enable slightly larger system sizes to be calculated, resulting in more accurate estimates for the gap by including larger system sizes. Note also that we have not considered data beyond Hilbert space sizes of $10^7$ for our comprehensive sampling of different geometries, while larger Hilbert spaces can be considered for single cases. Despite the fact that our data are not sufficient to ascertain the size of the gap in the thermodynamic continuum limit for these states, reassuringly, all of our simulations at these filling factors did identify the expected ground-state degeneracies (with $d= 3$ or $d=4$, respectively) and a clear separation of scales for the gap.

For the states at negative effective flux \cite{2005PhRvB..72d5344M, Davenport:2012el}, we see an interesting competition with the Read-Rezayi series, in line with the results for the continuum LLL \cite{Cooper:2007bw, Cooper:2008hx}.
The $r=-2$ state is interesting in that it occurs at the integer filling factor $\nu=2$, so it is a potential example of a bosonic integer quantum Hall effect (BIQHE) \cite{Senthil:2013kt,2009PhRvL.103j5303M, Regnault:2013ic, Moller:2015kg, He:2017gc}. As the BIQHE is not a fractionalized phase, it is associated with a singly degenerate ground state. However, our exact diagonalization calculations show higher ground-state degeneracies for our target Hamiltonian (\ref{eq:FullHamiltonian}), consisting of onsite repulsions projected to the lowest Hofstadter band. In particular, we find ground-state degeneracies of $d_{N=14}=2$ and $d_{N=16}= 2$ or $6$ for the $N=14$ and $N=16$ particle systems, respectively, while other system sizes are compatible with an interpretation as a singly degenerate ground state. The $N=14$ spectrum is shown in Fig.~\ref{fig:boson_spectra_C1_1}. The realized degeneracies are inconsistent with the interpretation as a BIQHE state. The $k=4$ Read-Rezayi state could be an alternative candidate for this filling, but it would have a $d=5$-fold degeneracy for $N$ divisible by $4$~\cite{Read:1999wx}. In this context, we note that in order to stabilize the Read-Rezayi state in the continuum LLL, a small amount of dipolar interaction is required~\cite{Cooper:2007bw}.
At any rate, our findings suggest that unlike the BIQHE in $|C|=2$ bands (see Sec.~\ref{subsubsec:FCI_C2_bosons}), the $\nu=2$ state is not realized in the single $|C|=1$ band of the band-projected Harper-Hofstadter-Hubbard Hamiltonian (\ref{eq:FullHamiltonian}). It therefore seems likely that the $\nu=2$ BIQHE state reported in a recent DMRG study for hardcore bosons requires at least the two lowest bands to be stabilized, which would be quite similar to the situation in two-flavor quantum Hall states~\cite{Regnault:2013ic} or Chern number $|C|=2$ bands~\cite{2009PhRvL.103j5303M, Moller:2015kg, He:2015ja}.

Finally, for the $r=-3$ series with $\nu=3/2$, we find a marked reduction of the gap above the second-lowest state, so the degeneracy of $d=2$ predicted by composite fermion theory does not describe this phase well. As observed in the continuum LLL ~\cite{Rezayi:2005em}, the $k=3$ Read-Rezayi state appears to be a good candidate for this filling, as a stable gap appears to form above the lowest $d=4$ states, in line with the expected ground-state degeneracy for this Read-Rezayi state. A full spectrum for the $N=12$ particle state is given in Fig.~\ref{fig:boson_spectra_C1_2}, and the finite-size scaling of the gap inferred for the Read-Rezayi states is shown as dotted lines in Fig.~\ref{fig:bosons_final_plot_C1}---these data are consistent with a finite gap in the thermodynamic continuum limit, even without the addition of long-range interactions~\cite{Rezayi:2005em}.

In order to further characterize the different candidate states, we calculate the density-density correlation functions $g(\br) = \langle \rho(\br) \rho(\mathbf{0})\rangle$ for the different ground states, as explained in Appendix~\ref{sec:CorrelationDerivation}. 
 In order to establish the accuracy of our code, we have further verified that correlations approach the exact continuum result for the corresponding state in the continuum Landau level on a torus (see Appendix~\ref{sec:CorrelationsAccuracy}). Our results show close agreement at short distances, while there are slight deviations at larger separations. We interpret these findings as being most likely a consequence of finite precision floating point arithmetic, as discussed in Appendix \ref{sec:CorrelationsAccuracy}.

\begin{figure}
\begin{minipage}[b]{\linewidth}

\raggedright

\begin{minipage}[b]{.5\linewidth}
\begin{tikzpicture}
\node at (0,0) {\centering\includegraphics{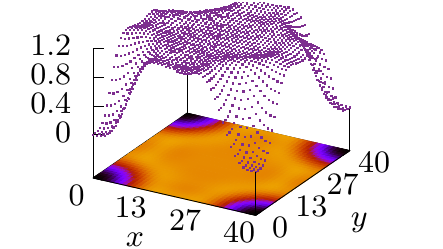}};
\node[overlay] at (-2.3,1.2) {(a)};
\phantomsubcaption
\label{fig:boson_corr_C1r1N8}
\end{tikzpicture}
\end{minipage}%
\begin{minipage}[b]{.5\linewidth}
\centering\includegraphics{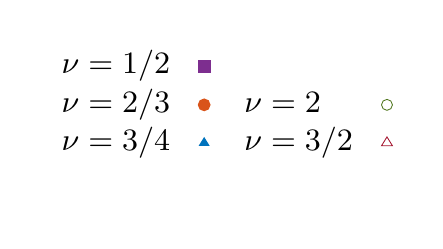}
\end{minipage}%

\vskip\baselineskip

\begin{minipage}[b]{.5\linewidth}
\begin{tikzpicture}
\node at (0,0) {\centering\includegraphics{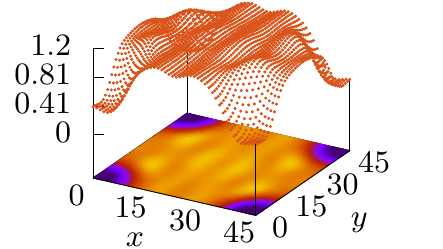}};
\node[overlay] at (-2.3,1.2) {(b)};
\phantomsubcaption
\label{fig:boson_corr_C1r2N10}
\end{tikzpicture}
\end{minipage}%
\begin{minipage}[b]{.5\linewidth}
\begin{tikzpicture}
\node at (0,0) {\centering\includegraphics{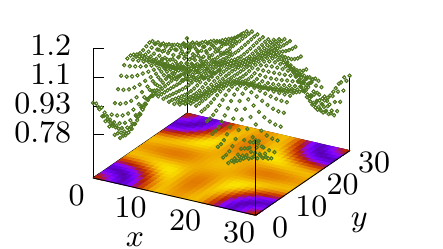}};
\node[overlay] at (-2.3,1.2) {(c)};
\phantomsubcaption
\label{fig:boson_corr_C1r-2N20}
\end{tikzpicture}
\end{minipage}%

\vskip\baselineskip

\begin{minipage}[b]{.5\linewidth}
\begin{tikzpicture}
\node at (0,0) {\centering\includegraphics{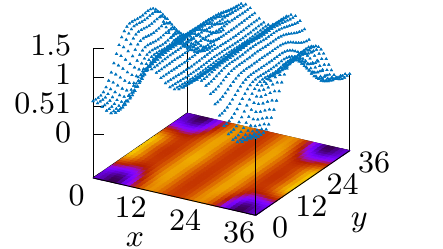}};
\node[overlay] at (-2.3,1.2) {(d)};
\phantomsubcaption
\label{fig:boson_corr_C1r3N9}
\end{tikzpicture}
\end{minipage}%
\begin{minipage}[b]{.5\linewidth}
\begin{tikzpicture}
\node at (0,0) {\centering\includegraphics{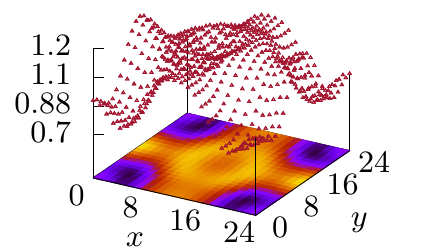}};
\node[overlay] at (-2.3,1.2) {(e)};
\phantomsubcaption
\label{fig:boson_corr_C1r-3N9}
\end{tikzpicture}
\end{minipage}%

\end{minipage}
\caption{Density-density correlation functions for bosonic states in the $|C|=1$ band. The plots are shown for lowest-lying ground state in the $(k_x,k_y)=(0,0)$ momentum sector, with (a)~$r=1$: $\nu=1/2$, $N=8$, $p=99$; (b)~$r=2$: $\nu=2/3$, $N=10$, $p=134$; (c)~$r=-2$: $\nu=2$, $N=20$, $p=91$; (d)~$r=3$: $\nu=3/4$, $N=9$, $p=107$; (e)~$r=-3$: $\nu=3/2$, $N=9$, $p=97$.}
\label{fig:boson_corr_C1}
\end{figure}
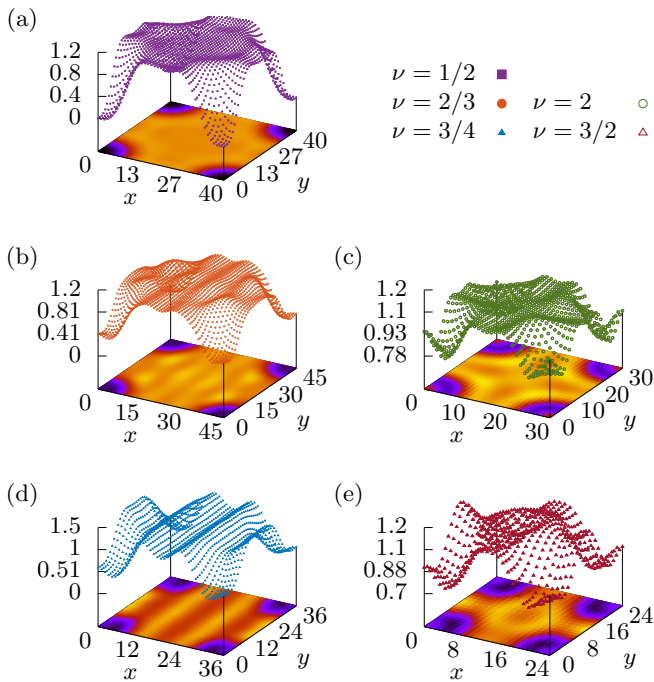

Correlation functions for all available filling factors are shown in Fig.~\ref{fig:boson_corr_C1}, based on the lowest-lying ground state at zero momentum. Note that the Laughlin state in Fig.~\ref{fig:boson_corr_C1r1N8} displays a correlation function that has saturated to a nearly constant value at large distances, with a zero correlation hole at zero separation. This is the expected form of the Laughlin correlation function and may be solved analytically for the torus, as discussed in Appendix~\ref{sec:CorrelationsAccuracy}. For all other states, we find oscillations that are generally stronger for the states with higher $|r|$ values. The relatively small isotropic fluctuations of the correlation profile in Fig.~\ref{fig:boson_corr_C1r1N8} may be an artifact of finite-size effects. However, the remaining oscillations in Figs.~\ref{fig:boson_corr_C1r-2N20},~\ref{fig:boson_corr_C1r3N9}, and~\ref{fig:boson_corr_C1r-3N9} are indicative of either states with longer correlation lengths, or potentially competing density wave instabilities. For example, the most marked oscillations are seen for the $N=9$, $\nu=3/4$ state in Fig.~\ref{fig:boson_corr_C1r-3N9}. These oscillations also break the rotational symmetry as they occur predominantly along the $x$ axis for this geometry, which may be a signature of an instability toward charge density wave formation. For the states at $\nu=3/2$ and $\nu=2$, the correlations show a local maximum at zero separation, followed by a shallow correlation hole, which could be consistent with the interpretation as clustered Read-Rezayi states.

To summarize, the $|C|=1$ boson data yields well-defined continuum limits $q\to\infty$, with negligible errors due to the extrapolation to large MUCs. From the cases considered, we can conclude that bosons in the $|C|=1$ Chern number bands obey the expected scaling relations for the gap, and we obtain a well-converged continuum limit with no exceptions. However, extrapolation of these values to the thermodynamic limit remains difficult to achieve, and is prone to finite-size effects. The predictions of composite fermion theory apply only to a subset of possible composite fermion states, due to both finite-size effects and the apparent competition with states of the Read-Rezayi series or other competing phases such as density wave instabilities.

\subsubsection{Fermionic States}
\label{subsubsec:FCI_C1_fermions}

Building on our analysis for bosonic states, we carry out a corresponding study for fermions in the $|C|=1$ band. As before, we consider cases with $|r|=1,2,3$ and typical values of $6\leq N \lesssim 12$ arising from the constraint on the Hilbert space dimension $\text{dim}\{\mathcal{H}\} < 10^7$. Note the Hilbert space of $N$ bosons in a $|C|=1$ band with $N_\phi$ flux maps to the Hilbert space of fermions at $N+N_\phi-1$ flux in the continuum, so the Hilbert space dimensions for the Jain states are identical for bosons at fermionic states of a given $r$ value. They are essentially the same on the lattice also, up to different numbers of conserved momenta. Hence, we are able to obtain a comparable sample of geometries as in the previous section. The $r=-1$ series is omitted because this corresponds to a band insulator, so composite fermion theory is not relevant. Overall, we have considered 18 different combinations of particle number and filling factor, with an average of $\sim 28$ different geometries for each, and a total of 498 different exact diagonalization calculations underlying the data in this section.
\begin{figure}
\begin{minipage}[b]{\linewidth}
\begin{minipage}[b]{.5\linewidth}
\begin{tikzpicture}
\node at (0,0) {\centering\includegraphics{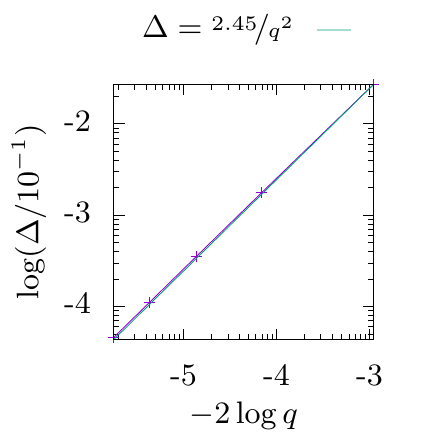}};
\node[overlay] at (-2,2) {(a)};
\phantomsubcaption
\label{fig:fermion_plots_C1_1}
\end{tikzpicture}
\end{minipage}%
\begin{minipage}[b]{.5\linewidth}
\begin{tikzpicture}
\node at (0,0) {\centering\includegraphics{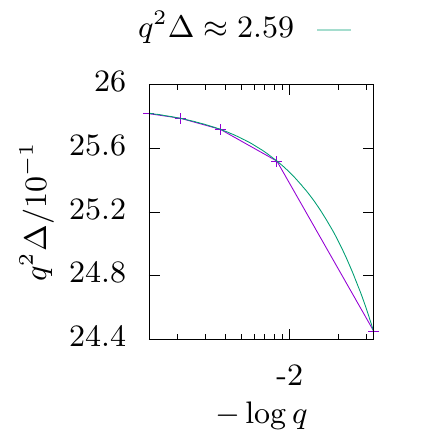}};
\node[overlay] at (-2,2) {(b)};
\phantomsubcaption
\label{fig:fermion_plots_C1_2}
\end{tikzpicture}
\end{minipage}%
\end{minipage}
\caption{Magnitude of the gap for fermionic 20-particle $\nu=2/3$ state in the $|C|=1$ band, as a function of MUC size, $q$. (a) Log-log plot of $\Delta$ vs $q^{-1}$. (b) Scaling of $q\Delta$ to a constant value in the continuum limit $n_\phi\to 0$.}
\label{fig:fermion_plots_C1}
\end{figure}

For each filling factor, we plot the energy gap, $\Delta$, against the MUC size, $q$, which reproduces the \emph{inverse-square} relation $\Delta \propto q^{-2}$ expected for fermions~\cite{Bauer:2016ju}, as illustrated in %Fig.~\ref{fig:fermion_plots_C1_1}. 
Fig.~\ref{fig:fermion_plots_C1} for the 20-particle $\nu=2/3$ data point.

From composite fermion theory, the expected ground-state degeneracy is $|krC+1|$ with $k=2$ for fermionic systems. This is indeed realized in all the observed energy spectra. As before, we find agreement with the value of the many-body gap in the continuum limit for the $N=8$ fermionic Laughlin state considered by Bauer \emph{et al.}~\cite{Bauer:2016ju}.

For all fermionic candidate states, we examine the continuum limit of $q^2 \Delta$ at large MUC size, as demonstrated in Fig.~\ref{fig:fermion_plots_C1_2}. As seen previously, finite-size effects may result in fluctuations at small $q$, so we neglect outliers at small $q$ when determining the continuum limit. 
\begin{figure}
\begin{center}
\includegraphics{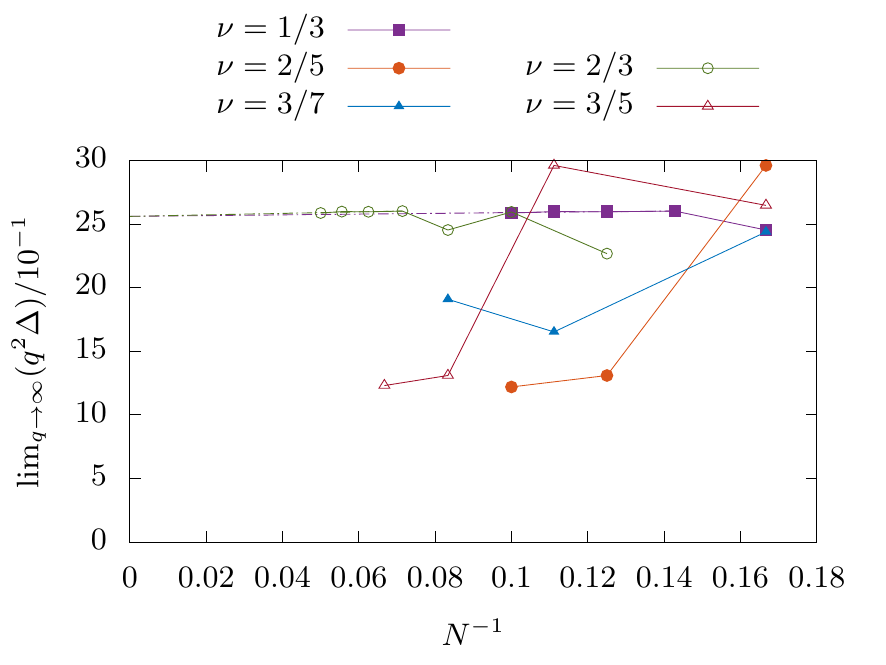}
\caption{Finite-size scaling of the gap to the thermodynamic continuum limit at fixed aspect ratio, for fermionic states in the $|C|=1$ band. The extrapolation to the $y$ axis is shown for the robust $\nu=1/3$ and $\nu=2/3$ states. Squares, circles, and triangles denote states with $|r|=1,2,3$, respectively, where the filled (hollow) symbols correspond to positive (negative) $r$. All of the error bars are smaller than the data points on the scale of the plot.}
\label{fig:fermions_final_plot_C1}
\end{center}
\end{figure}

Figure~\ref{fig:fermions_final_plot_C1} shows the thermodynamic continuum limiting behavior for the five filling factors under consideration. The sample size shown is comparable to that in Fig.~\ref{fig:bosons_final_plot_C1}. Notice that the $r=1$ series (this time corresponding to $\nu=1/3$) displays very minor finite-size effects.  In Appendix \ref{sec:fixed_n_phi}, we also compare this limit against the finite-size scaling at fixed flux density, followed by extrapolation of the thermodynamic values to the continuum limit. Both orders of limits agree well, as shown in Fig.~\ref{fig:fermions_C1r1_individual_plot}. Compared to the data for bosons in Fig.~\ref{fig:bosons_C1r1_individual_plot}, we note that finite-size corrections are more noticeable for smaller system sizes.  The gap in the continuum limit oscillates for small particle number but gradually settles to a well-defined thermodynamic continuum limit, which we extrapolate to be $\lim_{N,q\to\infty}(q\Delta)=2.56\pm0.02$\cite{Note1}. This nicely illustrates the dissipation of finite-size effects as system sizes exceed the correlation length.

The $\nu=2/3$ series also exhibits noteworthy behavior. We find that the many-body gaps are closely related to the values for $\nu=1/3$ states by particle-hole symmetry, via $\Delta(N) = \Delta(N_\phi-N)$, although lattice models typically break this symmetry~\cite{Wu:2012do}. Clearly, the particle-hole symmetry re-emerges as an exact symmetry in the limit of continuum Landau levels, so it is reassuring that it is also approximated closely for finite flux densities. In this case, we also extrapolate the thermodynamic continuum limit to be $\lim_{N,q\to\infty}(q\Delta)=2.56\pm0.02$ to two decimal places, perfectly matching the result for $\nu=1/3$\cite{Note1}.

For the $\nu=2/5$ state, we again find a particle-hole symmetric partner at $\nu=3/5$ with similar gaps. In both cases, the finite-size gap appears strongly enhanced for small system sizes, but settles to a relatively flat plateau for the last two system sizes that we have evaluated. These data are indicative of a gap in the thermodynamic continuum limit, in accordance with established numerical results for the LLL. Likewise, the $\nu=3/7$ state also yields finite gaps that are consistent with a nonzero thermodynamic continuum limit.

\begin{figure}
\begin{minipage}[b]{\linewidth}

\raggedright

\begin{minipage}[b]{.5\linewidth}
\begin{tikzpicture}
\node at (0,0) {\centering\includegraphics{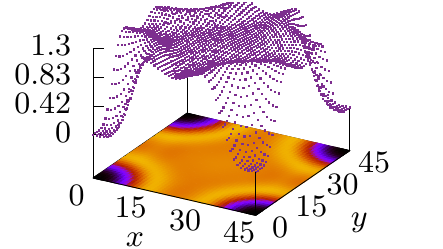}};
\node[overlay] at (-2.3,1.2) {(a)};
\phantomsubcaption
\label{fig:fermion_corr_C1r1N9}
\end{tikzpicture}
\end{minipage}%
\begin{minipage}[b]{.5\linewidth}
\centering\includegraphics{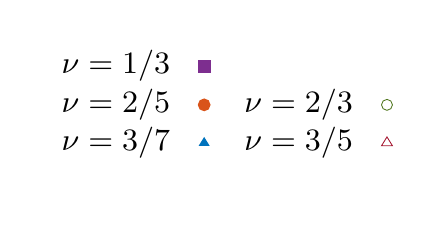}
\end{minipage}%

\vskip\baselineskip

\begin{minipage}[b]{.5\linewidth}
\begin{tikzpicture}
\node at (0,0) {\centering\includegraphics{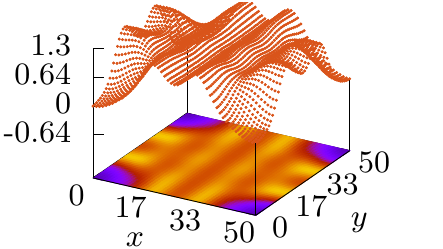}};
\node[overlay] at (-2.3,1.2) {(b)};
\phantomsubcaption
\label{fig:fermion_corr_C1r2N8}
\end{tikzpicture}
\end{minipage}%
\begin{minipage}[b]{.5\linewidth}
\begin{tikzpicture}
\node at (0,0) {\centering\includegraphics{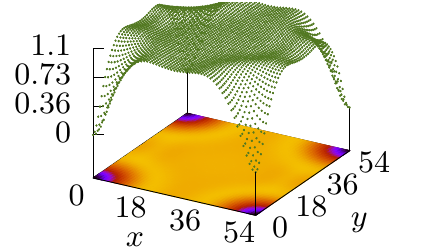}};
\node[overlay] at (-2.3,1.2) {(c)};
\phantomsubcaption
\label{fig:fermion_corr_C1r-2N18}
\end{tikzpicture}
\end{minipage}%

\vskip\baselineskip

\begin{minipage}[b]{.5\linewidth}
\begin{tikzpicture}
\node at (0,0) {\centering\includegraphics{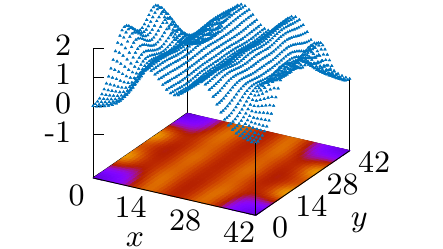}};
\node[overlay] at (-2.3,1.2) {(d)};
\phantomsubcaption
\label{fig:fermion_corr_C1r3N9}
\end{tikzpicture}
\end{minipage}%
\begin{minipage}[b]{.5\linewidth}
\begin{tikzpicture}
\node at (0,0) {\centering\includegraphics{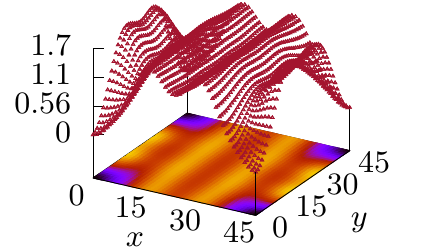}};
\node[overlay] at (-2.3,1.2) {(e)};
\phantomsubcaption
\label{fig:fermion_corr_C1r-3N9}
\end{tikzpicture}
\end{minipage}%

\end{minipage}
\caption{Density-density correlation functions for fermionic states in the $|C|=1$ band. The plots are shown for the lowest-lying ground state in the $(k_x,k_y)=(0,0)$ momentum sector, with (a)~$r=1$: $\nu=1/3$, $N=9$, $p=76$; (b)~$r=2$: $\nu=2/5$, $N=8$, $p=124$; (c)~$r=-2$: $\nu=2/3$, $N=18$, $p=107$; (d)~$r=3$: $\nu=3/7$, $N=9$, $p=85$; (e)~$r=-3$: $\nu=3/5$, $N=9$, $p=134$.}
\label{fig:fermion_corr_C1}
\end{figure}
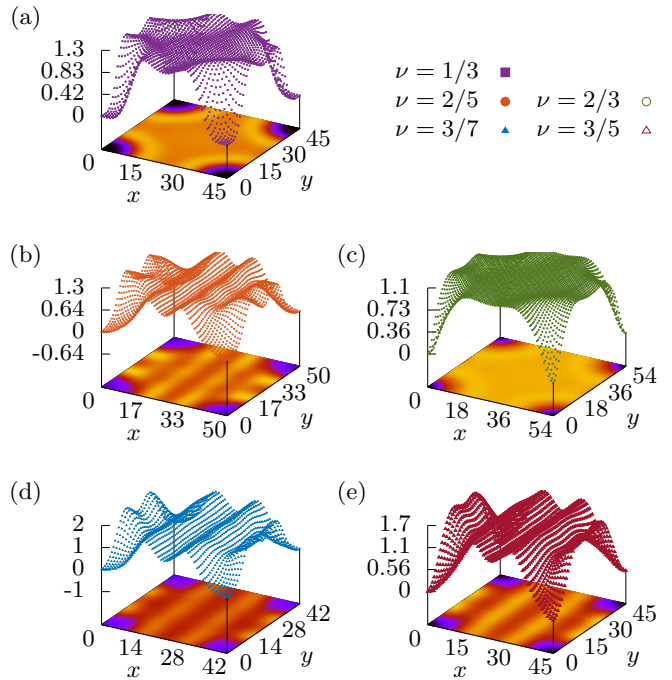

Density-density correlation functions for the ground states of the considered range of fillings are shown in Fig.~\ref{fig:fermion_corr_C1}. As expected for spinless fermions, the correlation at zero separation is identically zero due to Pauli exclusion. We note that the Laughlin state in Fig.~\ref{fig:fermion_corr_C1r1N9}, as well as the particle-hole symmetry-related $\nu=2/3$ state in Fig.~\ref{fig:fermion_corr_C1r-2N18}, tend to a constant correlation at large distances. However, the zero-separation correlation hole is more distinct for the Laughlin state in Fig.~\ref{fig:fermion_corr_C1r1N9}, analogous to that observed for the bosonic Laughlin state in Fig.~\ref{fig:boson_corr_C1r1N8}. As for $|C|=1$ bosons, the $r=2$ fermion state shows minor isotropic fluctuations at large distances, as shown in Fig.~\ref{fig:fermion_corr_C1r2N8}, which may be due to finite-size effects. The density-density correlation functions for the $|r|=3$ states in Figs.~\ref{fig:fermion_corr_C1r3N9} and~\ref{fig:fermion_corr_C1r-3N9}, however, show large anisotropic oscillations in the $y$ direction. For the $\nu=3/7$ state in Fig.~\ref{fig:fermion_corr_C1r3N9}, for example, we observe a global maximum of almost double the constant value at large distance observed for the Laughlin state in Fig.~\ref{fig:fermion_corr_C1r1N9}. Again, these directional oscillations in the $|r|=3$ states may be indicative of a charge density wave instability.

Overall, obtaining the fermion data is more computationally expensive than the corresponding data for bosons due to the higher ground-state degeneracies. However, for $|C|=1$, the Hilbert space dimensions are nearly identical, allowing a large number of geometries and system sizes. As before, we conclude that the scaling relations for the gap yield a well-defined continuum limit for large $q$ in all cases. As for bosons, the composite fermion prediction for the stability hierarchy is not observed, as the $\nu=3/7$ state appears to have a larger gap than the $\nu=2/5$ state, possibly signaling a different intervening phase. However, to the extent that our data are conclusive, they indicate that all examined states can have a finite gap in the thermodynamic continuum limit.

\subsection{FCIs in $|C|=2$ Bands}
\label{subsec:FCI_C2}

The preceding study of $|C|=1$ bands in the continuum limit provides a solid foundation from which to explore higher Chern number bands. However, naively extending the analysis in Sec.~\ref{subsec:FCI_C1} presents two major challenges. First, the Hilbert space dimension for systems with a higher Chern number is considerably larger, since the filling factor is reduced, and thus calculations at the same particle numbers are exponentially more expensive. Additionally, the systematic process of obtaining square configurations, outlined in Sec.~\ref{subsec:lattice_geometries}, is often too constricting to yield an adequate number of square configurations for higher Chern numbers. This is a geometric problem, which can be overcome by finding approximately square configurations for the problem cases. 

\subsubsection{Bosonic States}
\label{subsubsec:FCI_C2_bosons}

\begin{figure}
\begin{center}
\includegraphics{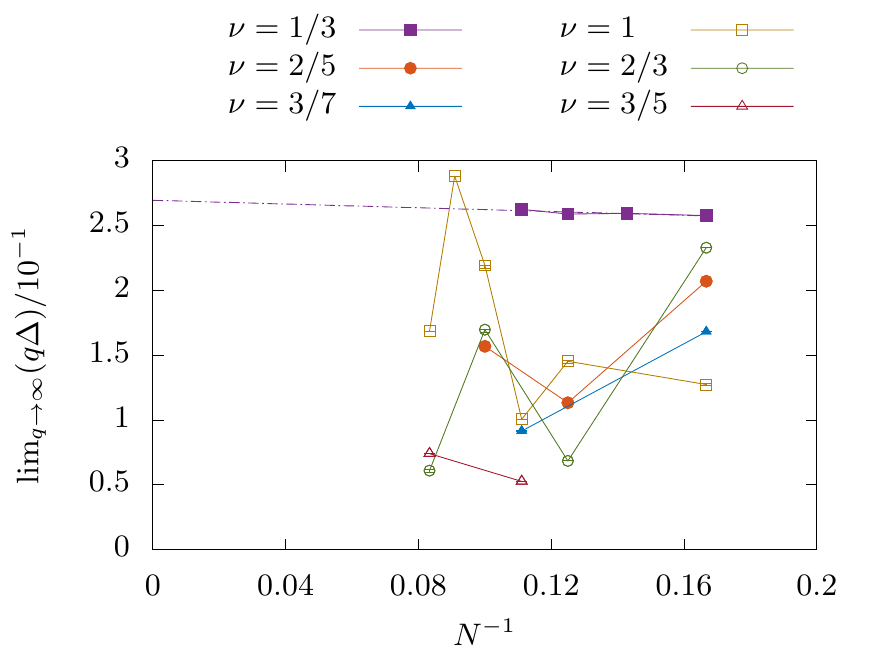}
\caption{Finite-size scaling of the gap to the thermodynamic effective continuum limit at fixed aspect ratio, for bosonic states in the $|C|=2$ band. The extrapolation to the $y$ axis is shown for the robust $\nu=1/3$ states. Squares, circles, and triangles denote states with $|r|=1,2,3$, respectively, where the filled (hollow) symbols correspond to positive (negative) $r$. All error bars due to the effective continuum limit are smaller than the symbols on the scale of the plot.}
\label{fig:bosons_final_plot_C2}
\end{center}
\end{figure}

As before, we start with bosonic systems with onsite interactions, considering filling factors of the series (\ref{eq:FillingFactorJainEquiv}) with $|r|=1,2,3$. Again, we include particle numbers with Hilbert space dimensions of $\text{dim}\{\mathcal{H}\}<10^7$. Overall, we have considered 23 different combinations of particle number and filling factor, with an average of $\sim 22$ different geometries for each, and a total of 510 different exact diagonalization calculations underlying the data in this section. The final data for the effective continuum limiting behavior for the six filling factors under consideration are shown in Fig.~\ref{fig:bosons_final_plot_C2}. Notice that the $q \Delta$ values are smaller than in the corresponding cases for $|C|=1$ bands in Fig.~\ref{fig:bosons_final_plot_C1}. The $r=1$ series is again almost completely unaffected by finite-size scaling, with an extrapolated thermodynamic effective continuum limit of $\lim_{N,q\to\infty}(q\Delta)=0.27\pm(4.4\times 10^{-3})$\cite{Note1}.  Finite-size effects are noticeable for all other series. We will first discuss the scaling to the effective continuum limit and then provide further discussion of the finite-size scaling for the different states.
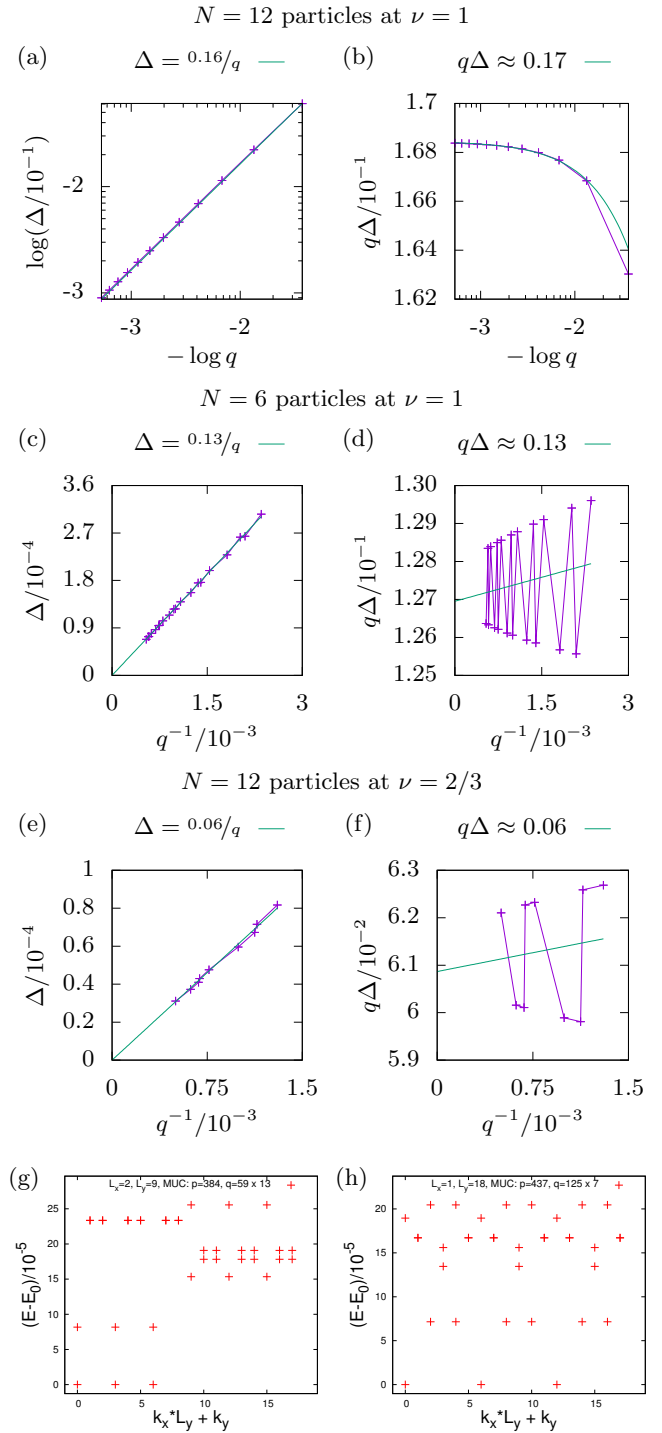
\begin{figure}
\begin{minipage}[b]{\linewidth}
$N=12$ particles at $\nu=1$
\begin{minipage}[b]{.5\linewidth}
\begin{tikzpicture}
\node at (0,0) {\centering\includegraphics{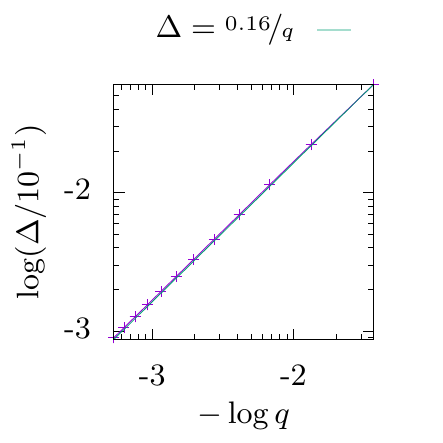}};
\node[overlay] at (-2,2) {(a)};
\phantomsubcaption
\label{fig:boson_plots_C2_1_a}
\end{tikzpicture}
\end{minipage}%
\begin{minipage}[b]{.5\linewidth}
\begin{tikzpicture}
\node at (0,0) {\centering\includegraphics{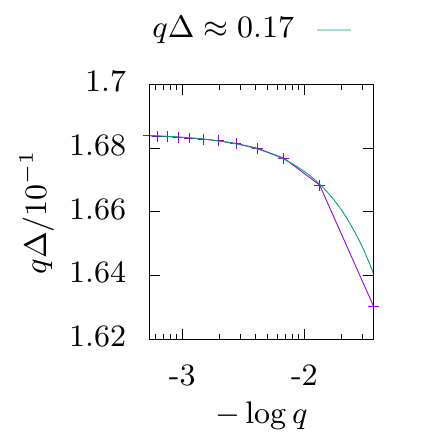}};
\node[overlay] at (-2,2) {(b)};
\phantomsubcaption
\label{fig:boson_plots_C2_1_b}
\end{tikzpicture}
\end{minipage}%
\end{minipage}
\begin{minipage}[b]{\linewidth}
$N=6$ particles at $\nu=1$
\begin{minipage}[b]{.5\linewidth}
\begin{tikzpicture}
\node at (0,0) {\centering\includegraphics{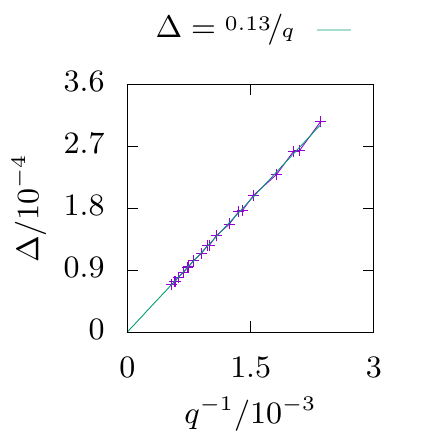}};
\node[overlay] at (-2,2) {(c)};
\phantomsubcaption
\label{fig:boson_plots_C2_2_c}
\end{tikzpicture}
\end{minipage}%
\begin{minipage}[b]{.5\linewidth}
\begin{tikzpicture}
\node at (0,0) {\centering\includegraphics{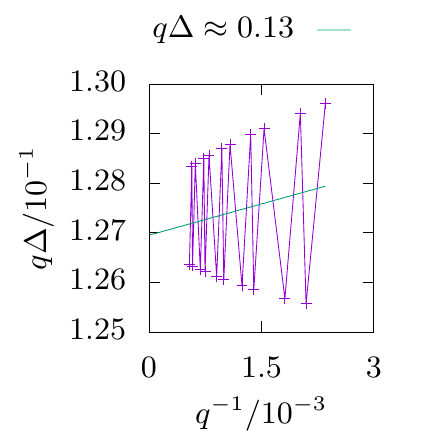}};
\node[overlay] at (-2,2) {(d)};
\phantomsubcaption
\label{fig:boson_plots_C2_2_d}
\end{tikzpicture}
\end{minipage}%
\end{minipage}
\begin{minipage}[b]{\linewidth}
$N=12$ particles at $\nu=2/3$
\begin{minipage}[b]{.5\linewidth}
\begin{tikzpicture}
\node at (0,0) {\centering\includegraphics{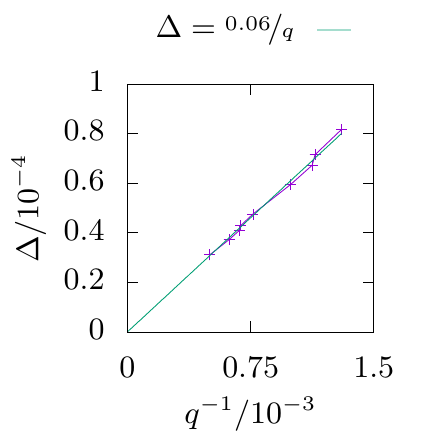}};
\node[overlay] at (-2,2) {(e)};
\phantomsubcaption
\label{fig:boson_plots_C2_3_e}
\end{tikzpicture}
\end{minipage}%
\begin{minipage}[b]{.5\linewidth}
\begin{tikzpicture}
\node at (0,0) {\centering\includegraphics{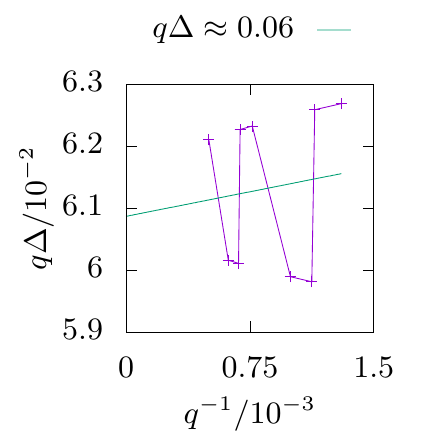}};
\node[overlay] at (-2,2) {(f)};
\phantomsubcaption
\label{fig:boson_plots_C2_3_f}
\end{tikzpicture}
\end{minipage}%
\\
\begin{minipage}[b]{.49\linewidth}
\begin{tikzpicture}
\node at (0,0) {\includegraphics[width=\linewidth]{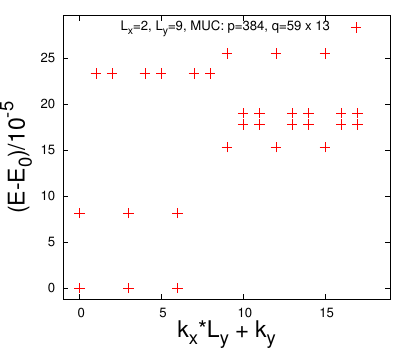}};
\node[overlay] at (-2,1.6) {(g)};
\phantomsubcaption
\label{fig:boson_spectra_C2_1}
\end{tikzpicture}
\end{minipage}
\begin{minipage}[b]{.49\linewidth}
\begin{tikzpicture}
\node at (0,0) {\includegraphics[width=\linewidth]{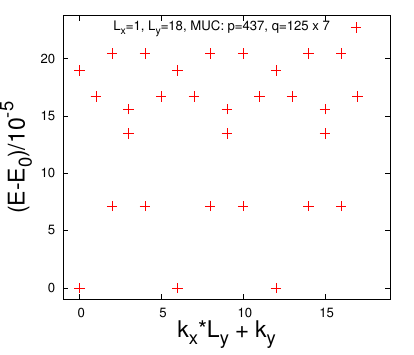}};
\node[overlay] at (-2,1.6) {(h)};
\phantomsubcaption
\label{fig:boson_spectra_C2_2}
\end{tikzpicture}
\end{minipage}
\end{minipage}
\caption{[(a)--(f)] Magnitude of the gap for bosonic states in the $|C|=2$ band, as a function of MUC size, $q$. [(g), (h)] Energy spectra for the bosonic 12-particle $\nu=2/3$ state in the $|C|=2$ band, at (g)~$p=384$, (h)~$p=437$. The plots are resolved to $n=2$ points per sector.}
\label{fig:boson_plots_C2}
\end{figure}

We find that the many-body gap scales inversely with $q$ for the $|C|=2$ bands, also. However, we find stronger fluctuations of the scaled gap around the limiting value, which is partly related to the absence of square geometries. We illustrate common behaviors by examining three examples in detail. 

In Figs.~\ref{fig:boson_plots_C2_1_a} and~\ref{fig:boson_plots_C2_1_b}, we display the 12-particle $\nu=1$ state. We choose this state as it has a high particle number and behaves in the familiar and expected way; i.e.,~it produces an adequate number of square configurations and its energy gap can be determined without any ambiguity. 

In Figs.~\ref{fig:boson_plots_C2_2_c} and~\ref{fig:boson_plots_C2_2_d}, we display the six-particle $\nu=1$ state. We choose this state as an example of a case which does not produce an adequate number of (or, indeed, any) square configurations, in accordance to our systematic method (see Sec.~\ref{subsec:lattice_geometries}). Therefore, for this case, we consider all configurations which are within an error $\epsilon\leq 2\%$ of being square, as this gives an adequate and comparable sample size of $\sim10$ configurations. This is noticeable by the deviations from a straight line in Fig.~\ref{fig:boson_plots_C2_2_c} and in the clear oscillations in Fig.~\ref{fig:boson_plots_C2_2_d}. The various rectangular configurations obey slightly different scaling relations with MUC size, which results in noticeable oscillations in the plots (note, however, the small scale on the $y$ axes). In the cases where we use approximately square configurations, the error in the effective continuum limit is no longer negligible on the scale of the thermodynamic limit plot and so must be taken into account. The precise determination of errors for the effective continuum limit of $q\to\infty$ is discussed in Appendix~\ref{sec:error_analysis}.

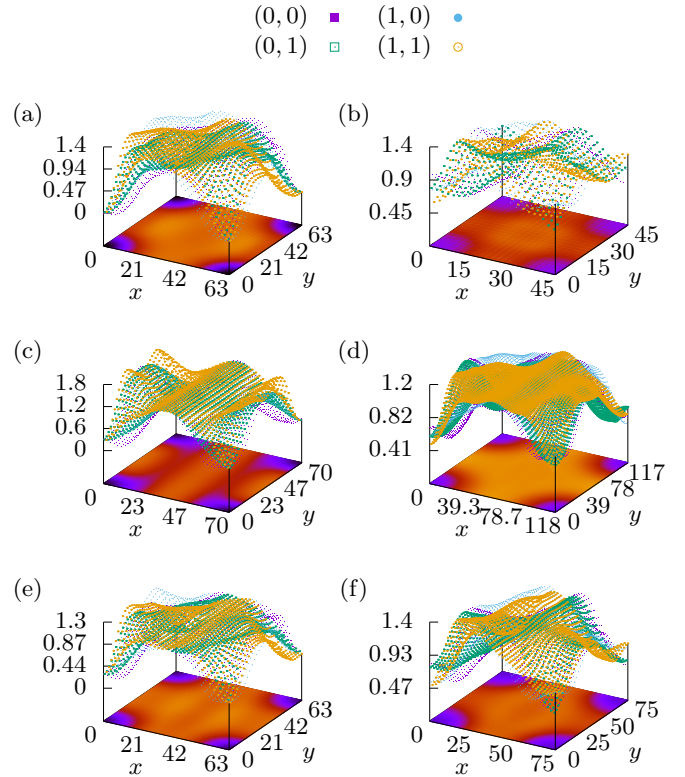
\begin{figure}
\begin{minipage}[b]{\linewidth}

\centering

\begin{minipage}[b]{\linewidth}
\centering\includegraphics{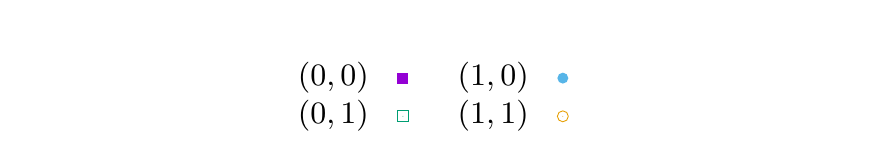}
\end{minipage}%

\vskip\baselineskip

\begin{minipage}[b]{.5\linewidth}
\begin{tikzpicture}
\node at (0,0) {\centering\includegraphics{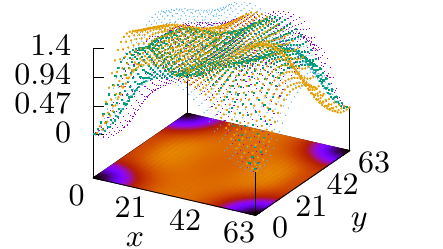}};
\node[overlay] at (-2.3,1.2) {(a)};
\phantomsubcaption
\label{fig:boson_corr_C2r1N7}
\end{tikzpicture}
\end{minipage}%
\begin{minipage}[b]{.5\linewidth}
\begin{tikzpicture}
\node at (0,0) {\centering\includegraphics{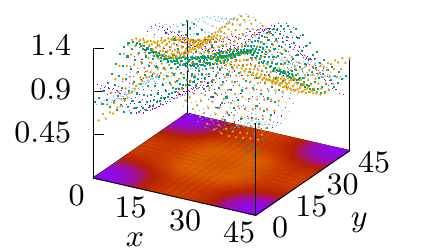}};
\node[overlay] at (-2.3,1.2) {(b)};
\phantomsubcaption
\label{fig:boson_corr_C2r-1N9}
\end{tikzpicture}
\end{minipage}%

\vskip\baselineskip

\begin{minipage}[b]{.5\linewidth}
\begin{tikzpicture}
\node at (0,0) {\centering\includegraphics{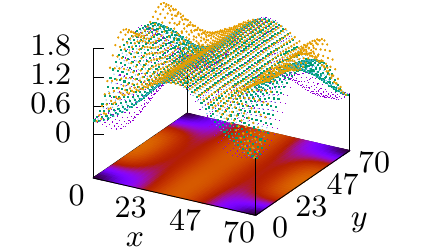}};
\node[overlay] at (-2.3,1.2) {(c)};
\phantomsubcaption
\label{fig:boson_corr_C2r2N8}
\end{tikzpicture}
\end{minipage}%
\begin{minipage}[b]{.5\linewidth}
\begin{tikzpicture}
\node at (0,0) {\centering\includegraphics{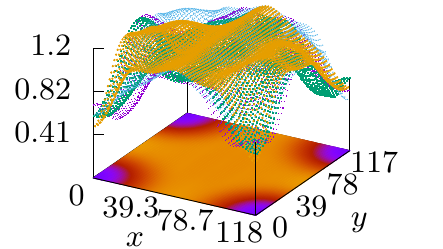}};
\node[overlay] at (-2.3,1.2) {(d)};
\phantomsubcaption
\label{fig:boson_corr_C2r-2N12}
\end{tikzpicture}
\end{minipage}%

\vskip\baselineskip

\begin{minipage}[b]{.5\linewidth}
\begin{tikzpicture}
\node at (0,0) {\centering\includegraphics{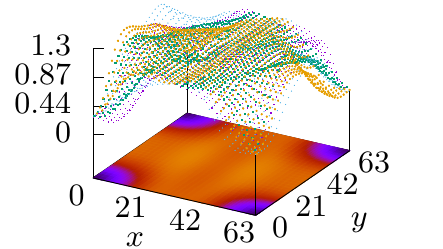}};
\node[overlay] at (-2.3,1.2) {(e)};
\phantomsubcaption
\label{fig:boson_corr_C2r3N9}
\end{tikzpicture}
\end{minipage}%
\begin{minipage}[b]{.5\linewidth}
\begin{tikzpicture}
\node at (0,0) {\centering\includegraphics{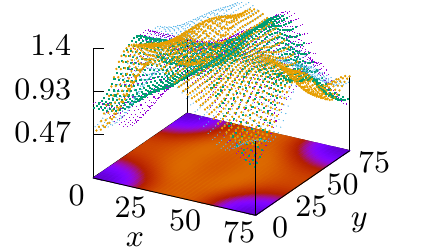}};
\node[overlay] at (-2.3,1.2) {(f)};
\phantomsubcaption
\label{fig:boson_corr_C2r-3N9}
\end{tikzpicture}
\end{minipage}%

\end{minipage}
\caption{Density-density correlation functions for bosonic states in the $|C|=2$ band. The plots are shown for the lowest-lying ground state in the $(k_x,k_y)=(0,0)$ momentum sector. The legend differentiates between the correlation functions at lattice positions $(x \bmod 2, y \bmod 2)$, as explained in the main text. We show data for (a)~$r=1$: $\nu=1/3$, $N=7$, $p=94$; (b)~$r=-1$: $\nu=1$, $N=9$, $p=112$; (c)~$r=2$: $\nu=2/5$, $N=8$, $p=123$; (d)~$r=-2$: $\nu=2/3$, $N=12$, $p=384$; (e)~$r=3$: $\nu=3/7$, $N=9$, $p=94$; and (f)~$r=-3$: $\nu=3/5$, $N=9$, $p=188$.}
\label{fig:boson_corr_C2}
\end{figure}

Finally, in Figs.~\ref{fig:boson_plots_C2_3_e} and~\ref{fig:boson_plots_C2_3_f}, we display the 12-particle $\nu=2/3$ state. We choose this state as a case of interest because it is the largest system size for the $\nu=2/3$ state, shown in Fig.~\ref{fig:bosons_final_plot_C2}. Yet, it retains a strong geometry dependency.
%This is a case of interest because of the clear oscillations in the filling factor series, which are known to continue to at least $N=14$~\cite{Moller:2015kg}. 
Note that these data are based on configurations which are within $\epsilon \leq 1\%$ of being square. Figure~\ref{fig:boson_plots_C2_3_e} shows the overall reciprocal scaling of the many-body gap with MUC size. However, from Fig.~\ref{fig:boson_plots_C2_3_f}, we can see that there seem to be two different rectangular configurations which have distinct scaling behaviors. By closely examining the energy spectra, this is indeed the case. Figures.~\ref{fig:boson_spectra_C2_1} and~\ref{fig:boson_spectra_C2_2} show the spectra for the two distinct rectangular configuration geometries present in the sample: the $L_x \times L_y = 2 \times 9$ and $L_x \times L_y = 1 \times 18$ cases (taking MUCs with the largest possible $L_y$ extension). In addition, the density-density correlation function corresponding to the $L_x \times L_y = 2 \times 9$ case is presented in Fig.~\ref{fig:boson_corr_C2r-2N12}. From composite fermion theory, we expect the degeneracy of the ground state to be 3, which is indeed what we observe; and the degeneracy is even clearly visible in the spectra, since the ground states happen to be in different momentum sectors. Yet, there is a discrepancy between the energy gaps for the two rectangular configurations, which is larger than the fluctuations of the previously discussed data at $\nu=1$ with $\epsilon \leq 2\%$ deviations from square geometries, in Fig.~\ref{fig:boson_plots_C2_2_d}. Nonetheless, any errors from the extrapolation to the effective continuum limit remain small compared to the finite-size fluctuations of the gap visible in Fig.~\ref{fig:bosons_final_plot_C2}. %The PES for both of these systems, shown in Fig.~\ref{fig:boson_entanglement_C2}, displays an entanglement gap, in both cases, which indicates the role of topological phases.

Overall, the $\nu=2/3$ state has the strongest finite-size effects, with smaller gaps for configurations with $N$ divisible by four: In its finite-size scaling, we see clear oscillations of the many-body gap under addition of pairs of particles. The next larger system size at $N=14$ was found to have a larger gap in the previous study at fixed flux density~\cite{Moller:2015kg}. Hence, the low value at $N=12$ should not be taken as an indication of the vanishing of the gap in the thermodynamic effective continuum limit. The correlation function for the 12-particle $\nu=2/3$ state corresponding to Fig.~\ref{fig:boson_spectra_C2_1} is shown in Fig.~\ref{fig:boson_corr_C2r-2N12}. Here we observe that charge density wave instabilities may also play a role. 

The $|C|=2$, $\nu=1$ state is the second candidate for a BIQHE state within the series (\ref{eq:FillingFactorJainEquiv}). Here, we consistently find a nondegenerate ground state in our numerical analysis, as predicted by composite fermion theory. The many-body gap in Fig.~\ref{fig:bosons_final_plot_C2} shows significant finite-size effects, precluding us from taking a quantitative extrapolation to the thermodynamic effective continuum limit. However, its magnitude is consistently nonzero for all available system sizes. We further note that the realizations of the BIQHE in $|C|=2$ optical flux lattices were likewise found to have a significant geometry dependency of the many-body gaps~\cite{Sterdyniak:2015jo}. Notwithstanding these finite-size effects, we stress that all geometries allow for a clear identification of a singly degenerate ground state. This is unlike the case of the potential $\nu=2$ BIQHE state in the lowest $|C|=1$ Hofstadter bands, where competing phases appear to dominate, as we have discussed in Sec.~\ref{subsubsec:FCI_C1_bosons}.

The $\nu=2/5$ state demonstrates a robust many-body gap for the states considered. The Hilbert space dimension of $|C|=2$ bosons is comparable to that of $|C|=1$ fermions in Fig.~\ref{fig:fermions_final_plot_C1} and so the data are limited due to computational expense. Nevertheless, the remaining filling factor series show the potential for a robust thermodynamic effective continuum limit. We note the lack of particle-hole symmetry for the $\nu=2/5$ and $\nu=3/5$ filling factor series, visible in Fig.~\ref{fig:bosons_final_plot_C2}, unlike for the $|C|=1$ fermions in Fig.~\ref{fig:fermions_final_plot_C1}. Additionally, we observe approximately the predicted composite fermion hierarchy of gaps for $r=-1,-2,-3$, as well as for $r=1,2,3$. Unfortunately, finite-size effects dominate extrapolation errors from $q\to \infty$, and so preclude a clear extrapolation to the thermodynamic effective continuum limit.

The correlation functions for the six filling factors under consideration are shown in Fig.~\ref{fig:boson_corr_C2}. Notice the appearance of four distinct correlation sheets. We differentiate between the sheets by labeling them corresponding to the $|C|^2$ possible solutions for $(x\bmod|C|,y\bmod|C|)$, where $x$ and $y$ denote the $x$- and $y$-axis lattice positions. Hence, the modulation along the $x$ and $y$ axes of period $|C|$ leads to the appearance of $|C|^2$ smooth correlation sheets. However, in a finite-size system, some of these sheets may be related by inversion symmetries of the type $x_i\leftrightarrow L_i-x_i$ whenever $L_i \bmod C \neq 0$. This observation seems to contradict models of higher Chern number $|C|$ bands as effective multilayer fractional quantum Hall systems composed of $|C|$ layers~\cite{Palmer:2006km,Palmer:2008cq,2012PhRvL.108y6809H,Moller:2014fa,Harper:2014fq}. It is unclear at present how to reconcile this view with our observations, as the conventional multilayer view allows for no more than $|C|$ distinct correlation functions. Although it is possible that a suitable basis could be found in which the number of sheets decreases, this would likely require a nonlocal transformation mixing several sites within the unit cell. On the other hand, it is plausible that a $|C|$-fold periodicity should appear along each axis, given that the single-particle wave functions of Harper's equation show such behavior~\cite{Harper:1955bj,Palmer:2008cq,2012PhRvL.108y6809H,Harper:2014fq}. For this single-particle problem in the Landau gauge, one singles out one of the axes for momentum conservation, so that $|C|$-fold oscillations in the eigenstates occur only in the perpendicular direction. However, as the problem is gauge invariant, either permutation of the two axes could be chosen to exhibit these behaviors. Furthermore, the correlation function should be isotropic in space in the infinite system. Hence, it appears natural that the correlation functions display $|C|$-fold periodicity along both axes.

The state at $\nu=1/3$ in Fig.~\ref{fig:boson_corr_C2r1N7} shows features reminiscent of the Laughlin states in the $|C|=1$ band, since this state also has positive flux attachment with one filled band in the composite fermion spectrum ($r=1$). We refer to such states as primary composite fermion states. The zero-separation correlation hole is most pronounced here and converges to zero for all of the correlation sheets; this is observed for all of the states with positive $r$ in Fig.~\ref{fig:boson_corr_C2}. Furthermore, the isotropic fluctuations at large distances show signs of settling, although it is hard to discern the limiting value of the correlation function in this case. Note that pairs of sheets are related by inversion symmetry for the specific geometry shown in the figure. This is a recurring feature for higher Chern bands. In the present case, we see that the $\{(0,0),(1,0)\}$ and $\{(0,1),(1,1)\}$ pairs are related along the $x$ axis; and the $\{(0,0),(0,1)\}$ and $\{(1,0),(1,1)\}$ pairs are related along the $y$ axis. Due to the large number of data points and intricacy of these figures, the data are available, along with this paper, to view interactively online as Supplementary Material~\footnote{Supplemental Material at \url{http://bartholomewandrews.com/data/correlation_plots.zip} includes the original data for the correlation functions and instructions to view these interactively.}. 

The correlation functions for the $\nu=1$ and $\nu=2/3$ fillings in Figs.~\ref{fig:boson_corr_C2r-1N9} and~\ref{fig:boson_corr_C2r-2N12} are similarly isotropic at large distances with comparable global maxima. However, the correlation sheets for these separate cases do not converge to a unique value at the correlation hole. In contrast, the correlation functions for the $\nu=2/5$, $\nu=3/7$, and $\nu=3/5$ filling factor series, in Figs.~\ref{fig:boson_corr_C2r2N8},~\ref{fig:boson_corr_C2r3N9}, and~\ref{fig:boson_corr_C2r-3N9}, show signs of anisotropy with directional oscillations, which may be indicative of competing charge density wave instabilities. Note that signs of charge density waves were also observed for the corresponding $r$ values ($r=-3,2,3$) for fermions in $|C|=1$ bands, shown above in Fig.~\ref{fig:fermion_corr_C1}.

\begin{figure}
\centering
\begin{minipage}[b]{.49\linewidth}
\begin{tikzpicture}
\node at (0,0) {\includegraphics[width=\linewidth]{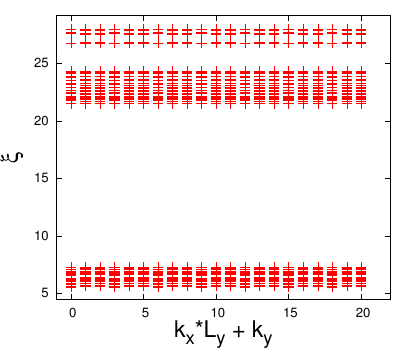}};
\node[overlay] at (-2,1.6) {(a)};
\phantomsubcaption
\label{fig:boson_entanglement_C2_1}
\end{tikzpicture}
\end{minipage}
\begin{minipage}[b]{.49\linewidth}
\begin{tikzpicture}
\node at (0,0) {\includegraphics[width=\linewidth]{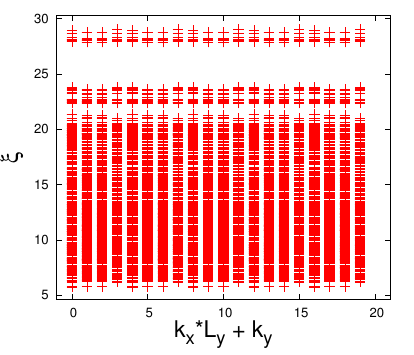}};
\node[overlay] at (-2,1.6) {(b)};
\phantomsubcaption
\label{fig:boson_entanglement_C2_2}
\end{tikzpicture}
\end{minipage}
\caption{PES for bosonic states in the $|C|=2$ band. We show data for: a) $r=1$: $\nu=1/3$, $N=7$, $p=94$; b) $r=2$: $\nu=2/5$, $N=8$, $p=123$. In both cases we take $N_A=\lfloor{N/2}\rfloor$. The counts of eigenstates from the bottom of the spectra up to the principal entanglement gaps, in each of the momentum sectors, is a) 31, 30, 30 (repeated for 21 sectors), b) 441, 430, 430, 430 (repeated for 20 sectors), respectively.}
\label{fig:boson_entanglement_C2}
\end{figure}

Next, we examine the particle entanglement spectra of the examined quantum liquids. In general, the PES for these series are gapped confirming the existence of a topological phase. For instance, the PES for the selected states in Fig.~\ref{fig:boson_corr_C2} have principal entanglement gaps, $\Delta_\xi$, of (a) 14.25, (b) 1.37, (c) 4.09, (d) 1.12, and (e) 1.85, after tracing out $\lfloor{N/2}\rfloor$ particles. The count of eigenstates below the principal entanglement gaps for these states, in each of the momentum sectors, are (a) 31, 30, 30 (repeated for 21 sectors), (b) 53 (repeated for 9 sectors), (c) 441, 430, 430, 430 (repeated for 20 sectors), (d) 5605, 5586, 5601, 5583, 5601, 5583 (repeated for 18 sectors), (e) 504 (repeated for 21 sectors), and (f) 198 (repeated for 15 sectors), respectively. The spectra corresponding to the correlation functions in Figs.~\ref{fig:boson_corr_C2r1N7} and~\ref{fig:boson_corr_C2r2N8} are shown in Fig.~\ref{fig:boson_entanglement_C2}. Here, the primary composite fermion $\nu=1/3$ state in Fig.~\ref{fig:boson_entanglement_C2_1} shows the largest and clearest gap by a significant margin, as expected. All other states have a smaller principal entanglement gap higher in the spectrum, as in Fig.~\ref{fig:boson_entanglement_C2_2}. Unlike the primary composite fermion states, other states of the composite fermion series (\ref{eq:FillingFactorJainEquiv}) are not characterized by a generalized exclusion principle \cite{Bernevig:2012ka}, so they obey no simple counting rule for these numbers of quasiparticle states.

Overall, the bosonic series for the second Chern band presented some of the expected difficulties owing to the commensurability of several constraints on the geometries; however, these problems were largely overcome by allowing for a scaling in $q$. The only noticeable drawbacks, compared to the $|C|=1$ band, are the reduced number of data points, particularly for higher particle numbers, and the correspondingly larger uncertainty in the extrapolation as $q\to\infty$. As emphasized in the previous discussion, although considering approximately square configurations undoubtedly introduces error bars in the data, the deviation from square systems is not directly proportional to the error observed in the effective continuum limit. Rather, the subsequent error in the effective continuum limit is principally determined by the specific variation in the spectra for a given state.

\subsubsection{Fermionic States}
\label{subsubsec:FCI_C2_fermions}

We now extend our analysis to  fermions with NN interactions. For $|C|>1$, the Hilbert space dimensions for fermionic states are higher than those of the corresponding bosonic states due to the smaller filling factors and we thus expect to be able to compute fewer fermion states due to computational limitations. 
\begin{figure}
\begin{center}
\begin{tikzpicture}
\node at (0,0){\includegraphics{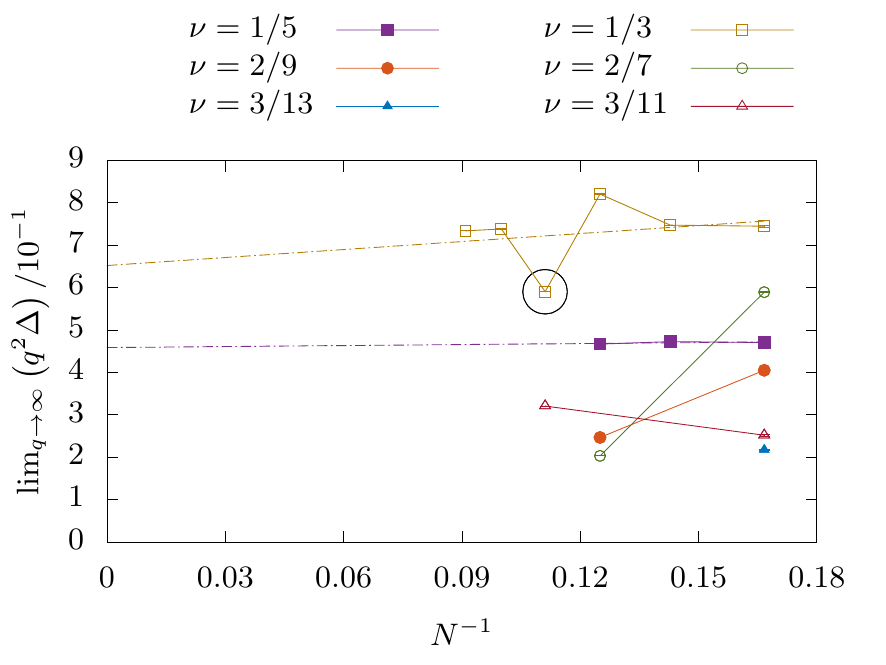}};
%\draw[overlay, black, thin] (1.04,0.4) circle(0.2cm);
%\draw[overlay, black, thick] [->] (0,0.4) -- (0.85,0.4);
\end{tikzpicture}
\caption{Finite-size scaling of the gap to the thermodynamic effective continuum limit at fixed aspect ratio, for fermionic states in the $|C|=2$ band. The extrapolation to the $y$ axis is shown for the robust $\nu=1/5$ and $\nu=1/3$ states. The $N=9$ data point for the $\nu=1/3$ series is circled to indicate that there is a competing topological phase present with $d=2$. Squares, circles, and triangles denote states with $|r|=1,2,3$, respectively, where the filled(hollow) symbols correspond to positive(negative) $r$. All error bars are smaller than the data points on the scale of the plot.}
\label{fig:fermions_final_plot_C2}
\end{center}
\end{figure}

Figure~\ref{fig:fermions_final_plot_C2} shows the data for the gap in the effective continuum limit for the six filling factors under consideration as a function of the inverse system size. Because of computational limitations, substantial data were only obtained for the $|r|=1$ series, while we have few data points for the other filling factors. Again, the $r=1$ primary composite fermion state shows the smallest finite-size effects on the many-body gap, and we extrapolate the thermodynamic effective continuum limit to be $\lim_{N,q\to\infty}(q^2\Delta)=0.46\pm0.02$\cite{Note1}. Note also that the magnitude of $q^2 \Delta$ values is lower than in the corresponding $|C|=1$ fermion plot in Fig.~\ref{fig:fermions_final_plot_C1}. Finite-size effects are noticeable for all series and the $q\to\infty$ extrapolation errors are much larger compared to the $|C|=2$ boson data. All of the fermion data was obtained using systems which were within $\delta R \leq 1\%$ of square simulation cells. Some of these systems were exactly square, but no filling fraction yields enough such geometries to use exact square systems exclusively, throughout the scaling procedure. More specifically, we have considered 18 different combinations of particle number and filling factor, with an average of $\sim 24$ different geometries for each. There are a total of 433 different exact diagonalization calculations underlying the data in this section.

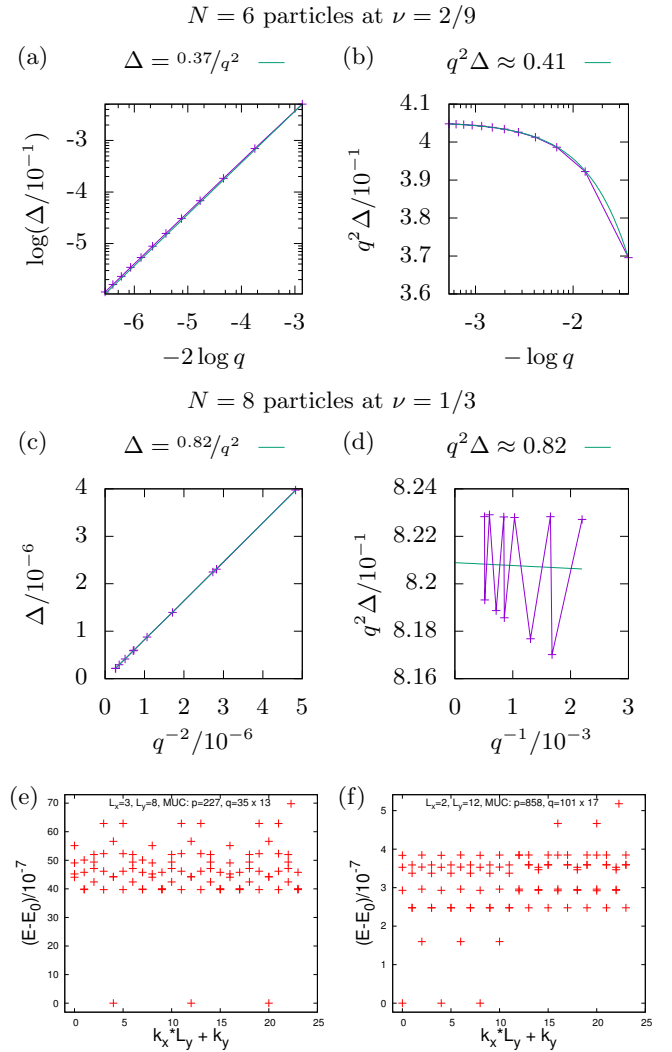
\begin{figure}
\begin{minipage}[b]{\linewidth}
$N=6$ particles at $\nu=2/9$
\begin{minipage}[b]{.5\linewidth}
\begin{tikzpicture}
\node at (0,0) {\centering\includegraphics{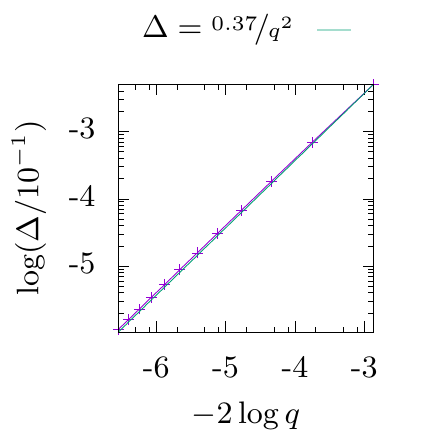}};
\node[overlay] at (-2,2) {(a)};
\phantomsubcaption
\label{fig:fermion_plots_C2_1_a}
\end{tikzpicture}
\end{minipage}%
\begin{minipage}[b]{.5\linewidth}
\begin{tikzpicture}
\node at (0,0) {\centering\includegraphics{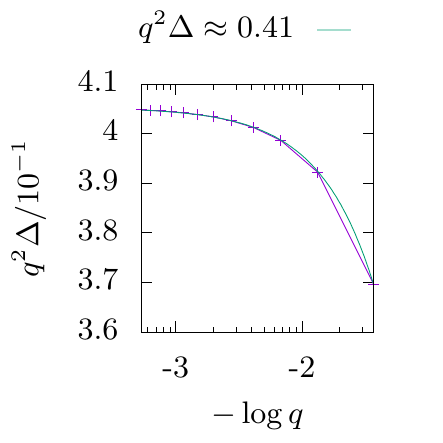}};
\node[overlay] at (-2,2) {(b)};
\phantomsubcaption
\label{fig:fermion_plots_C2_1_b}
\end{tikzpicture}
\end{minipage}%
\end{minipage}
\begin{minipage}[b]{\linewidth}
$N=8$ particles at $\nu=1/3$
\begin{minipage}[b]{.5\linewidth}
\begin{tikzpicture}
\node at (0,0) {\centering\includegraphics{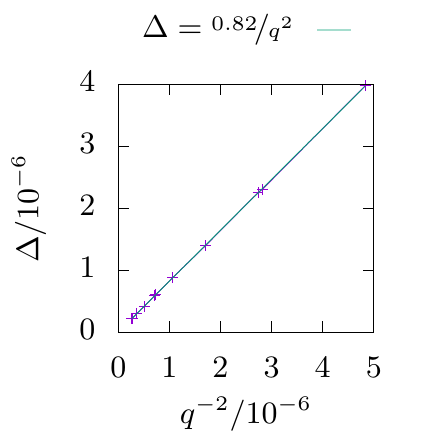}};
\node[overlay] at (-2,2) {(c)};
\phantomsubcaption
\label{fig:fermion_plots_C2_2_c}
\end{tikzpicture}
\end{minipage}%
\begin{minipage}[b]{.5\linewidth}
\begin{tikzpicture}
\node at (0,0) {\centering\includegraphics{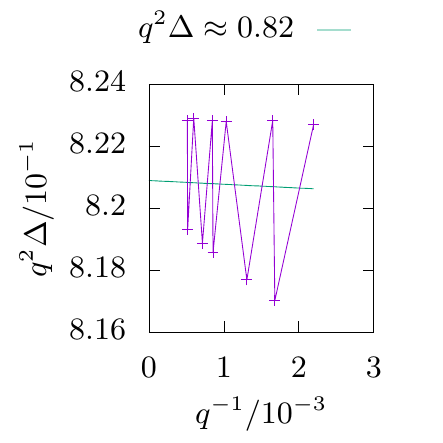}};
\node[overlay] at (-2,2) {(d)};
\phantomsubcaption
\label{fig:fermion_plots_C2_2_d}
\end{tikzpicture}
\end{minipage}%
\\
\begin{minipage}[b]{.49\linewidth}
\begin{tikzpicture}
\node at (0,0) {\includegraphics[width=\linewidth]{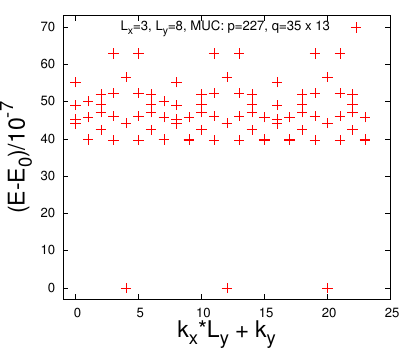}};
\node[overlay] at (-2,1.6) {(e)};
\phantomsubcaption
\label{fig:fermion_spectra_C2_1}
\end{tikzpicture}
\end{minipage}
\begin{minipage}[b]{.49\linewidth}
\begin{tikzpicture}
\node at (0,0) {\includegraphics[width=\linewidth]{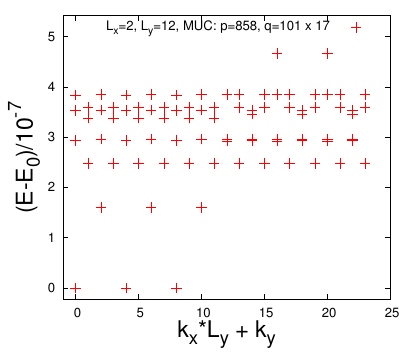}};
\node[overlay] at (-2,1.6) {(f)};
\phantomsubcaption
\label{fig:fermion_spectra_C2_2}
\end{tikzpicture}
\end{minipage}
\end{minipage}
\caption{[(a)--(d)] Magnitude of the gap for fermionic states in the $|C|=2$ band, as a function of MUC size, $q$. [(e), (f)] Energy spectra for the fermionic 8-particle $\nu=1/3$ state in the $|C|=2$ band, at (e)~$p=227$ and (f)~$p=858$. The plots are resolved to $n=4$ points per sector.}
\label{fig:fermion_plots_C2}
\end{figure}

Figures~\ref{fig:fermion_plots_C2_1_a} and~\ref{fig:fermion_plots_C2_1_b} show the plots for the 6-particle $\nu=2/9$ state. This is selected as an example of a state which has a clean scaling limit. The plots of $\Delta$ and $q^2 \Delta$ for the 8-particle $\nu=1/3$ state in Figs.~\ref{fig:fermion_plots_C2_2_c} and~\ref{fig:fermion_plots_C2_2_d} show slight oscillations due to the $\delta R \leq 1\%$ approximation in square configurations, similar to the bosonic states in Figs.~\ref{fig:boson_plots_C2_2_c} and~\ref{fig:boson_plots_C2_2_d}. However, these deviations are not as large as in the $|C|=2$ bosonic problem case in Fig.~\ref{fig:boson_plots_C2_3_f}. The effective continuum limit can be determined with a reasonable error. The spectra in Figs.~\ref{fig:fermion_spectra_C2_1} and~\ref{fig:fermion_spectra_C2_2} show the origin of the oscillations in Fig.~\ref{fig:fermion_plots_C2_2_d}. As with the $|C|=2$ bosons in Figs.~\ref{fig:boson_spectra_C2_1} and~\ref{fig:boson_spectra_C2_2}, we see a competition between two distinct rectangular geometries. The higher lying bands are more densely packed for the $L_x\times L_y=3\times 8$ system in Fig.~\ref{fig:fermion_spectra_C2_1} than for the $L_x\times L_y=2\times 12$ spectrum in Fig.~\ref{fig:fermion_spectra_C2_2}.

The $\nu=1/3$ series obtained for negative flux attachment ($r=-1$) has an exceptionally large gap with moderate finite-size effects and allows for a clear scaling of the many-body gap to the thermodynamic effective continuum limit. We extrapolate a limit of $\lim_{N,q\to\infty}(q\Delta)=0.65\pm0.16$ in this case\cite{Note1}. Note that at this filling factor, for $N=9$ we find that some lattice geometries realize a competing phase with $d=2$ instead of the degeneracy $d=3$ predicted by composite fermion theory. This competing phase appears to be topological, with a large entanglement gap of $\Delta_\xi = 6.40$ for $p=73$ ($N_A=4$), for example, and a corresponding eigenstate count of 385 (repeated for 27 sectors). As we find only few lattice geometries at this single system size showing this behavior, we do not attempt to further characterize this competing state.  For the purposes of the effective continuum limit shown in Fig.~\ref{fig:fermions_final_plot_C2}, only the geometries with the predicted threefold degeneracy were taken into account.

The data series for the remaining filling factors in Fig.~\ref{fig:fermions_final_plot_C2} show few points due to the steep Hilbert space dimension scaling with particle number for $|C|=2$ fermions. However, these series produce the correct ground-state degeneracies and the initial data have the potential for a robust gap in the thermodynamic effective continuum limit.

\begin{figure}
\begin{minipage}[b]{\linewidth}

\centering

\begin{minipage}[b]{\linewidth}
\centering\includegraphics{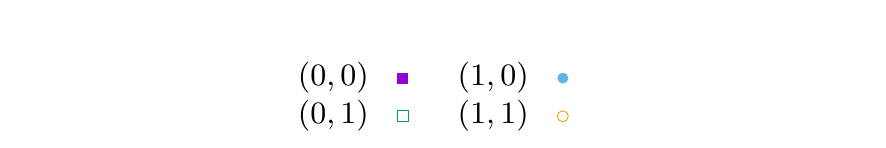}
\end{minipage}%

\vskip\baselineskip

\begin{minipage}[b]{.5\linewidth}
\begin{tikzpicture}
\node at (0,0) {\centering\includegraphics{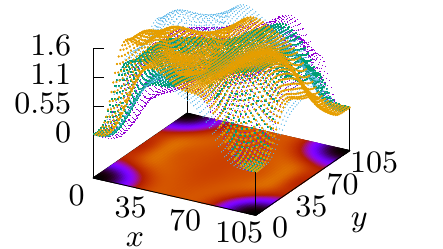}};
\node[overlay] at (-2.3,1.2) {(a)};
\phantomsubcaption
\label{fig:fermion_corr_C2r1N7}
\end{tikzpicture}
\end{minipage}%
\begin{minipage}[b]{.5\linewidth}
\begin{tikzpicture}
\node at (0,0) {\centering\includegraphics{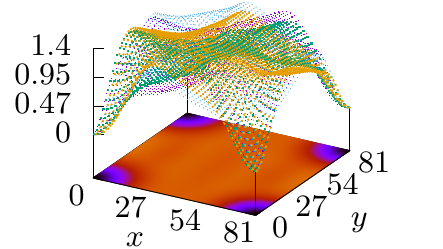}};
\node[overlay] at (-2.3,1.2) {(b)};
\phantomsubcaption
\label{fig:fermion_corr_C2r-1N9}
\end{tikzpicture}
\end{minipage}%

\vskip\baselineskip

\begin{minipage}[b]{.5\linewidth}
\begin{tikzpicture}
\node at (0,0) {\centering\includegraphics{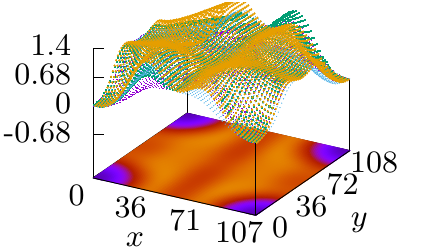}};
\node[overlay] at (-2.3,1.2) {(c)};
\phantomsubcaption
\label{fig:fermion_corr_C2r2N8}
\end{tikzpicture}
\end{minipage}%
\begin{minipage}[b]{.5\linewidth}
\begin{tikzpicture}
\node at (0,0) {\centering\includegraphics{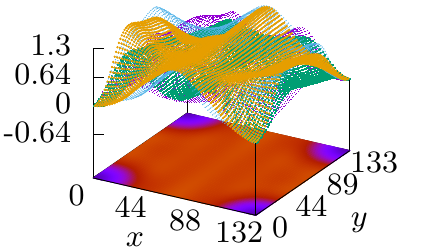}};
\node[overlay] at (-2.3,1.2) {(d)};
\phantomsubcaption
\label{fig:fermion_corr_C2r-2N8}
\end{tikzpicture}
\end{minipage}%

\vskip\baselineskip

\begin{minipage}[b]{.5\linewidth}
\begin{tikzpicture}
\node at (0,0) {\centering\includegraphics{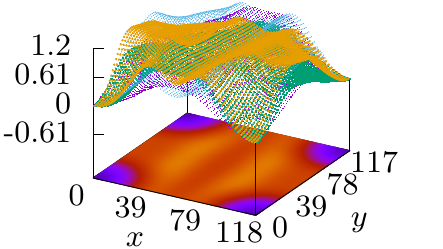}};
\node[overlay] at (-2.3,1.2) {(e)};
\phantomsubcaption
\label{fig:fermion_corr_C2r3N6}
\end{tikzpicture}
\end{minipage}%
\begin{minipage}[b]{.5\linewidth}
\begin{tikzpicture}
\node at (0,0) {\centering\includegraphics{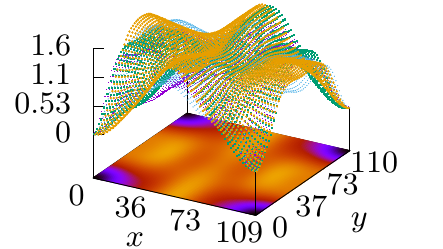}};
\node[overlay] at (-2.3,1.2) {(f)};
\phantomsubcaption
\label{fig:fermion_corr_C2r-3N6}
\end{tikzpicture}
\end{minipage}%

\end{minipage}
\caption{Density-density correlation functions for fermionic states in the $|C|=2$ band. The plots are shown for the lowest-lying ground state in the $(k_x,k_y)=(0,0)$ momentum sector, with sheets colored as in Fig.~\ref{fig:boson_corr_C2}. We show data for (a)~$r=1$: $\nu=1/5$, $N=7$, $p=157$; (b)~$r=-1$: $\nu=1/3$, $N=9$, $p=121$; (c)~$r=2$: $\nu=2/9$, $N=8$, $p=160$; (d)~$r=-2$: $\nu=2/7$, $N=8$, $p=313$; (e)~$r=3$: $\nu=3/13$, $N=6$, $p=265$; and (f)~$r=-3$: $\nu=3/11$, $N=6$, $p=272$.}
\label{fig:fermion_corr_C2}
\end{figure}
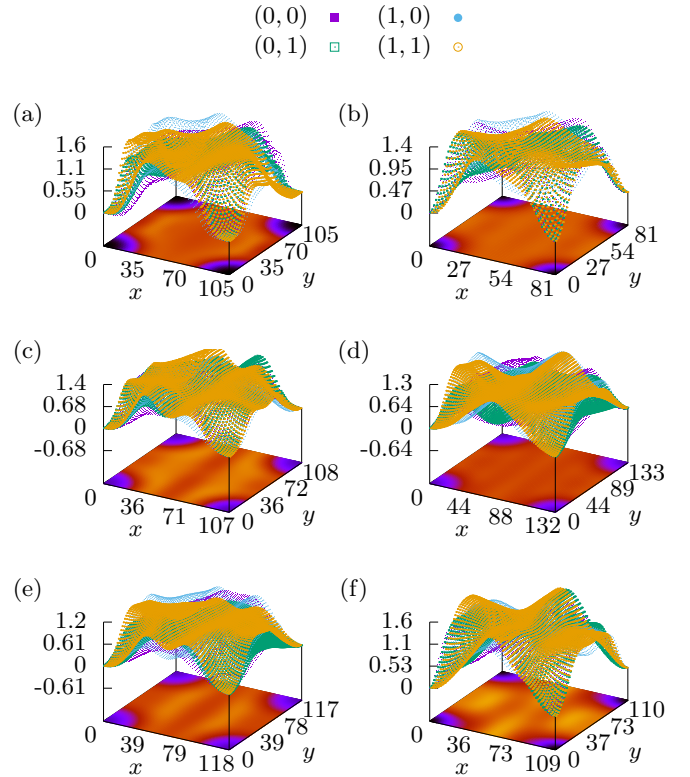

The correlation functions for the available filling factors are shown in Fig.~\ref{fig:fermion_corr_C2}. We note a few repeating characteristics that resemble features of states for the $|C|=1$ bands in Figs.~\ref{fig:boson_corr_C1} and~\ref{fig:fermion_corr_C1}, as well as the $|C|=2$ bosons in Fig.~\ref{fig:boson_corr_C2}. The primary composite fermion state in Fig.~\ref{fig:fermion_corr_C2r1N7} shows a pronounced correlation hole at zero separation and isotropic fluctuations at large distances. The fluctuations in this case are, however, larger than those in the corresponding boson plot in Fig.~\ref{fig:boson_corr_C2r1N7}. The correlation plots for the flux densities at $r=-1$ and $r=-2$ in Figs.~\ref{fig:fermion_corr_C2r-1N9} and~\ref{fig:fermion_corr_C2r-2N8} again show some degree of rotational symmetry and isotropy at large distances, whereas the plots with $r=-3,2,3$ in Figs.~\ref{fig:fermion_corr_C2r-3N6},~\ref{fig:fermion_corr_C2r2N8}, and~\ref{fig:fermion_corr_C2r3N6} show directional oscillations, potentially indicative of an instability due to charge density wave order. Recall that this was also observed for the $|C|=2$ bosons in Fig.~\ref{fig:boson_corr_C2} and the $|C|=1$ fermions in Fig.~\ref{fig:fermion_corr_C1}. The smooth correlation functions are again visibly split into $|C|^2$ sheets.     

%Figures~\ref{fig:fermion_plots_C2_2_c} \&~\ref{fig:fermion_plots_C2_2_d}, on the other hand, show the plots for the 8-particle $\nu=1/3$ state. This state is selected not simply because of the size of its error bars, but rather because its spectra exhibit interesting physical behavior. The plots in Fig.~\ref{fig:fermion_plots_C2_2_c} show slight fluctuations throughout, with larger fluctuations at small $q$, as expected. However, on this occasion there are also significant fluctuations at large $q$, as shown in Fig.~\ref{fig:fermion_plots_C2_2_d}, possibly due to numerical instability. Upon closer examination of the spectra of a particular rectangular configuration ($L_x\times L_y=3\times 8$), shown in Fig.~\ref{fig:fermion_spectra_C2_1}, we notice that low-lying states become almost degenerate, with a ground state degeneracy of 48, when the band geometry is made flatter by the larger MUC. This may be an indication that the state is in some topological phase. The $L_x\times L_y=3\times 8$ case was taken as an example, however this behavior is also observed for other rectangular geometries at this filling factor and particle number.

\begin{figure}
\centering
\begin{minipage}[b]{.49\linewidth}
\begin{tikzpicture}
\node at (0,0) {\includegraphics[width=\linewidth]{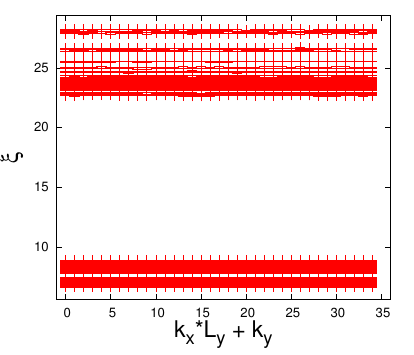}};
\node[overlay] at (-2,1.6) {(a)};
\phantomsubcaption
\label{fig:fermion_entanglement_C2_1}
\end{tikzpicture}
\end{minipage}
\begin{minipage}[b]{.49\linewidth}
\begin{tikzpicture}
\node at (0,0) {\includegraphics[width=\linewidth]{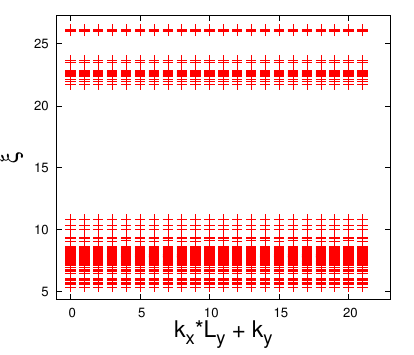}};
\node[overlay] at (-2,1.6) {(b)};
\phantomsubcaption
\label{fig:fermion_entanglement_C2_2}
\end{tikzpicture}
\end{minipage}
\caption{PES for fermionic states in the $|C|=2$ band. We show data for (a) $r=1$: $\nu=1/5$, $N=7$, $p=157$; and (b) $r=-3$: $\nu=3/11$, $N=6$, $p=272$. In both cases, we take $N_A=\lfloor{N/2}\rfloor=3$. The counts of eigenstates from the bottom of the spectrum up to the principal entanglement gap, in each of the momentum sectors, are (a) 77 and (b) 51.}
\label{fig:fermion_entanglement_C2}
\end{figure}

The PES for the fermionic series have notably large and clear gaps overall. For example, the spectra for the states in Fig.~\ref{fig:fermion_corr_C2} have $\Delta_{\xi}$ values of (a) 13.76, (b) 7.58, (c) 0.82, (d) 6.06, (e) 9.46, and (f) 10.84, after tracing out $\lfloor{N/2}\rfloor$ particles. The corresponding eigenstate counts from the bottom of the spectra up to the principal entanglement gaps, in each of the momentum sectors, are (a) 77 (repeated for 35 sectors), (b) 385 (repeated for 27 sectors), (c) 1117, 1110, 1118, 1110 (repeated for 36 sectors), (d) 445, 440, 446, 440 (repeated for 28 sectors), (e) 77 (repeated for 26 sectors), and (f) 51 (repeated for 22 sectors), respectively. The PES corresponding to Figs.~\ref{fig:fermion_corr_C2r1N7} and~\ref{fig:fermion_corr_C2r-3N6} are shown in Fig.~\ref{fig:fermion_entanglement_C2}. Each of the fermionic states, with the exception of the PES corresponding to Fig.~\ref{fig:fermion_corr_C2r2N8}, show PES with large gaps and relatively uniform eigenstate counts across the momentum sectors. These are features which we otherwise found to be realized only for the primary composite fermion state within the bosonic series. For the fermionic series under examination, the primary composite fermion $\nu=1/5$ state remains distinguished predominantly by the magnitude of the gap. 

Overall, the $|C|=2$ fermion series produces robust results for the gaps of the states (\ref{eq:FillingFactorJainEquiv}). While we have not generated enough data to ascertain a nonzero gap in the thermodynamic limit for all members of the family, all observed finite-size gaps are nonzero, and we find a clear thermodynamic effective continuum limit for the $r=\pm 1$ states. Several high-particle-number points are omitted but the error bars in the data obtained are reasonable. The $r=1$ series is again the most stable and the range of $q^2 \Delta$ limits is lower than in Fig.~\ref{fig:fermions_final_plot_C1}. With the exception of one competing topological phase at $\nu=1/3$, the ground-state degeneracy follows the predictions of composite fermion theory throughout.

\subsection{FCIs in $|C|=3$ Bands}
\label{subsec:FCI_C3}

For $|C|=3$, we find stronger finite-size effects than in $|C|=2$ bands. The Hilbert space dimension of the states is higher still for given $N$ and thus, fewer high-particle-number systems are computationally accessible. Coupled with this, the energy spectra are difficult to analyze. Not only is the ground-state gap often ambiguous, but the spectra in general are complex, showing a plethora of competing geometric and topological physical effects. For these reasons, the analysis of the $|C|=3$ fermionic states is omitted and we focus on the bosonic systems with contact interactions. Note that, just as in Sec.~\ref{subsubsec:FCI_C2_fermions}, all the systems in this section are within 1\% of square geometries. Some of the systems were exactly square, but all filling factors required the use of some approximately square geometries within the scaling procedure.

\subsubsection{Bosonic States}

\begin{figure}
\begin{center}
\includegraphics{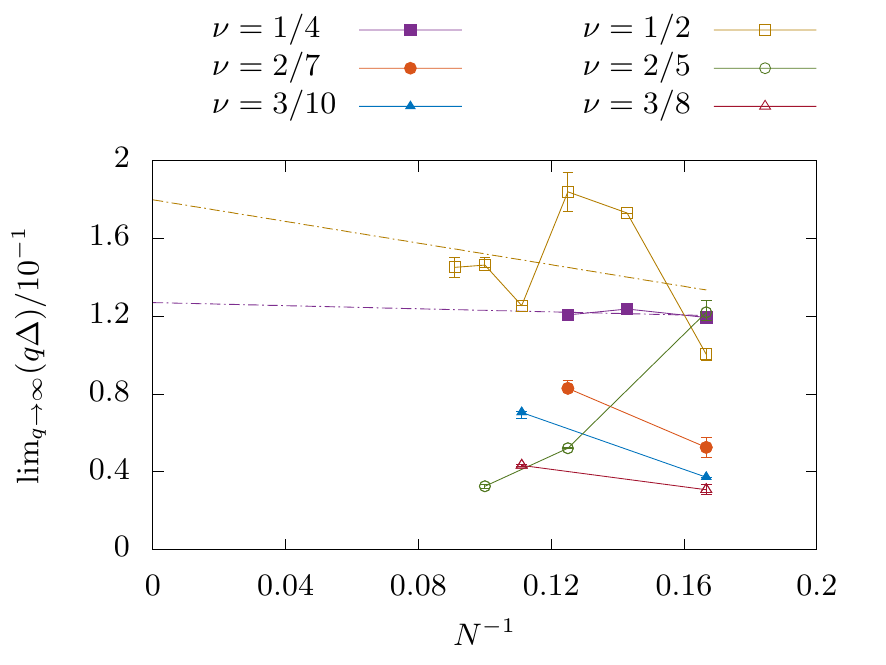}
\caption{Finite-size scaling of the gap to the thermodynamic effective continuum limit at fixed aspect ratio, for bosonic states in the $|C|=3$ band. The extrapolation to the $y$ axis is shown for the robust $\nu=1/4$ and $\nu=1/2$ states. Squares, circles, and triangles denote states with $|r|=1,2,3$, respectively, where the filled (hollow) symbols correspond to positive (negative) $r$.}
\label{fig:bosons_final_plot_C3}
\end{center}
\end{figure}

As in Secs.~\ref{subsubsec:FCI_C1_bosons} and ~\ref{subsubsec:FCI_C2_bosons}, we continue our analysis in a similar fashion and examine the effective continuum limit, followed by finite-size scaling to the thermodynamic limit where possible. Figure~\ref{fig:bosons_final_plot_C3} shows the effective continuum limiting behavior for the six filling factors under consideration. As expected, due to computational limitations, fewer high-particle-number states are analyzed, compared to the $|C|=2$ boson data in Fig.~\ref{fig:bosons_final_plot_C2}. Nevertheless, a reasonable sample is obtained, comparable to that of the $|C|=2$ fermion data in Fig.~\ref{fig:fermions_final_plot_C2}. Overall, we have considered 18 different combinations of particle number and filling factor, with an average of $\sim 25$ different geometries for each. There are a total of 460 different exact diagonalization calculations underlying the data in this section.

We find smaller values for the $q \Delta$ limits, when compared to the lower Chern number ($|C|=2$) scaling shown in Fig.~\ref{fig:boson_corr_C2}. This is a general trend with increasing Chern number, which we discuss later. 

A stable $r=1$ series with $\nu=1/4$ is observed. In this case, the thermodynamic effective continuum limit is extrapolated to be $\lim_{N,q\to\infty}(q\Delta)=0.13\pm0.01$\cite{Note1}. The corresponding negative flux attached version of this series with $\nu=1/2$ at $r=-1$ is also found to be exceptionally stable, with the gap exceeding that for $\nu=1/4$, extrapolated as $\lim_{N,q\to\infty}(q\Delta)=0.18\pm0.07$\cite{Note1}. The error bars due to the effective continuum limit are significant yet adequate, and more noticeable than those in Fig.~\ref{fig:bosons_final_plot_C2}. 

The remaining data series are insufficient to make any comments on scaling to the thermodynamic effective continuum limit; however, the predicted ground-state degeneracies from composite fermion theory are observed at our finite $N$, and the finite many-body gaps show the potential for a robust gap in the thermodynamic effective continuum limit.  

\begin{figure}
\begin{minipage}[b]{\linewidth}
\begin{minipage}[b]{.5\linewidth}
\begin{tikzpicture}
\node at (0,0) {\centering\includegraphics{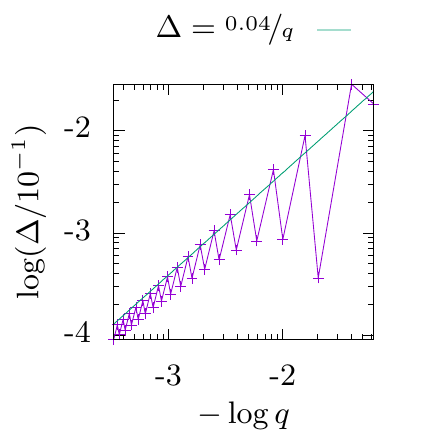}};
\node[overlay] at (-2,2) {(a)};
\phantomsubcaption
\label{fig:boson_plots_C3_1}
\end{tikzpicture}
\end{minipage}%
\begin{minipage}[b]{.5\linewidth}
\begin{tikzpicture}
\node at (0,0) {\centering\includegraphics{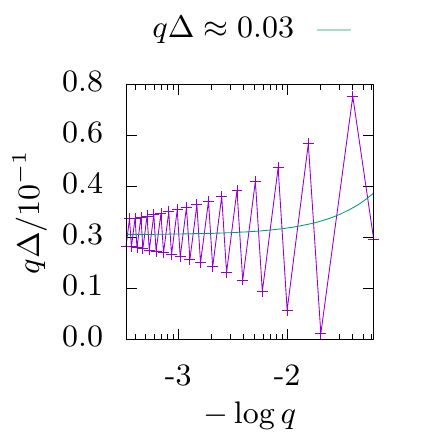}};
\node[overlay] at (-2,2) {(b)};
\phantomsubcaption
\label{fig:boson_plots_C3_2}
\end{tikzpicture}
\end{minipage}%
\end{minipage}
\begin{minipage}[b]{.49\linewidth}
\begin{tikzpicture}
\node at (0,0) {\includegraphics[width=\linewidth]{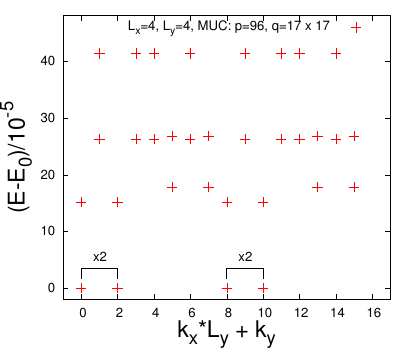}};
\node[overlay] at (-2,1.6) {(c)};
\phantomsubcaption
\label{fig:boson_spectra_C3_1}
\end{tikzpicture}
\end{minipage}
\begin{minipage}[b]{.49\linewidth}
\begin{tikzpicture}
\node at (0,0) {\includegraphics[width=\linewidth]{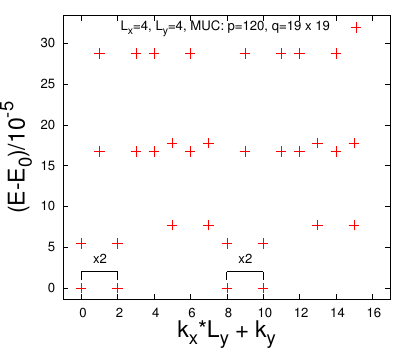}};
\node[overlay] at (-2,1.6) {(d)};
\phantomsubcaption
\label{fig:boson_spectra_C3_2}
\end{tikzpicture}
\end{minipage}
\begin{minipage}[b]{.49\linewidth}
\begin{tikzpicture}
\node at (0,0) {\includegraphics[width=\linewidth]{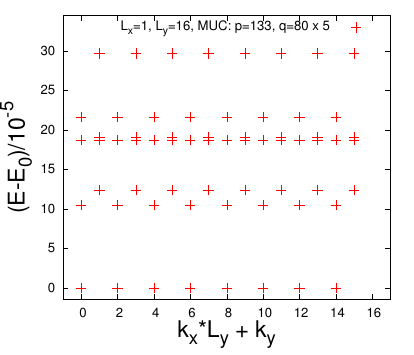}};
\node[overlay] at (-2,1.6) {(e)};
\phantomsubcaption
\label{fig:boson_spectra_C3_3}
\end{tikzpicture}
\end{minipage}
\begin{minipage}[b]{.49\linewidth}
\begin{tikzpicture}
\node at (0,0) {\includegraphics[width=\linewidth]{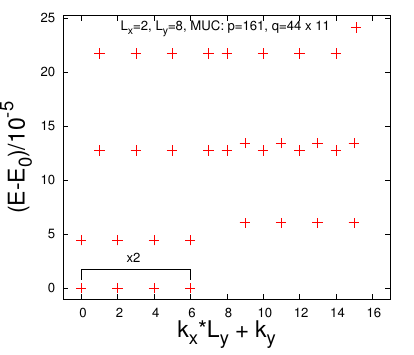}};
\node[overlay] at (-2,1.6) {(f)};
\phantomsubcaption
\label{fig:boson_spectra_C3_4}
\end{tikzpicture}
\end{minipage}
\caption{[(a), (b)] Magnitude of the gap for bosonic six-particle $\nu=3/8$ states in the $|C|=3$ band, as a function of MUC size, $q$. [(c)--(f)] Energy spectra for the bosonic six-particle $\nu=3/8$ state in the $|C|=3$ band, at (c)~$p=96$, (d)~$p=120$, (e)~$p=133$, (f)~$p=161$. The plots are resolved to $n=4$ points per sector.}
\label{fig:boson_plots_C3}
\end{figure}
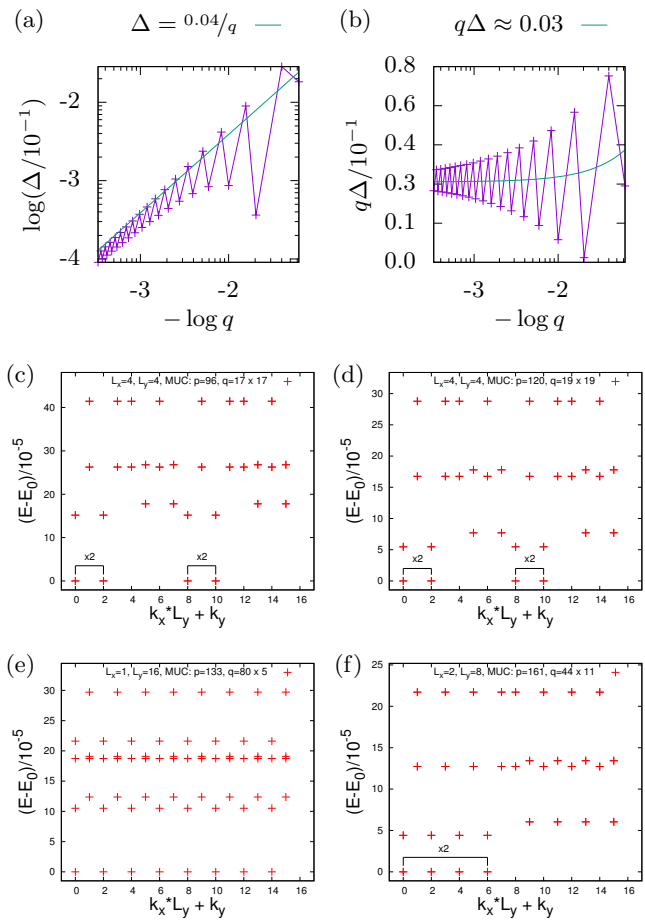

Figure~\ref{fig:boson_plots_C3} shows the plots for the six-particle $\nu=3/8$ state. This system is selected as a case of interest, since it has a large ground-state degeneracy, and we obtain significant error bars for its effective continuum limit. The plot of the scaling of the gap in Fig.~\ref{fig:boson_plots_C3_1} shows the expected reciprocal relation, with some slight deviations due to the 1\% square approximation of configurations. The plot of $q \Delta$ vs $1/q$ given in Fig.~\ref{fig:boson_plots_C3_2} shows these deviations in more detail. As previously mentioned, the small-$q$ deviations may be attributed to finite-size effects and they stabilize as the MUC size is increased. 

\begin{figure}
\centering
\begin{minipage}[b]{.49\linewidth}
\begin{tikzpicture}
\node at (0,0) {\includegraphics[width=\linewidth]{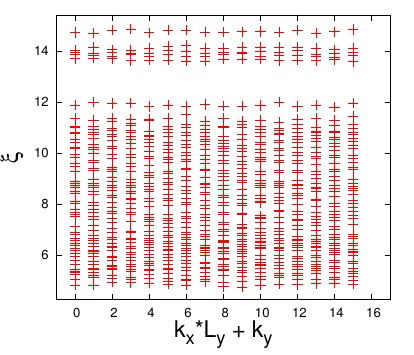}};
\node[overlay] at (-2,1.6) {(a)};
\phantomsubcaption
\label{fig:boson_entanglement_C3_1}
\end{tikzpicture}
\end{minipage}
\begin{minipage}[b]{.49\linewidth}
\begin{tikzpicture}
\node at (0,0) {\includegraphics[width=\linewidth]{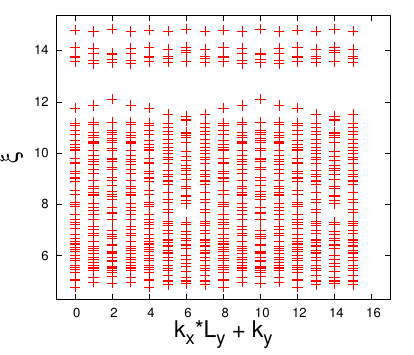}};
\node[overlay] at (-2,1.6) {(b)};
\phantomsubcaption
\label{fig:boson_entanglement_C3_2}
\end{tikzpicture}
\end{minipage}
\caption{PES for the bosonic six-particle $\nu=3/8$ state in the $|C|=3$ band with $N_A=\lfloor{N/2}\rfloor=3$, at (a)~$p=120$, and (b)~$p=133$. The count of eigenstates from the bottom of the spectrum up to the principal entanglement gap is 46 per momentum sector.}
\label{fig:boson_entanglement_C3}
\end{figure}

Four distinct energy spectra for different geometries realizing the six-particle $\nu=3/8$ state are shown in Figs.~\ref{fig:boson_spectra_C3_1},~\ref{fig:boson_spectra_C3_2},~\ref{fig:boson_spectra_C3_3}, and~\ref{fig:boson_spectra_C3_4}. These cases differ in the realized shape of the MUC. Notice that the spectra shown in Figs.~\ref{fig:boson_spectra_C3_1} and~\ref{fig:boson_spectra_C3_2} correspond to the same $L_x\times L_y=4\times4$ square configuration, and yield similar spectra. The other two spectra in Figs.~\ref{fig:boson_spectra_C3_1} \&~\ref{fig:boson_spectra_C3_2} correspond to geometries with $L_x\times L_y=1\times 16$ and $L_x\times L_y =2\times 8$ MUCs, respectively, and yield qualitatively distinct features. (Again, these geometries are chosen with the maximum possible value of $L_y$ consistent with the lattice size.) As a result of such distinct geometries, the fluctuations of the gap persist up to large values of $q$ in the scaling shown in Fig.~\ref{fig:boson_plots_C3_2}. Geometric effects such as this give rise to the significant error bars in Fig.~\ref{fig:bosons_final_plot_C3}. 
The entanglement gaps for these systems are shown in Fig.~\ref{fig:boson_entanglement_C3}. While the numerical value of $\Delta_\xi$ is relatively small, the opening of the gap confirms the topological nature of this state.

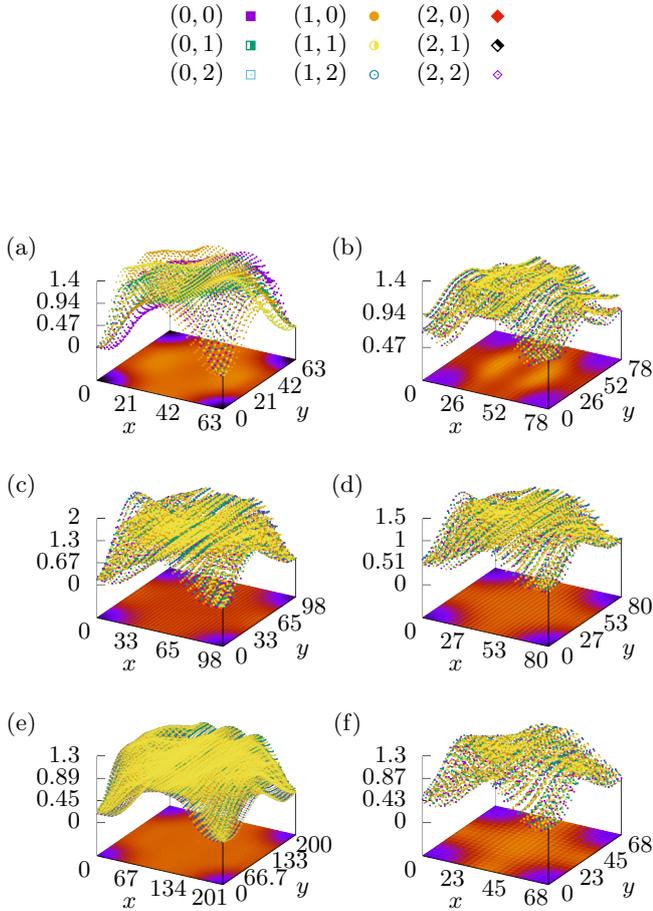
\begin{figure}
\begin{minipage}[b]{\linewidth}

\centering

\begin{minipage}[b]{\linewidth}
\centering\includegraphics{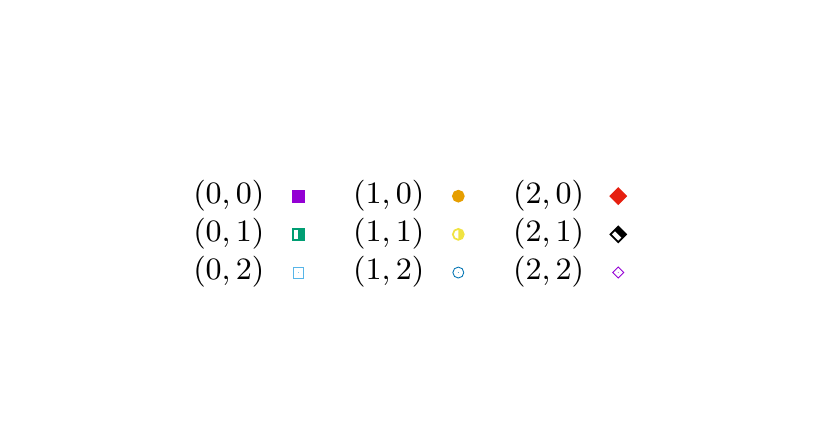}
\end{minipage}%

\vskip\baselineskip

\begin{minipage}[b]{.5\linewidth}
\begin{tikzpicture}
\node at (0,0) {\centering\includegraphics{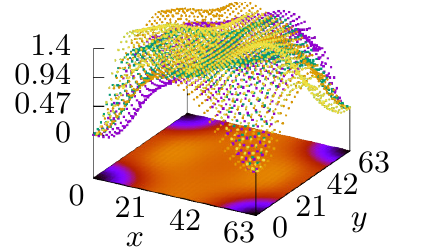}};
\node[overlay] at (-2.3,1.2) {(a)};
\phantomsubcaption
\label{fig:boson_corr_C3r1N7}
\end{tikzpicture}
\end{minipage}%
\begin{minipage}[b]{.5\linewidth}
\begin{tikzpicture}
\node at (0,0) {\centering\includegraphics{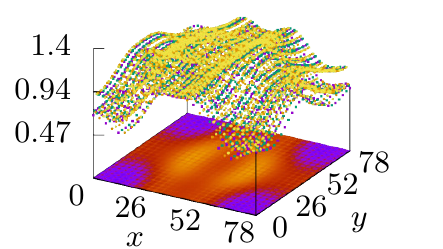}};
\node[overlay] at (-2.3,1.2) {(b)};
\phantomsubcaption
\label{fig:boson_corr_C3r-1N9}
\end{tikzpicture}
\end{minipage}%

\vskip\baselineskip

\begin{minipage}[b]{.5\linewidth}
\begin{tikzpicture}
\node at (0,0) {\centering\includegraphics{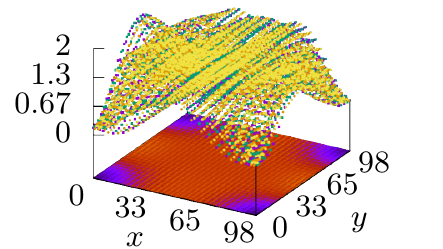}};
\node[overlay] at (-2.3,1.2) {(c)};
\phantomsubcaption
\label{fig:boson_corr_C3r2N8}
\end{tikzpicture}
\end{minipage}%
\begin{minipage}[b]{.5\linewidth}
\begin{tikzpicture}
\node at (0,0) {\centering\includegraphics{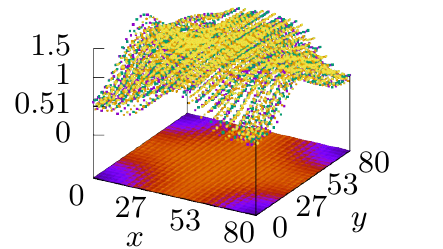}};
\node[overlay] at (-2.3,1.2) {(d)};
\phantomsubcaption
\label{fig:boson_corr_C3r-2N8}
\end{tikzpicture}
\end{minipage}%

\vskip\baselineskip

\begin{minipage}[b]{.5\linewidth}
\begin{tikzpicture}
\node at (0,0) {\centering\includegraphics{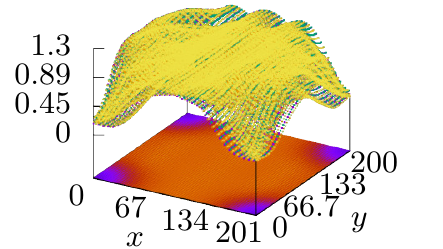}};
\node[overlay] at (-2.3,1.2) {(e)};
\phantomsubcaption
\label{fig:boson_corr_C3r3N9}
\end{tikzpicture}
\end{minipage}%
\begin{minipage}[b]{.5\linewidth}
\begin{tikzpicture}
\node at (0,0) {\centering\includegraphics{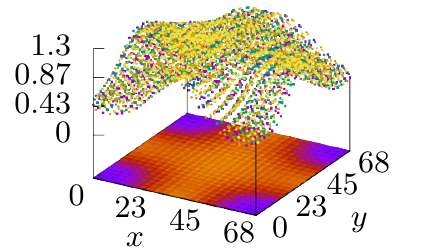}};
\node[overlay] at (-2.3,1.2) {(f)};
\phantomsubcaption
\label{fig:boson_corr_C3r-3N6}
\end{tikzpicture}
\end{minipage}%

\end{minipage}
\caption{Density-density correlation functions for bosonic states in the $|C|=3$ band. The plots are shown for the lowest-lying ground state in the $(k_x,k_y)=(0,0)$ momentum sector. The legend differentiates between correlation functions at lattice positions $(x\bmod 3, y\bmod 3)$, as explained in the main text. We show data for (a)~$r=1$: $\nu=1/4$, $N=7$, $p=114$; (b)~$r=-1$: $\nu=1/2$, $N=9$, $p=113$; (c)~$r=2$: $\nu=2/7$, $N=8$, $p=114$; (d)~$r=-2$: $\nu=2/5$, $N=8$, $p=107$; (e)~$r=3$: $\nu=3/10$, $N=9$, $p=447$; and (f)~$r=-3$: $\nu=3/8$, $N=6$, $p=96$.}
\label{fig:boson_corr_C3}
\end{figure}

Correlation functions for the discussed filling factors are shown in Fig.~\ref{fig:boson_corr_C3}. As for the bosons in the $|C|=1,2$ bands [Figs.~\ref{fig:boson_corr_C1r1N8} and~\ref{fig:boson_corr_C2r1N7}], only the primary composite fermion state with a flux attachment of $r=1$ in Fig.~\ref{fig:boson_corr_C3r1N7} has a fully formed correlation hole at zero separation. The correlation functions are again modulated by the Chern number, giving rise to $|C|^2$ sheets, which now is visible even in the color plots of our figures, for example, in Fig.~\ref{fig:boson_corr_C3r-3N6}. In this Chern band, all of the correlation functions seem to show isotropy in the large-distance limit. However, small scale features are hard to discern. As with the $|C|=2$ bosons in Fig.~\ref{fig:boson_corr_C2}, for the cases with negative flux attachment ($r<0$)  the correlation function sheets do not converge to the same value at zero separation. This is shown in Figs.~\ref{fig:boson_corr_C3r-1N9},~\ref{fig:boson_corr_C3r-2N8}, and~\ref{fig:boson_corr_C3r-3N6} for this Chern band, mirroring the behaviors seen in Figs.~\ref{fig:boson_corr_C2r-1N9},~\ref{fig:boson_corr_C2r-2N12}, and~\ref{fig:boson_corr_C2r-3N9} for $|C|=2$.

The PES for the remaining bosonic series in the $|C|=3$ band have small but distinct gaps. Considering the spectra for the states in Fig.~\ref{fig:boson_corr_C3}, we find $\Delta_{\xi}$ values of (a) 12.84, (b) 1.02, (c) 1.49, (d) 1.71, (e) 1.41, and (f) 1.92, after tracing out $\lfloor{N/2}\rfloor$ particles. The corresponding eigenstate counts from the bottom of the spectra up to the principal entanglement gaps, in each of the momentum sectors, are (a) 51 (repeated for 28 sectors), (b) 323, 323, 323, 318, 318, 318 (repeated for 18 sectors), (c) 1127, 1112, 1112, 1112 (repeated for 28 sectors), (d) 438, 432, 437, 432 (repeated for 20 sectors), (e) 1364, 1364, 1364, 1356, 1356, 1356 (repeated for 30 sectors), and (f) 46 (repeated for 16 sectors), respectively. In addition, these spectra typically show several smaller gaps higher in the spectrum. The primary composite fermion $\nu=1/4$ state is again the largest and most distinct, with a uniform count of eigenstates below the principal entanglement gap, across the momentum spectrum.  

\begin{figure}[t]
\begin{minipage}[b]{\linewidth}
\begin{minipage}[b]{.5\linewidth}
\centering
\begin{tikzpicture}
\node at (0,0) {\includegraphics{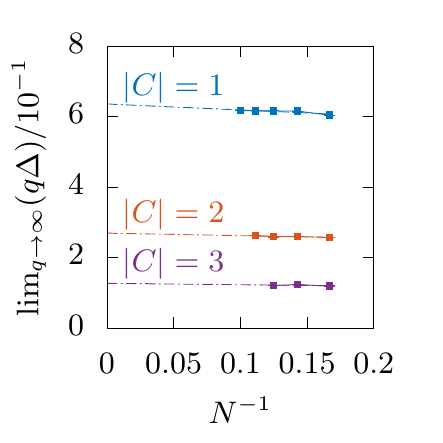}};
\node[overlay] at (-2,1.9) {(a)};
\phantomsubcaption
\label{fig:combined_r1_1}
\end{tikzpicture}
\end{minipage}%
\begin{minipage}[b]{.5\linewidth}
\centering
\begin{tikzpicture}
\node at (0,0) {\includegraphics{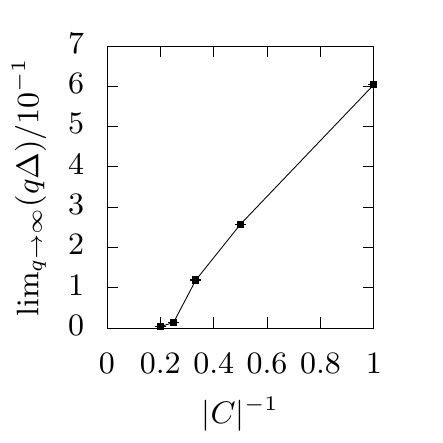}};
\node[overlay] at (-2,1.9) {(b)};
\phantomsubcaption
\label{fig:combined_r1_2}
\end{tikzpicture}
\end{minipage}%
\end{minipage}
\caption{(a) Finite-size scaling of the gap to the thermodynamic (effective) continuum limit at fixed aspect ratio, for robust $r=1$ bosonic states. The filling factors are $\nu=\nicefrac{1}{2},\nicefrac{1}{3},\nicefrac{1}{4}$ for Chern numbers $|C|=1,2,3$, respectively. (b) Finite-size scaling of the (effective) continuum limit of the gap at fixed aspect ratio, against Chern number, for robust $r=1$ bosonic states with $N=6$ particles. In both cases, all of the error bars are smaller than the data points on the scale of the plots.}
\label{fig:combined_r1}
\end{figure}

\section{Thermodynamic limits and scaling of the effective continuum limit with Chern number}
\label{sec:limit}

In this section, we consolidate our analyses of the $|C|=1,2,3$ bands in order to comment on the behavior of the thermodynamic limits that we could extrapolate from the effective continuum limits at finite system sizes. 

Extrapolated thermodynamic limits for bosons are presented in Table~\ref{tab:bosons}. One overarching characteristic of the plots in Figs.~\ref{fig:bosons_final_plot_C1},~\ref{fig:bosons_final_plot_C2}, and~\ref{fig:bosons_final_plot_C3} is the robust $r=1$ series. The corresponding gaps are extracted and shown in Fig.~\ref{fig:combined_r1_1}. Up to the $|C|=3$ system, we find that $\lim_{q\to\infty}(q \Delta)$ for $N=6$ scales approximately inversely with Chern number, as seen in Fig.~\ref{fig:combined_r1_2}. In addition, we show that this (approximate) reciprocal relation does not hold precisely for the $|C|=4,5$ bands. However, we caution that our data are very limited in those cases. 

We also highlight again the stable $r=-1$ series with filling $\nu=1/2$ in $|C|=3$ bands for which the gap is extrapolated to the thermodynamic effective continuum limit in Fig.~\ref{fig:bosons_final_plot_C3}. Since the larger $N$ systems should intuitively be given more weight when taking the limit, this value is perhaps an overestimate of the true thermodynamic effective continuum limit. This is captured by the larger error bars.

The thermodynamic effective continuum limits for the gaps of fermionic states are summarized in Table~\ref{tab:fermions}. As for bosons, we find that the gap decreases with Chern number. However, due to computational expense, we did not consider enough Chern numbers to postulate a scaling relation. As seen before in Fig.~\ref{fig:fermions_final_plot_C1}, the $\nu=1/3$ and $\nu=2/3$ series yield the same thermodynamic effective continuum limit due to particle-hole symmetry. For the $\nu=1/3$ series in the $|C|=2$ band, shown in Fig.~\ref{fig:fermions_final_plot_C1}, we note intuitively that the extrapolated limit is perhaps an underestimate since larger $N$ systems should be given greater weight. Again, this is accounted for in the uncertainty.

Our studies of the density-density correlation functions for the higher Chern bands show some common features for the states with successful thermodynamic extrapolations. Compared to the other states, the correlation functions corresponding to the successfully extrapolated series, shown in Figs.~\ref{fig:boson_corr_C1r1N8},~\ref{fig:boson_corr_C2r1N7},~\ref{fig:boson_corr_C3r1N7}, and~\ref{fig:boson_corr_C3r-1N9} for bosons and in Figs.~\ref{fig:fermion_corr_C1r1N9},~\ref{fig:fermion_corr_C1r-2N18},~\ref{fig:fermion_corr_C2r1N7}, and~\ref{fig:fermion_corr_C2r-1N9} for fermions are characterized by smaller oscillations in the large distance limit and are more likely to be fully isotropic. This is consistent with small correlation lengths for these cases, as plausibly expected for states with small composite fermion filling factors $|r|$. In addition, the correlation functions of higher $|r|$ values also show some of the features expected for quantum Hall liquids such as a small-distance correlation hole. Most series for which we could not find a satisfactory thermodynamic (effective) continuum limit show visible oscillations throughout the simulation cell, which may be either indications of finite-size effects, or competing charge density wave orders. 

All of the correlation functions for higher Chern bands show a characteristic modulation of the magnitude of correlations as a function of $x$ and $y$ positions modulo the Chern number and so give the appearance of $|C|^2$ correlation sheets. This modulation may also explain the continued sensitivity of the states to the geometry of the system, as simulation cell sizes which are multiples of the Chern number, i.e., geometries $N_x \bmod |C|  =0$ and $N_y \bmod |C|  =0$, are special but are generally difficult to realize in conjunction with all other constraints.
% Consequently, the correlation functions for the higher Chern bands are more intricate and harder to analyze.

\begin{table}[H]
\centering
\begin{ruledtabular}
\subcaption{\label{tab:bosons}bosons}
\begin{tabular}{c c c c}
$|C|$ & $r$ & $\nu$ & $\lim_{N,q\to\infty}(q\Delta)$ \\
\hline
$1$ & $1$ & $1/2$ & $0.64\pm0.01$ \\
$2$ & $1$ & $1/3$ & $0.27\pm0.004$ \\
$3$ & $1$ & $1/4$ & $0.13\pm0.01$ \\
   & $-1$ & $1/2$ & $0.18\pm0.07$
\end{tabular}
\end{ruledtabular}
%\centering
\begin{ruledtabular}
\vspace{.2em}
\subcaption{\label{tab:fermions}fermions}
\vspace{.3em}
\begin{tabular}{c c c c}
$|C|$ & $r$ & $\nu$ & $\lim_{N,q\to\infty}(q^2\Delta)$ \\
\hline
$1$ & $1$ & $1/3$ & $2.56\pm0.02$ \\
       & $-2$ & $2/3$ & $2.56\pm0.02$ \\
$2$ & $1$ & $1/5$ & $0.46\pm0.02$ \\
       & $-1$ & $1/3$ & $0.65\pm0.16$
\end{tabular}
\end{ruledtabular}
\caption{\label{tab:summary}Summary of states with (effective) continuum limits that could be extrapolated to the thermodynamic limit, given to two decimal places, for (a) bosons and (b) fermions\cite{Note1}. The uncertainty quoted for the limit is the asymptotic standard error from a linear regression of $q\Delta$ against $1/N$.}
\end{table}

\section{Discussion \& Conclusions}
\label{sec:conc}

%Further investigations should focus on the stability of fermionic states, the role of long-range interactions and the detailed analysis of the ground-states and excitations in terms of the composite fermion trial wavefunctions.
%}

In this paper, we have quantitatively analyzed the composite fermion series of states for higher Chern number bands in the Harper-Hofstadter model~\cite{Kol:1993wv, 2009PhRvL.103j5303M, Moller:2015kg}. Exact diagonalization calculations of these fractional quantum Hall liquids in the Hofstadter model are challenging, owing to numerous Diophantine constraints relating filling factor, flux density, and lattice geometry. We exploit the scaling of the energy scales in the size of the MUC, first observed by Bauer \emph{et al.}~\cite{Bauer:2016ju}, to resolve some of these commensuration issues. We are thus able to extract finite-size data exclusively for nearly square systems, leading to more reliable determination of the many-body gaps as compared to finite-size scaling at fixed flux density.
%We have presented new data for the scaling towards the effective continuum limit at fixed aspect ratio, and we have cross-validated our results against scaling towards the effective continuum limit at fixed flux density, as well as quasi-continuum limits obtained for the states in the $|C|=1$ band in Bauer, Jackson, \& Roy. 

We confirm that the prediction of composite fermion theory for the ground-state degeneracy is correct at all filling factors that we examined, with few exceptions due to competing phases. Several states were shown to have stable gaps in the thermodynamic (effective) continuum limit. Among these---as expected---the primary composite fermion states with filling factor $\nu=1/(k|C|+1)$ are the most robust, and we find that they have an (effective) continuum limit that is largely independent of particle number. We found several other states that allow for a reliable finite-size scaling of the gap, as summarized in Table \ref{tab:summary}.
However, for many candidate phases predicted by composite fermion theory, we have found that scaling toward the (effective) continuum limit does not sufficiently alleviate finite-size effects to draw firm conclusions about their stability in the thermodynamic (effective) continuum limit. In part, this is due to the system-size limitations used in our study. The topological character of the different target phases has been clearly shown through the use of entanglement spectroscopy, which reveals the existence of entanglement gaps.

Our data also shed light on the fate of two potential BIQHE states in the Hofstadter model. A first candidate arises in $|C|=1$ bands at filling $\nu=2$, for $r=-2$ filled composite fermion levels. However, this state is clearly not realized within the lowest-band-projected Harper-Hofstadter model examined in our paper, as we do not find the correct ground-state degeneracy of one for all system sizes. We therefore conjecture that the recently reported $\nu=2$ state of hardcore bosons~\cite{He:2017gc} likely requires filling of (at least) the lowest two Landau levels, which would bring it in line with other realizations of the BIQHE that require two flavors of bosons. The second candidate is the $\nu=1$ state in $C=2$ bands. Here, we find conclusively a large gap above a nondegenerate ground state for all system sizes. While the magnitude of the gap shows important variations with system size even after taking the effective continuum limit, our data are consistent with the existence of a gapped phase in the thermodynamic effective continuum limit, subject to the known generic caveats~\cite{Cubitt:2015ch}.

%The majority of cases considered do not show signs of stabilizing in this limit, which is due to finite-size effects and, occasionally, competing phases. For the $|C|=1$ band, certain non-Laughlin states did show signs of stabilizing, such as the $\nu=2/3$ states for fermions or the $\nu=3/4$ states for bosons. For higher Chern bands, there were also a select few cases, such as the $\nu=1/2$ states for bosons in the $|C|=3$ band, however more large-$N$ data is needed to confirm this conclusively. 

In addition to spectral properties, we have studied the two-particle correlation functions of the Hofstadter model, revealing their unexpected structure which resembles a total of $|C|^2$ continuous sheets. This result is in disagreement with suggestions that Chern number $C$ bands can be regarded as $|C|$-layer quantum Hall systems. In this multilayer picture, we would only expect $|C|$ distinct correlation functions, so we hope that our results will stimulate further research that will clarify the origin of this discrepancy.
%In exceptional cases, we analyzed the two-particle correlation functions and PES of the states to show evidence of competing topological phases and charge density wave excitations. This consideration offers an explanation as to why outlying states presented difficulties in the analysis.

We have shown that approximately square geometries stabilize some of the expected isotropic quantum liquid phases predicted by composite fermion theory. In general, we find that variations of the gap due to a small change in aspect ratio are smaller than the finite-size effects but still remain significant. Hence, the sensitivity of the problem to details of the geometry seems to indicate that competing phases are likely to exist. % Alternatively, finding may indicate that our target phases have longer correlation lengths, but we cannot rule out the presence of competing phases, including density wave instabilities.
Indeed, in addition to the isotropic quantum Hall liquids discussed in our work, several candidates for symmetry broken phases~\cite{Natu:2016fp, Hugel:2017kt} or phases combining a broken symmetry and topological response \cite{Spanton:2017vf} have recently been proposed. We hope that the rich interplay of these competing phases will stimulate further active research in the physics of fractional topological insulators in Hofstadter models. Future research should focus on experimental probes for these regimes, as well as on specific realizations that can favor the various candidate phases, for example, via the effect of longer range interactions or anisotropy.

\begin{acknowledgments}

We acknowledge useful discussions with Nicolas Regnault, Nigel Cooper and Michael Zaletel as well as related collaborations with Nigel Cooper, Rahul Roy, and Antoine Sterdyniak. B.A.~also thanks Chaitanya Mangla, Andrew Fowler, Philipp Verpoort, Michael Rutter, and Be{\~n}at Mencia for technical help with the computations. Our numerical results were produced using \textsc{DiagHam}. B.A.~acknowledges support from the Engineering and Physical Sciences Research Council (EPSRC) under Grant No.~EP/M506485/1. G.M.~acknowledges support from the Royal Society under Grant No.~UF120157 and from the University of Kent. He also thanks the TCM Group, Cavendish Laboratory and Trinity Hall for their hospitality.
\end{acknowledgments}

%\clearpage
%\newpage

\appendix
\setcounter{secnumdepth}{2}

\onecolumngrid

\section{Periodic Landau Gauge Vector Potential for Rectangular Lattices}
\label{sec:RectangularLandau}

Consider a general rectangular lattice with $\mathbf{l}_x=l_x \hat{\mathbf{e}}_x$ and $\mathbf{l}_y=l_y \hat{\mathbf{e}}_y$. In this basis, the absolute position vector may be written as
\begin{equation}
\mathbf{r}=
\begin{pmatrix}
x \\
y
\end{pmatrix}
= \xi_x \mathbf{l}_x + \xi_y \mathbf{l}_y
\end{equation}
with $\xi_x = x / l_x$ and $\xi_y = y / l_y$. Following from Hasegawa and Kohmoto~\cite{2013PhRvB..88l5426H}, we know that the periodic Landau gauge phase is given as
\begin{equation}
\chi(\mathbf{r})=-SB\lfloor \xi_x \rfloor \xi_y,
\end{equation}
where $S=|\mathbf{l}_x \times \mathbf{l}_y|=l_x l_y$. Hence, the phase may be written as
\begin{equation}
\label{eq:chi}
\chi(\mathbf{r})=-Bl_x y \left\lfloor \frac{x}{l_x} \right\rfloor.
\end{equation} 
Ultimately, we would like to calculate the periodic Landau gauge vector potential for rectangular lattices $\mathbf{A}^{\text{(p,rect)}}$, which may be expressed in terms of the Landau gauge vector potential $\mathbf{A}^{\text{(L,rect)}}$ for rectangular lattices as
\begin{equation}
\label{eq:Arect}
\mathbf{A}^{\text{(p,rect)}} = \mathbf{A}^{\text{(L,rect)}} + \nabla \chi(\mathbf{r}).
\end{equation}
From Eq.~(\ref{eq:chi}), we may write
\begin{equation}
\label{eq:delchi}
\nabla\chi(\mathbf{r})=-B
\begin{pmatrix}
y \sum_{n=-\infty}^{n=\infty} \delta (x/l_x-n+\epsilon) \\
\lfloor x/l_x \rfloor \\
0
\end{pmatrix},
\end{equation}
where $\epsilon$ is an infinitesimal, added to avoid an ambiguity of the phase factor at lattice site positions. Now, given that the Landau gauge vector potential for rectangular lattices is 
\begin{equation}
\label{eq:ALandau}
\mathbf{A}^{\text{(L,rect)}}=\frac{SB}{2\pi}\xi_x\mathbf{F}_y, 
\end{equation}
where
\begin{equation}
\mathbf{F}_y\equiv 2\pi \left( \frac{\hat{\mathbf{e}}_z\times \mathbf{l}_x}{(\mathbf{l}_x \times \mathbf{l}_y)\cdot\hat{\mathbf{e}}_z}\right) = \frac{2\pi}{l_y}
\mathbf{e}_y,
\end{equation}
we may substitute Eqs.~(\ref{eq:ALandau}) and~(\ref{eq:delchi}) into Eq.~(\ref{eq:Arect}), which yields
\begin{equation}
\mathbf{A}^{\text{(p,rect)}}=B
\begin{pmatrix}
-y \sum_{n=-\infty}^{n=\infty} \delta (x/l_x-n+\epsilon)  \\
x-\lfloor x/l_x \rfloor \\
0
\end{pmatrix}.
\end{equation}
Note that this potential reproduces the same discrete implementation of the finite-size Harper-Hofstadter Hamiltonian (\ref{eq:Hamiltonian}), as would be obtained by applying the magnetic translation algebra with a basis of $\{T_M(\mathbf{l}_x), T_M(\mathbf{l}_y)\}$ (see, e.g., the supplementary material of Ref.~\onlinecite{Moller:2015kg}).

%\onecolumngrid

\section{Periodic Landau Gauge Transformation in Fourier Space}
\label{sec:PeriodicLandauTransform}

As a gauge transform of the electromagnetic vector potential $\mathbf{A}(\mathbf{r})\to \mathbf{A}(\mathbf{r}) + \nabla \chi(\mathbf{r})$  acts multiplicatively on the wave function in position space via $\psi(\mathbf{r}) \to \exp[i\chi(\mathbf{r})] \psi(\mathbf{r})$, its action in reciprocal space takes the form of a convolution with the gauge function. Let us therefore consider the Fourier transform of the gauge transforms between a periodic Landau gauge with respect to the standard Landau gauge to establish how momenta are transformed.

Consider a system with a total of $N_xN_y$ sites, $q=l_x l_y$ sites in each MUC, and $L_x L_y$ MUCs in the system. Let the system be pierced with a perpendicular magnetic field $\mathbf{B}=2\pi n_\phi\hat{\mathbf{e}}_z$, where the lattice constant is set to one and sites in the MUC are labeled with a sublattice index 
$\alpha = 0,\dots , (q -1)$. In the Landau gauge, the MUC is naturally $q \times 1$. To realize this gauge in a finite-size geometry, we require $N_x \bmod q = 0$, and hence we obtain momenta $k_x^{(\text{L})} = 2\pi n^\text{(L)}/ N_x$, with $n^\text{(L)} = 0, \dots, N_x/q-1$ and $k_y^{(\text{L})} = 2\pi m^\text{(L)}/ N_y$, with $m^\text{(L)} = 0, \dots, N_y$. By contrast, the set of allowed momentum vectors in the periodic gauge are
\begin{equation}
\{\mathbf{k}^{(\text{p})}\}=\left\{ \left( \frac{2\pi}{N_x}n, \frac{2\pi}{N_y}m \right) \right\},
\end{equation}
with momentum indices $n=0,\dots,L_x-1$ and $m=0,\dots,L_y-1$. The resulting Brillouin zones (BZ) have different shapes, with the BZ for the Landau gauge spanning a narrow tall rectangle $\mathbf{k} \in [-\pi/q, \pi/q] \times [-\pi, \pi]$, whereas the periodic gauge yields a wider and shorter BZ geometry.

The absolute position vector $\mathbf{r}_{st\alpha}=\mathbf{R}_{st}+\boldsymbol{\rho}_\alpha$ may be written as
\begin{equation}
\mathbf{r}_{st\alpha}=sl_x \hat{\mathbf{e}}_x + t l_y \hat{\mathbf{e}}_y + \boldsymbol{\rho}_\alpha,
\end{equation} 
with spatial indices $s=0,\dots,L_x-1$ and $t=0,\dots,L_y-1$, and corresponding sublattice vectors
\begin{equation}
\boldsymbol{\rho}_{\alpha}=
\begin{pmatrix}
\alpha \bmod l_x \\
\lfloor \alpha / l_x \rfloor
\end{pmatrix}.
\end{equation}
The magnetic field may be written as $\mathbf{B}=\nabla \times \mathbf{A}$,
with a vector potential in the Landau gauge
\begin{equation}
\mathbf{A}^{(\mathrm{L})}=Bx\hat{\mathbf{e}}_y
\end{equation}
that is independent of $y$. Other vector potentials may be obtained via a gauge transformation
\begin{equation}
\mathbf{A}\to \mathbf{A}+\nabla \chi(\mathbf{r}).
\end{equation}
To ensure gauge periodicity, we take
\begin{equation}
\chi(\mathbf{r})=-B\lfloor x \rfloor y,
\end{equation}
which, with an arbitrary rectangular lattice basis $\{l_x \hat{\mathbf{e}}_x$, $l_y \hat{\mathbf{e}}_y\}$, becomes
\begin{equation}
\chi(\mathbf{r})=-B l_x \lfloor x/l_x \rfloor y,
\end{equation}
as discussed by Hasegawa and Kohmoto~\cite{2013PhRvB..88l5426H}. 
We are interested in transforming to some arbitrary periodic Landau gauge, such that
\begin{equation}
\psi^{(\mathrm{p})}=G_\alpha \psi^{(\mathrm{L})},
\end{equation}
where the gauge factor $G_\alpha\equiv e^{\mathrm{i}\chi}$. In Fourier space, this may be written as
\begin{equation}
\label{eq:wfpl}
\hat{\psi}^{(\mathrm{p})}=\hat{G}_\alpha \ast \hat{\psi}^{(\mathrm{L,p})},
\end{equation}
where $\ast$ denotes the convolution, and $\hat{\psi}^{(\mathrm{L,p})}$ indicates the wave function in the original Landau gauge Fourier transformed with respect to the BZ of the periodic gauge. Specifically, the Fourier transform with respect to the MUC in periodic gauge is defined as
\begin{equation}
\hat{f}(x,y)=\sum_{s=0}^{L_x-1}\sum_{t=0}^{L_y-1} e^{-\mathrm{i}\mathbf{k}^{(\text{p})}\cdot\mathbf{r}_{st\alpha}}f(x,y),
\end{equation}
where $f(x,y)$ is an arbitrary function of $x$ and $y$ positions. The corresponding Fourier transform of the Landau gauge wave function in Eq.~(\ref{eq:wfpl}) is of a general form, with functions given by solutions to the Harper equation. However, the Fourier transform of the gauge factor is analytically calculable, and we proceed by evaluating it here. 
%Fourier transforming with respect to the absolute position, we find that
%%\begin{widetext}
%\begin{equation}
%\hat{G}_\alpha = \sum_{s=0}^{L_x -1} \sum_{t=0}^{L_y -1} \exp\left\{-\mathrm{i} B l_x (tl_y + \lfloor \alpha / l_x \rfloor ) \left\lfloor \frac{s l_x + \alpha  \bmod l_x}{l_x} \right\rfloor  \right\} \exp\left\{ \mathrm{i} 
%\begin{pmatrix}
%k^{(\text{p})}_x \\
%k^{(\text{p})}_y
%\end{pmatrix}\cdot
%\begin{pmatrix}
%sl_x + \alpha \bmod l_x \\
%tl_y + \lfloor \alpha / l_x \rfloor
%\end{pmatrix}
%\right\}.
%\end{equation}
%%\end{widetext}
%%
Noting that
\begin{equation}
\left\lfloor \frac{s l_x + \alpha \bmod l_x}{l_x} \right\rfloor = s, 
\end{equation}
and taking out constant factors, we find that
\begin{equation}
\hat{G}_\alpha = e^{-\mathrm{i}\mathbf{k}^{(\text{p})}\cdot\boldsymbol{\rho}_\alpha} \sum_{s,t} \left( e^{-\mathrm{i}B q} \right)^{st} \left( e^{-\mathrm{i}(k^{(\text{p})}_x l_x + B l_x \lfloor \alpha / l_x \rfloor)}  \right)^{s} \left( e^{-\mathrm{i}k^{(\text{p})}_y l_y}  \right)^{t}.
\end{equation}
Since $B=2\pi n_\phi = 2\pi p / q, \;\forall p\in\mathbb{Z}$, we make the simplification
\begin{equation}
\left( e^{-\mathrm{i}B q} \right)^{st} = 1,
\end{equation}
which allows us to separate the summation, such that
\begin{equation}
\hat{G}_\alpha = e^{-\mathrm{i}\mathbf{k}^{(\text{p})}\cdot\boldsymbol{\rho}_\alpha} \sum_{s=0}^{L_x-1} \left( e^{-\mathrm{i} (k^{(\text{p})}_x l_x + B l_x \lfloor \alpha / l_x \rfloor)}  \right)^{s} \sum_{t=0}^{L_y-1} \left( e^{-\mathrm{i}k^{(\text{p})}_y l_y}  \right)^{t}.
\end{equation}
Since $k^{(\text{p})}_y=2\pi m/l_y L_y$ for $m=0,\dots,L_y-1$, we deduce that
\begin{equation}
\sum_{t=0}^{L_y-1} \left( e^{-\mathrm{i}k^{(\text{p})}_y l_y}  \right)^{t}=L_y \delta_{k^{(\text{p})}_y,0}.
\end{equation}
Hence, our expression reduces to
\begin{equation}
\hat{G}_\alpha(\mathbf{k}^{(\text{p})}) = L_y e^{-\mathrm{i}k^{(\text{p})}_x (\alpha \bmod l_x)}\sum_{s=0}^{L_x-1}e^{-\mathrm{i} \frac{2\pi}{l_y} (l_y n + p L_x \lfloor \alpha / l_x \rfloor)\frac{s}{L_x}} \delta_{k^{(\text{p})}_y,0}.
\end{equation}
Furthermore, since the total number sites in the $x$ direction is necessarily a multiple of $q$, it follows that $L_x\propto l_y$ in all cases. This allows us to make the simplification
%
%\begin{widetext}
\begin{equation}
\label{eq:finalGalpha}
\hat{G}_\alpha(\mathbf{k}^{(\text{p})}) = 
L_x L_y \exp\left\{-\mathrm{i}k^{(\text{p})}_x (\alpha \bmod l_x)\right\} \delta_{k^{(\text{p})}_xN_x/2\pi+p\kappa\lfloor \alpha / l_x \rfloor, 0} \delta_{k^{(\text{p})}_y,0},
\end{equation}
%\end{widetext}
%
where $\kappa$ is the constant of proportionality such that $L_x = \kappa l_y$. Hence, the gauge factor may be explicitly expressed as a function of periodic gauge momentum in the $x$ direction. The $k_y$ dependence in $\hat{\psi}^{(\text{p})}$ comes solely from $\hat{\psi}^{(\text{L,p})}$. Consequently, the $k_y$ momentum in the periodic gauge equals the original $k_y$ momentum in the Landau gauge modulo $2\pi / l_y$, while the transformation on the $k_x$ dependence is nontrivial as ensues from Eq.~(\ref{eq:finalGalpha}).

\section{Scaling to the Continuum Limit at Fixed Flux Density}
\label{sec:fixed_n_phi}

To cross-validate our scaling to the effective continuum at fixed aspect ratio, we additionally perform scaling for select cases at fixed flux density, $n_{\phi}$.

In this procedure, we select a set of $q$ values approximately geometrically distributed with common ratio $2$, in the range $10\lesssim q \lesssim 10^3$. This provides a spread of $q$ values which reflects the distribution used in the scaling at fixed aspect ratio.\footnote{Furthermore, $q$ values with multiple factors are preferred during the selection process, and prime $q$ immediately rejected, so as to maximize the chances of approximately square configurations for comparison.} $q$ defines the number of sites in each MUC, $l_x l_y$, which we factorize into all distinct pairs of factors. For each $q$ value, we study $N$ values in the range $N_{\text{min}}\lesssim N \lesssim N_{\text{max}}$, where $N_{\text{min}}$ and $N_{\text{max}}$ are the minimum and maximum number of particles studied in the fixed aspect ratio scaling. Here $N/\nu$ defines the total number of MUCs in the system, $L_x L_y$, which we also factorize into all distinct pairs of factors. At this point, for each $q,N$ configuration, we select the $(l_x,l_y)$ and $(L_x,L_y)$ pairs so as to minimize the deviation from a square system, $\epsilon$. This minimization is performed only as a subsidiary constraint to improve the comparison with the fixed aspect ratio scaling in the bulk of the paper. In practice, $\epsilon$ may be as high as $50\%$ for this scaling procedure. 

\begin{figure}
        \centering
        \begin{subfigure}[b]{0.475\textwidth}
            \centering
\begin{tikzpicture}
\node at (0,0) {\includegraphics{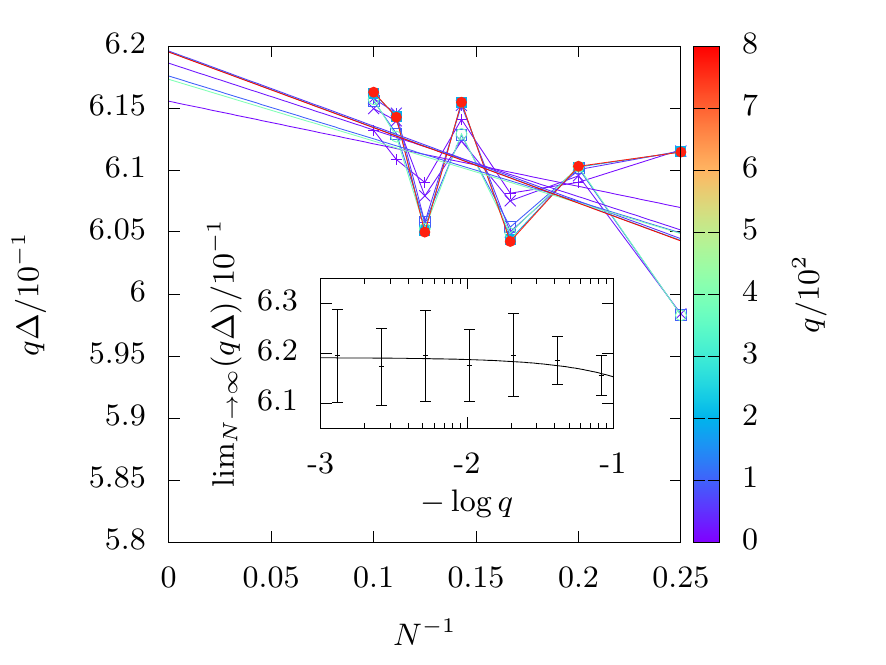}};
\node[overlay] at (-4.5,3) {(a)};
\phantomsubcaption
\label{fig:bosons_C1r1_individual_plot}
\end{tikzpicture}
        \end{subfigure}
        \hfill
        \begin{subfigure}[b]{0.475\textwidth}  
            \centering 
\begin{tikzpicture}
\node at (0,0) {\includegraphics{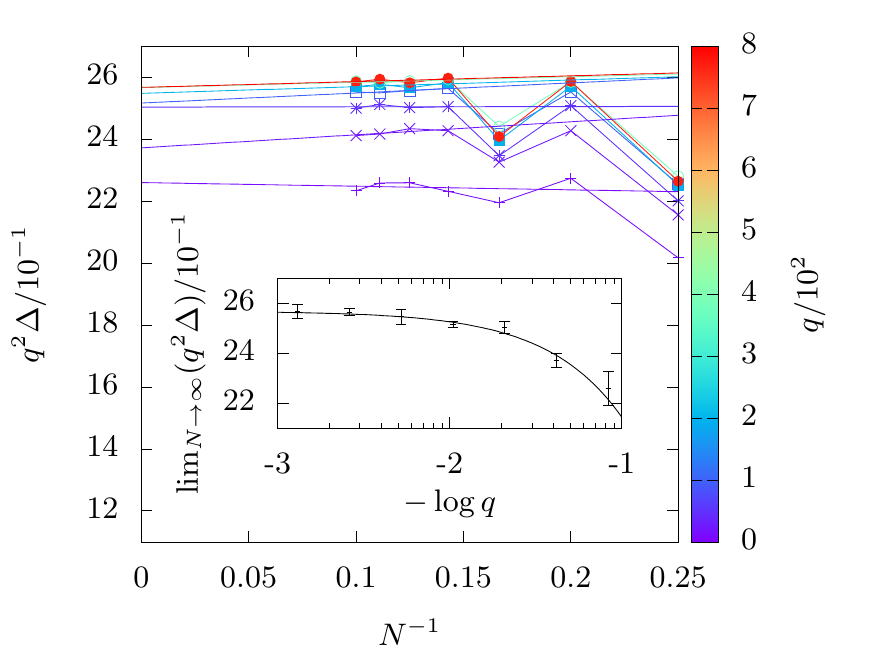}};
\node[overlay] at (-4.5,3) {(b)};
\phantomsubcaption
\label{fig:fermions_C1r1_individual_plot}
\end{tikzpicture}
        \end{subfigure}
        \caption{Finite-size scaling of the gap at fixed flux density. We show the finite-size gaps for flux densities $n_\phi=p/q$ with increasing values of denominator $q$ given by the color scale as a function of the inverse system size for (a) the bosonic $\nu=1/2$ states, and (b) the fermionic $\nu=1/3$ states, in the $|C|=1$ band. The extrapolations to the thermodynamic limit excludes outliers at small system sizes, i.e., excluding $N=4$ in panel (a) and $N=4,5,6$ in panel (b). The corresponding plots for the thermodynamic extrapolated gaps, for each finite $q$, are inset.}
        \label{fig:individual_plots}
    \end{figure}

\begin{figure*}
        \centering
        \begin{subfigure}[b]{0.475\textwidth}
            \centering
\begin{tikzpicture}
\node at (0,0) {\includegraphics{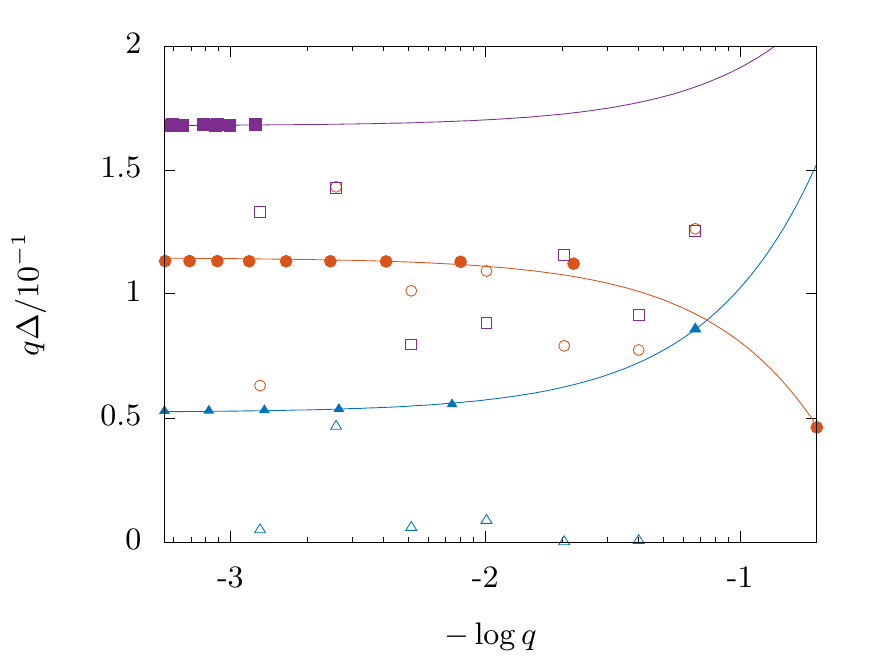}};
\node[overlay] at (-4.5,3) {(a)};
\phantomsubcaption
\label{fig:bosons_C2r-3N9}
\end{tikzpicture}
        \end{subfigure}
        \hfill
        \begin{subfigure}[b]{0.475\textwidth}  
            \centering 
\begin{tikzpicture}
\node at (0,0) {\includegraphics{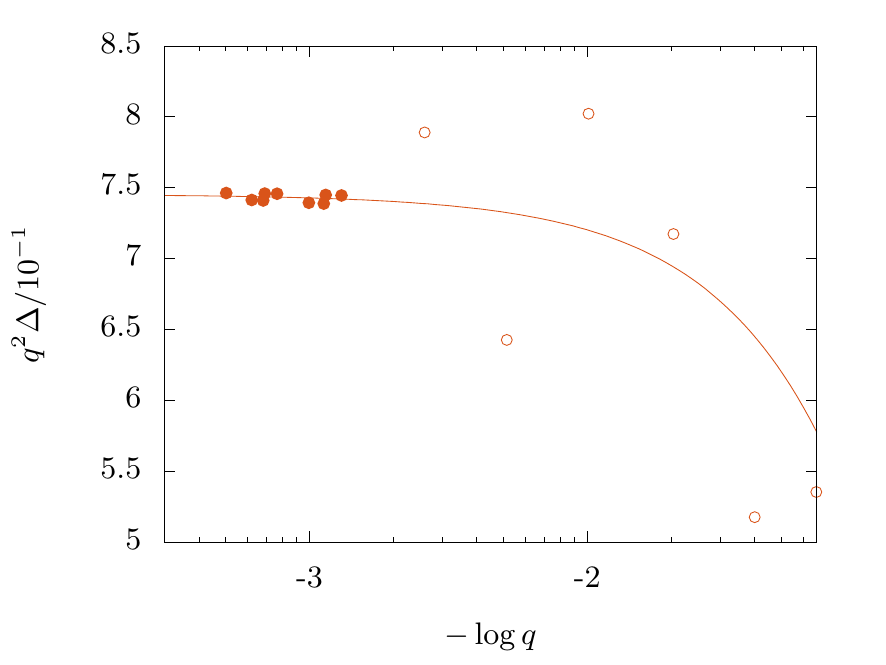}};
\node[overlay] at (-4.5,3) {(b)};
\phantomsubcaption
 \label{fig:fermions_C2r2N6}
\end{tikzpicture}
        \end{subfigure}
        \caption{Finite-size scaling of the gap in the $C=2$ band for (a) bosonic states, where squares, circles, and triangles denote states with $\{r=-3, N=9; r=2, N=8; r=3, N=6\}$, respectively; and (b) fermionic states with $\{r=-1, N=6\}$. The filled (hollow) symbols correspond to scaling with fixed aspect ratio (flux density). The linear trend-lines are shown for the scaling with fixed aspect ratio.} 
        \label{fig:fixed_comparison}
    \end{figure*}

To illustrate the mutual consistency of the scaling at fixed flux density ($\lim_{q,N\to \infty}$) and the scaling at fixed aspect ratio ($\lim_{N,q\to \infty}$), we provide data on the $r=1$ Laughlin states, for both bosons [Fig.~\ref{fig:bosons_C1r1_individual_plot}] and fermions [Fig.~\ref{fig:fermions_C1r1_individual_plot}] in a $|C|=1$ band. Here, we find extrapolated values of $\lim_{q,N\to \infty}(q\Delta)=0.62\pm(7.0\times 10^{-4})$ for bosons and $\lim_{q,N\to \infty}(q^{2}\Delta)=2.56\pm(7.2\times 10^{-3})$ for fermions, which is in close agreement with Sec.~\ref{subsec:FCI_C1}.
We also examine the scaling at finite flux density for more fragile states in higher Chern number bands. Generally in these cases, we find that the effective continuum limit at constant aspect ratio provides a much smoother extrapolation that minimizes finite-size effects. Examples are shown in Fig.~\ref{fig:fixed_comparison}.

\section{Derivation of the Correlation Function}
\label{sec:CorrelationDerivation}

The two-particle correlation function may be written as the expectation value of the density operator, $\rho$, of a particle at site $i$ with the density operator of a particle at site $j$:
\begin{equation}
\braket{\rho_i \rho_j}=\braket{c_i^\dagger c_i c_j^\dagger c_j},
\end{equation}
where $c^\dagger$, $c$ are the creation and annihilation operators, respectively. We may normal order the expression such that, for bosons or fermions, 
\begin{equation}
\braket{\normord{\rho_i \rho_j}}=\braket{c_i^\dagger c_j^\dagger c_j c_i + c_i^\dagger c_i \delta_{ij}}.
\end{equation}
From here, we substitute in the expression for the Fourier transform with respect to absolute position
\begin{equation}
c_{\mathbf{r}}=\frac{1}{\sqrt{N_{\text{c}}}} \sum_{n, \mathbf{k}} u_{n, \alpha} (\mathbf{k})e^{\mathrm{i}\mathbf{k}\cdot\mathbf{r}}c_{n,\mathbf{k}},
\end{equation}
where $N_\text{c}$ is the number of MUCs, $n$ is the band index, $\alpha$ is the sublattice index corresponding to position $\mathbf{r}$, and $\mathbf{k}$ is the momentum. This substitution yields
%
%\begin{widetext}
\begin{equation}
\begin{split}
\braket{\normord{\rho_i \rho_j}}&= \frac{1}{N_\text{c}^2} \sum_{\{n\},\{\mathbf{k}\}}u_{n_1,\alpha_i}^*(\mathbf{k}_1) u_{n_2,\alpha_j}^*(\mathbf{k}_2) u_{n_3,\alpha_j}(\mathbf{k}_3) u_{n_4,\alpha_i}(\mathbf{k}_4) e^{\mathrm{i}(-\mathbf{k}_1\cdot\mathbf{r}_i-\mathbf{k}_2\cdot\mathbf{r}_j + \mathbf{k}_3\cdot\mathbf{r}_j + \mathbf{k}_4\cdot\mathbf{r}_i)} \braket{c_{n_1,\mathbf{k}_1}^\dagger c_{n_2,\mathbf{k}_2}^\dagger c_{n_3,\mathbf{k}_3}c_{n_4,\mathbf{k}_4}} \\
&+\frac{1}{N_\text{c}} \sum_{\substack{n_1,n_4\\ \mathbf{k}_1,\mathbf{k}_4}} u_{n_1,\alpha_i}^* (\mathbf{k}_1) u_{n_4,\alpha_i}(\mathbf{k}_4) e^{\mathrm{i}(-\mathbf{k}_1\cdot\mathbf{r}_i+\mathbf{k}_4\cdot\mathbf{r}_i)}\braket{c_{n_1,\mathbf{k}_1}^\dagger c_{n_4,\mathbf{k}_4}}\delta_{ij}.
\end{split}
\end{equation}
%\end{widetext}
%
Introducing the single-particle wave function, $\phi_{n,\mathbf{k}}(\mathbf{r}_i)=u_{n,\alpha_i}(\mathbf{k})e^{\mathrm{i}\mathbf{k}\cdot\mathbf{r}_i}$, this expression reduces to
%
%\begin{widetext}
\begin{equation}
\begin{split}
\braket{\normord{\rho_i \rho_j}}&= \frac{1}{N_\text{c}^2} \sum_{\{n\},\{\mathbf{k}\}} \phi_{n_1,\mathbf{k}_1}^*(\mathbf{r}_i) \phi_{n_2,\mathbf{k}_2}^*(\mathbf{r}_j) \phi_{n_3,\mathbf{k}_3}(\mathbf{r}_j) \phi_{n_4,\mathbf{k}_4}(\mathbf{r}_i) \braket{c_{n_1,\mathbf{k}_1}^\dagger c_{n_2,\mathbf{k}_2}^\dagger c_{n_3,\mathbf{k}_3}c_{n_4,\mathbf{k}_4}} \\
&+\frac{1}{N_\text{c}} \sum_{\substack{n_1,n_4\\ \mathbf{k}_1,\mathbf{k}_4}} \phi_{n_1,\mathbf{k}_1}^*(\mathbf{r}_i) \phi_{n_4,\mathbf{k}_4}(\mathbf{r}_i) \underbrace{\braket{c_{n_1,\mathbf{k}_1}^\dagger c_{n_4,\mathbf{k}_4}}}_{\propto \delta_{\mathbf{k}_1,\mathbf{k}_4}}\delta_{ij}.
\end{split}
\end{equation}
%\end{widetext}
%
Because of the proportionality relation of the density expectation value, the last sum reduces to a sum over a single momentum.

\section{Accuracy of Correlation Functions}
\label{sec:CorrelationsAccuracy}
 
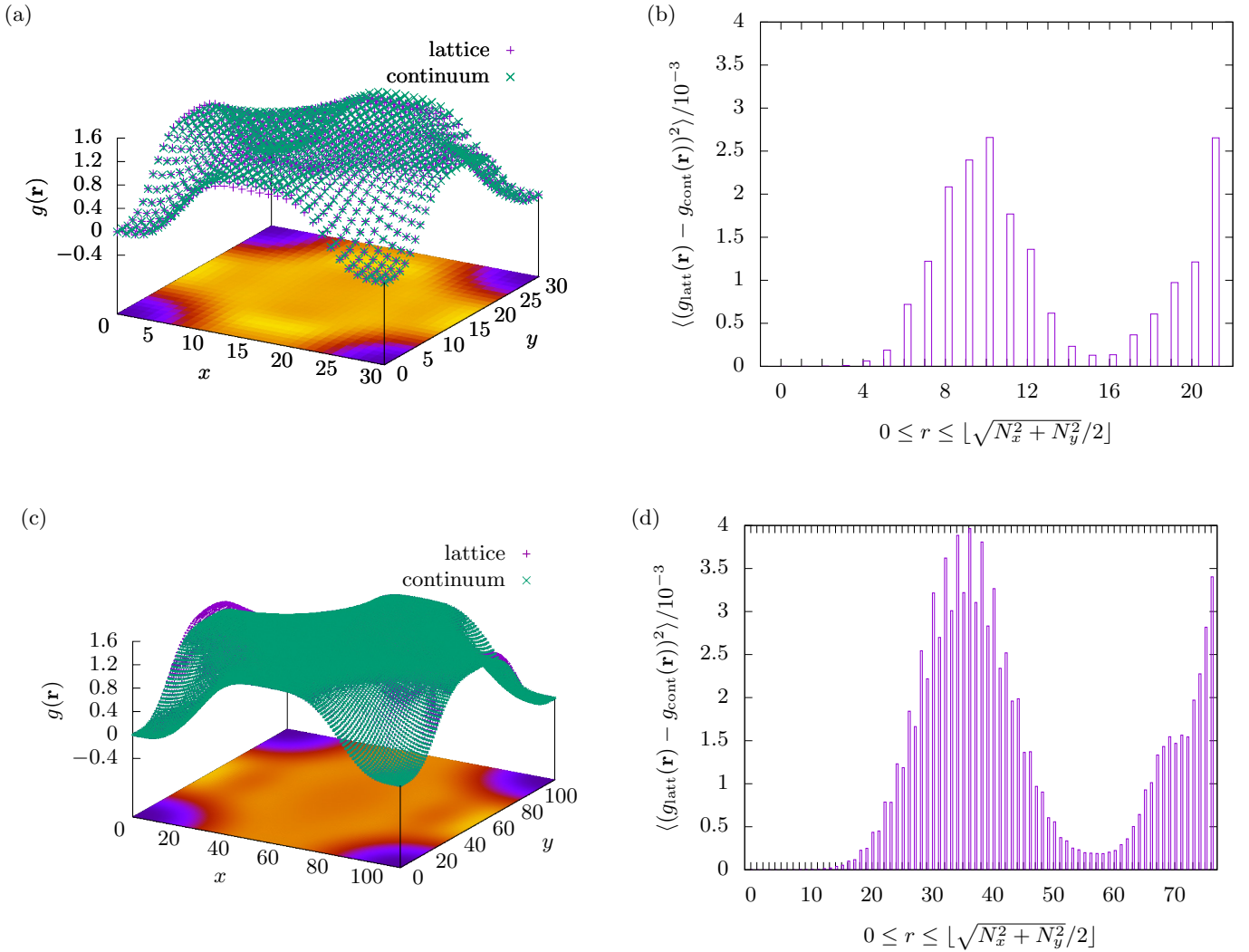
\begin{figure*}
        \centering
        \begin{subfigure}[b]{0.475\textwidth}
            \centering
\begin{tikzpicture}
\node at (0,0) {\includegraphics{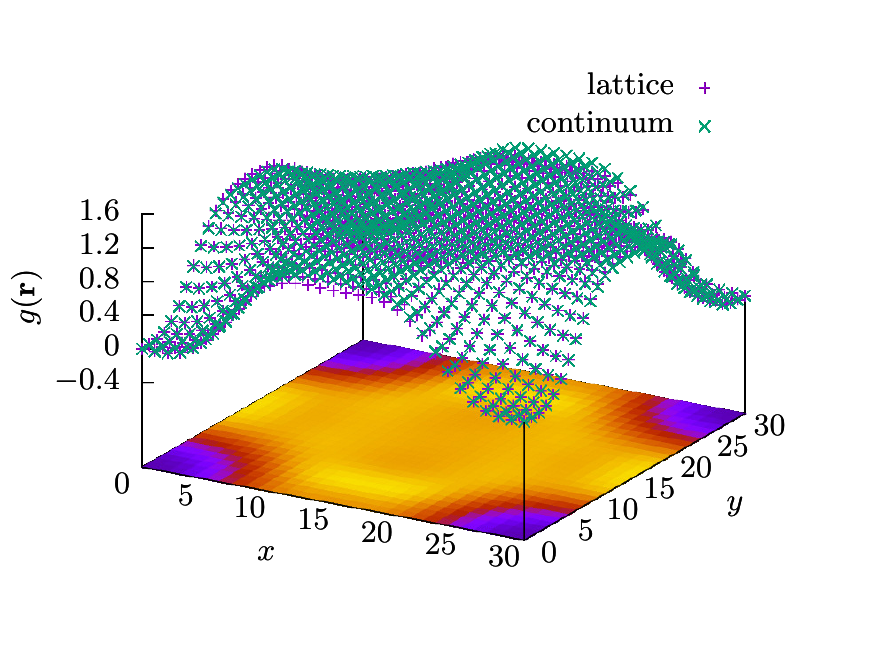}};
\node[overlay] at (-4.5,3) {(a)};
\phantomsubcaption
\label{fig:3d_q74}
\end{tikzpicture}
        \end{subfigure}
        \hfill
        \begin{subfigure}[b]{0.475\textwidth}  
            \centering 
\begin{tikzpicture}
\node at (0,0) {\includegraphics{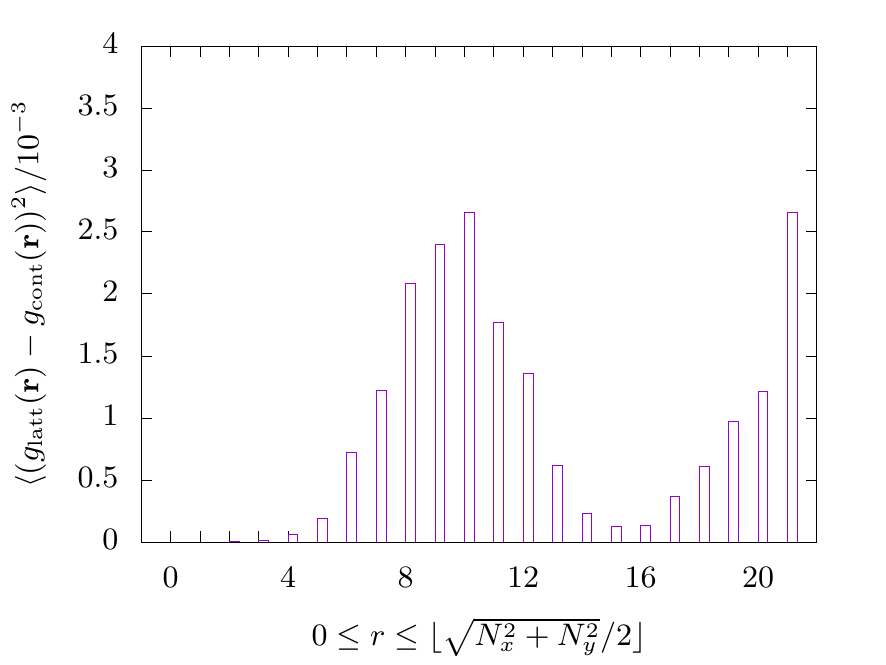}};
\node[overlay] at (-4.5,3) {(b)};
\phantomsubcaption
 \label{fig:hist_q74}
\end{tikzpicture}
        \end{subfigure}
        \vskip\baselineskip
        \begin{subfigure}[b]{0.475\textwidth}   
            \centering
\begin{tikzpicture}
\node at (0,0) {\includegraphics{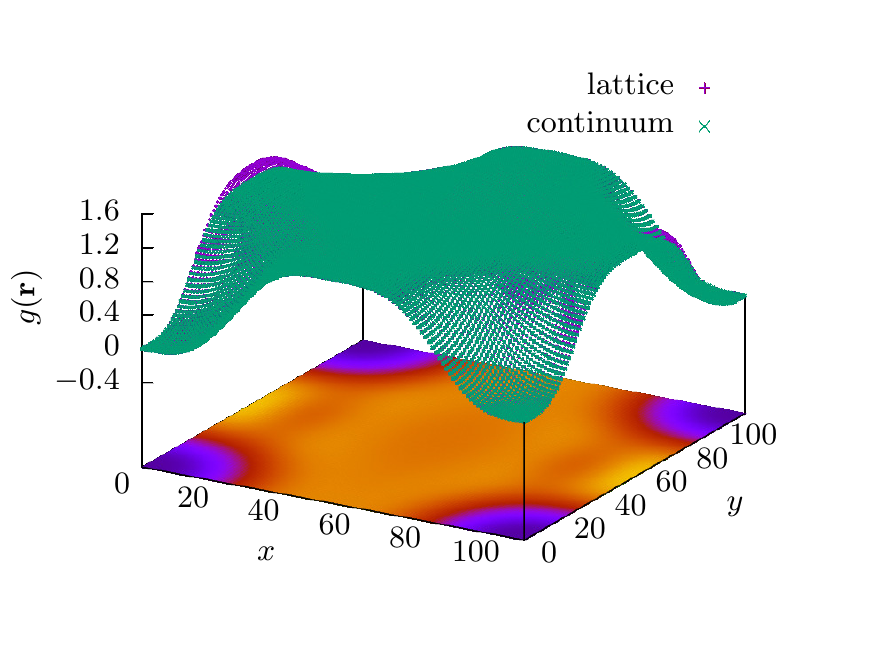}};
\node[overlay] at (-4.5,3) {(c)};
\phantomsubcaption
 \label{fig:3d_q971}
\end{tikzpicture}
        \end{subfigure}
        \quad
        \begin{subfigure}[b]{0.475\textwidth}   
            \centering 
\begin{tikzpicture}
\node at (0,0) {\includegraphics{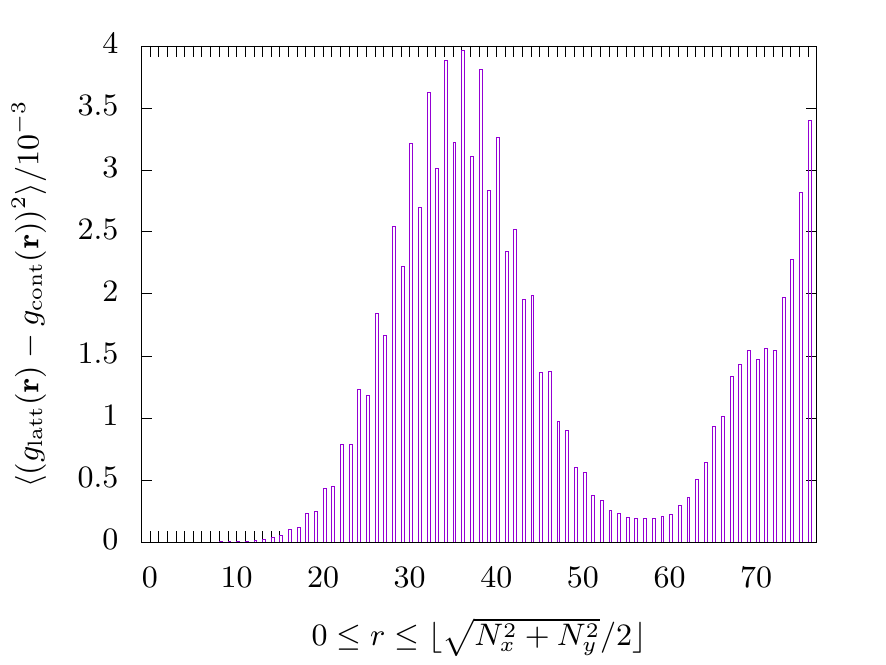}};
\node[overlay] at (-4.5,3) {(d)};
\phantomsubcaption
\label{fig:hist_971}
\end{tikzpicture}
        \end{subfigure}
        \caption{[(a), (c)] Two-particle correlation function for the bosonic six-particle $\nu=1/2$ (Laughlin) state in the $|C|=1$ band and $(k_x,k_y)=(0,0)$ momentum sector, with (a) $p=74$ and (c) $p=971$. The lattice result is additionally projected to the base, and the exact continuum solution is plotted for comparison. [(b), (d)] Variance between the continuum and lattice results with (b) $p=74$ and (d) $p=971$. The average is taken with respect to the points enclosed in origin-centric annuli of width $1.5a$, where $a$ is the lattice spacing.} 
        \label{fig:boson_C1r1N6_correlations}
    \end{figure*}
In order to verify the accuracy of the density-density correlation functions used in this paper, we compare the correlation for the robust six-particle $\nu=1/2$ state with the exact continuum result for a torus, shown in Fig.~\ref{fig:boson_C1r1N6_correlations}. The derivation of the lattice correlation function in terms of single-particle eigenstates is shown in Appendix~\ref{sec:CorrelationDerivation}, and the exact form of the correlation function on the continuum torus is discussed in many sources, for example, by Yoshioka \emph{et al.}~\cite{Yoshioka:1984ev}. Note the slight deviation of the lattice result from the exact solution. Figures~\ref{fig:3d_q74} and~\ref{fig:3d_q971} show plots of the density-density correlation function with $p=71$ and $p=971$, respectively, whereas Figs.~\ref{fig:hist_q74} and~\ref{fig:hist_971} show the corresponding variance between the continuum and lattice results. We observe an agreement at the zero-separation correlation hole which oscillates with distance [note the small scale of the variance in Figs.~\ref{fig:hist_q74} \&~\ref{fig:hist_971}]. Because of the lack of scaling with radius, this discrepancy is attributed to computational imprecision of the single-particle eigenvectors, which we obtain with standard diagonalization routines of the \textsc{LAPACK} library.
 \begin{figure*}
        \centering
        \begin{subfigure}[b]{0.475\textwidth}
            \centering
\begin{tikzpicture}
\node at (0,0) {\centering\includegraphics{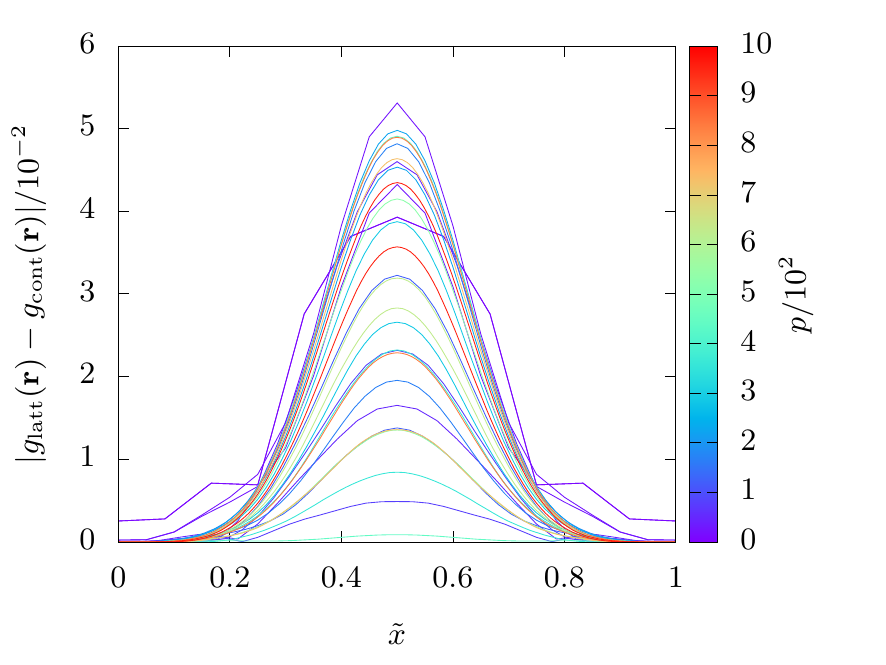}};
\node[overlay] at (-4.5,3.5) {(a)};
\phantomsubcaption
 \label{fig:sym_plot_x_1}
\end{tikzpicture}
        \end{subfigure}
        \hfill
        \begin{subfigure}[b]{0.475\textwidth}  
            \centering 
\begin{tikzpicture}
\node at (0,0) {\centering\includegraphics{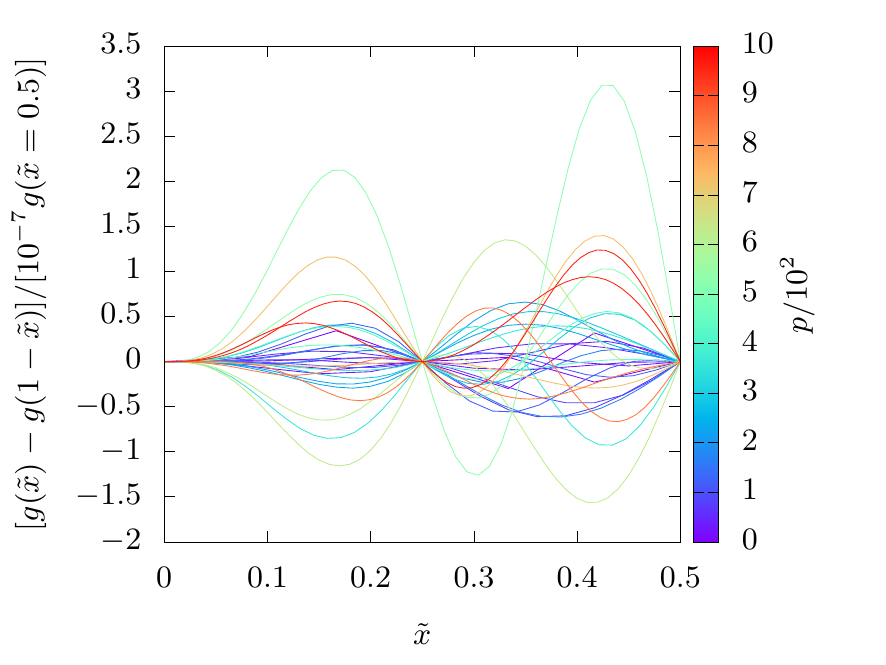}};
\node[overlay] at (-4.5,3.5) {(b)};
\phantomsubcaption
 \label{fig:sym_plot_x_2}
\end{tikzpicture}
        \end{subfigure}
\caption{(a) Absolute deviation from the exact continuum torus solution and (b) asymmetry in the two-particle correlation function for the bosonic eight-particle $\nu=1/2$ (Laughlin) state in the $|C|=1$ band and $(k_x,k_y)=(0,0)$ momentum sector, with $L_x=L_y=4$. A $y=0$ cross section is rescaled such that $\tilde{x}=x/N_x$.}
\label{fig:sym_plot_x}
\end{figure*}

The same analysis is performed for the robust eight-particle state in the same Laughlin series. The asymmetry of the lattice results as well as their deviation from the continuum is shown in Fig.~\ref{fig:sym_plot_x}. Since the continuum torus correlation function is symmetric by construction, we confirm that the lattice results obey the fundamental symmetry also, up to the scale of Fig.~\ref{fig:sym_plot_x_1}. In Fig.~\ref{fig:sym_plot_x_2}, we explicitly plot the asymmetry in the lattice results (note the small scale of the plot). The fact that we do not find a monotonic behavior of the asymmetry with $p$, supporting the view that the deviations are due to the numerical accuracy of the single-particle eigenstates.

\section{Error Analysis}
\label{sec:error_analysis}

In order to obtain the thermodynamic (effective) continuum limit, we linearly extrapolated the data for the $q^{(2)} \Delta$ vs $1/q$ and $q^{(2)} \Delta$ vs $1/N$ plots. To determine this scaling, we rejected low-$q$/-$N$ outliers and focused only on high-$q$/-$N$ data points, since they are closer to the mode and also the limit of the distribution. As illustrated in the above discussion, occasionally data for the limit is not precise. We define error bars relative to the linear trendline. For most cases, this is the asymptotic standard error for the $y$-intercept fit parameter of a standard linear regression in $N^{-1}$. However, for $|C|>1$ in the $q\to\infty$ limit, the error bars were read off on a case-by-case basis, by inspection, since they were often asymmetric and larger than the asymptotic standard error estimate.

\twocolumngrid

\bibliographystyle{apsrev4-1}
\bibliography{Hofstadter}

\end{document}